\documentclass[10pt]{amsart}
\usepackage{graphicx,color}
\newtheorem{theorem}{Theorem}[section]
\newtheorem{lemma}[theorem]{Lemma}

\theoremstyle{definition}
\newtheorem{definition}[theorem]{Definition}

\theoremstyle{remark}

\numberwithin{equation}{section}

\usepackage{color}


\title[Inverse scattering on the quantum graph]{Inverse scattering on the quantum graph \\
 --- Edge model for graphene}

\author{Kazunori ANDO}
\address[K. Ando]{Department of Electrical and Electronic Engineering and Computer Science, Ehime University, Matsuyama, 790-8577, Japan}
\email{ando@cs.ehime-u.ac.jp}
\author{Hiroshi ISOZAKI}
\address[H. Isozaki]{Graduate School of Pure and Applied Sciences, Professor Emeritus,
University of Tsukuba, 
Tsukuba, 305-8571, Japan}
\email{isozakih@math.tsukuba.ac.jp}
\author{Evgeny KOROTYAEV}
\address[E. Korotyaev]{
Department of  Math. Analysis,
Saint-Petersburg State University,  Universitetskaya nab. 7/9, St. Petersburg, 199034, Russia
}
\email{e.korotyaev@spbu.ru}
\author{Hisashi MORIOKA}
\address[H. Morioka]{Department of Electrical and Electronic Engineering and Computer Science, Ehime University, Matsuyama, 790-8577, Japan}
\email{morioka@cs.ehime-u.ac.jp}

\date{\today}

\begin{document}
\maketitle

\begin{abstract}
We consider the inverse scattering on the quantum graph associated with the hexagonal lattice. Assuming that the potentials on the edges are compactly supported, we show that the S-matrix for all energies in any open set in the continuous spectrum determines the potentials.
\end{abstract}

\section{Introduction}
\subsection{Assumptions and main results}
In this article, we study the spectral and inverse scattering theory associated with quantum graph, which is by definition a graph endowed with metric on its edges and equipped with differential operators on them. The quantum graph was introduced in 1930s as a simple model  for free electrons in organic molecules \cite{Paul36}, and   its role has been  increasing in physics, chemistry and engineering with particular interest in material science. 
A physical example we have in mind in this paper is the graphene, for which there are two mathematical models, both being based on the periodic hexagonal lattice. One model considers the propagation of waves only on the vertices, and deals with the discrete Laplacian on the vertex set, while the other  focuses on the propagation of waves generated by Schr{\"o}dinger operators defined on each edge and scattered by vertices and potentials. 
This latter  is the quantum graph, and the former is often called the discrete graph.
In this paper, we call the former the {\it vertex model} and the latter the {\it edge model}. 

The edge model thus deals with a family of one-dimensional Schr{\"o}dinger operators
$$
- \frac{d^2}{dz^2} + q_{\bf e}(z)
$$
defined on the edges of the hexagonal lattice assuming the Kirchhoff condition on the vertices. Here, $z$ varies over the interval $(0,1)$ and ${\bf e} \in \mathcal E$, $\mathcal E$ being the set of all edges of the hexagonal lattice.  
The following assumptions are imposed on the potentials.

\medskip
\noindent
{\bf (Q-1)} \ \ \ \ \ \ {\it $q_{\bf e}(z)$ is real-valued, and $q_{\bf e} \in L^2(0,1)$}.

\medskip
\noindent
{\bf (Q-2)} \ \ \ \ \ \   {\it $q_{\bf e}(z) = 0$ on $(0,1)$ except for a finite  number of edges.}

\medskip
\noindent
{\bf (Q-3)} \ \ \ \ \ \  $q_{\bf e}(z) = q_{\bf e}(1-z)$ for $z \in (0,1)$.

\medskip
Under these assumptions, the Schr{\"o}dinger operator 
$$
\widehat H_{\mathcal E} = \Big\{
- \frac{d^2}{dz^2} + q_{\bf e}(z)\, ; \, {\bf e} \in \mathcal E 
\Big\}
$$
is self-adjoint and its essential spectrum $\sigma_e(\widehat H_{\mathcal E})$ is equal to $[0,\infty)$. 
We are interested in the spectral properties of this operator from two-sided view points, the forward and the inverse problems. We start from the forward problem. We show that  there exists a discrete (but infinte) subset $\mathcal T \subset {\bf R}$ such that $\sigma_e(\widehat H_{\mathcal E})\setminus \mathcal T$ is absolutely continuous. We construct a complete family of generalized eigenfunctions  describing the continuous spectrum, and represent Heisenberg's S-matrix.   
Based on these results on the forward problem, we turn to the inverse problem. The following two theorems are our main aim.

\begin{theorem}
\label{Maintheorem1}
Assume (Q-1), (Q-2) and (Q-3). Then,
given any open interval $I \subset (0,\infty)\setminus \mathcal T$, and the S-matrix $S(\lambda)$ for all $\lambda \in I$, one can uniquely reconstruct the potential $q_{\bf e}(z)$ for all ${\bf e} \in \mathcal E$.
\end{theorem}

Under our assumptions (Q-1), (Q-2), (Q-3), the S-matrix is 
meromorphic in the complex domain $\{{\rm Re}\,\lambda > 0\}$ with possible branch points at $\mathcal T$. Therefore, the assumption of Theorem \ref{Maintheorem1} is equivalent to the condition that we are given the S-matrix for all energies in the continuous spectrum except for the set of exceptional points  $\mathcal T$.   

One can also deal with a perturbation of periodic edge potentials.

\begin{theorem}
\label{Maintheorem2}
Assume (Q-1) and (Q-3). 
Assume that there exists a real $q_0(z) \in L^2(0,1)$ satisfying $q_0(z) = q_0(1-z)$ and that $q_{\bf e}(z) = q_0(z)$ on $(0,1)$ except for a finite number of edges ${\bf e} \in \mathcal E$. Given an open interval $I \subset \sigma_e(\widehat H_{\mathcal E}) \setminus \mathcal T$ and the S-matrix $S(\lambda)$ for all $\lambda \in I$, one can uniquely reconstruct the potential $q_{\bf e}(z)$ for all edges ${\bf e}\in \mathcal E$.
\end{theorem}

Note that under the assumption of Theorem \ref{Maintheorem2}, $\sigma_e(\widehat H_{\mathcal E})$ is a union of intervals with possible gaps 
between them.

Our proof gives not only the identification but also the reconstruction procedure of the potential, although some parts are transcendental.
Our results are not restricted to the hexagonal lattice. Theorems \ref{Maintheorem1} and \ref{Maintheorem2} also hold for the square lattice and the triangular lattice. In fact,  the forward problem part, i.e. the study of the spectral structure of the quantum graph encompasses a larger class of quantum graphs. The method for solving the inverse problem, however,  leans over geometric features of the graph, hence should be discussed individually for the graph in question.

\subsection{Basic strategy}
There is a close similarity between  Schr{\"o}dinger operators in the continuous model and those in the discrete model or in the quantum graph. Let us  explain it 
by reviewing the basic strategy of the stationary scattering theory. 

For the Schr{\"o}dinger operator $- \Delta + V(x)$ in ${\bf R}^d$, 
where $V(x)$ decays sufficiently rapidly at infinity, there exists a generalized eigenfunction
$\varphi(x,\xi)$, $x, \xi \in {\bf R}^d$, satisfying the equation
$$
( - \Delta + V(x) - |\xi|^2)\varphi(x,\xi) = 0
$$
having the following behavior at infinity
\begin{equation}
\varphi(x,\xi) = e^{ix\cdot\xi} + \frac{e^{ikr}}{r^{(d-1)/2}}
a(k,\theta,\omega) + o(r^{-(d-1)/2}), \quad r = |x| \to \infty,
\label{Eigenfunxtionforthecontinuousmodel}
\end{equation}
where $k = |\xi|, \theta = x/r, \omega = \xi/k$. This family of generalized eigenfunctions $\big\{\varphi(x,\xi) ;$ $\xi \in {\bf R}^d\big\}$ 
defines a generalized Fourier transform
$$
\big(\mathcal F f\big)(\xi) = (2\pi)^{-{d/2}}\int_{{\bf R}^d}
\overline{\varphi(x,\xi)}f(x)dx,
$$
which is a unitary operator from the absolutely continuous subspace for $- \Delta + V(x)$ to $L^2({\bf R}^d)$ diagonalizing $- \Delta + V(x)$. Heisenberg's {\it S-matrix} is defined to be $S(k) = I + C_0(k)A(k)$, where $C_0(k)$ is a suitable constant and $A(k)$, called the {\it scattering amplitude}, is an integral operator  on $L^2(S^{d-1})$ with kernel $a(k,\theta,\omega)$. This S-matrix is unitary  for each $k > 0$ 
and is believed to contain all the information of the physical system in question. 
Note that $\omega, \theta$ in (\ref{Eigenfunxtionforthecontinuousmodel}) correspond to the incoming and outgoing directions of scattered particles, and, by passing to the Fourier transformation, $kS^{d-1} = \{k\omega \, ; \, \omega \in S^{d-1}\}$ is the characteristic surface of  the unperturbed operator $- \Delta - k^2$. This surface should be called the manifold at infinity, since it parametrizes the vectors representing the incoming and outgoing scattering states.

 One can draw the same picture for discrete Schr{\"o}dinger operators (in the vertex model) on perturbed periodic lattices of rank $d$. In this case, by the Floquet (or Floquet-Bloch) theory, the manifold at infinity is the characteristic surface of the discrete Laplacian (difference operator) and is called the {\it Fermi surface},
$M_{\mathcal V,\lambda}$,
which is a submanifold of codimension 1 in the torus ${\bf T}^d$. As for the quantum graph (the edge model), the action of ${\bf Z}^d$ transforms the vertex set $\mathcal V$ into the torus ${\bf T}^d$ as in the case of vertex model,  and, in addition,  the edge set into the fundamental graph $\mathcal E_{\ast}$, which is a graph on the torus ${\bf T}^d$ :
$$
{\bf Z}^d : \mathcal V \to {\bf T}^d, \quad \mathcal E \to \mathcal E_{\ast}.
$$
Therefore, in the case of  quantum graph, the underlying Hilbert space is unitarily equivalent to $L^2({\bf T}^d\times \mathcal E_{\ast})$. 
The notion of the characteristic surface is not obvious for the case of the quantum graph. However, the Kirchhoff condition induces the  Laplacian on the set of vertices, which is the source of the continuous spectrum of $
\widehat H_{\mathcal E}$, and the Laplacian $- (d/dz)^2 + q_{\bf e}(z)$ on each edge ${\bf e}$ gives rise to the eigenvalues embedded in the continuous spectrum. Observing the resolvent of the free Hamiltonian, i.e. the one for the case $q_{\bf e}=0$ for all ${\bf e} \in \mathcal E$ 
(cf. Lemma \ref{S3RElambda+i0Formula1} below), 
we see that the manifold at infinity for  the Schr{\"o}dinger operator   on the quantum graph is $M_{\mathcal V,-\cos\sqrt{\lambda}}$, 
i.e. the Fermi surface of the vertex Laplacian. The key of our arguments lies in the fact that the continuous spectrum of the edge model is determined by that of the vertex model. The stationary scattering theory for the edge model is thus developed as in the case of the vertex model, which in turn can be discussed in a way parallel to the continuous model.

The procedure of the inverse scattering is as follows.
Since the perturbation is compactly supported, the S-matrix for the quantum graph is shown to be equivalent to the Dirichlet-to-Neumann map (D-N map in short) for an interior boundary value problem in the lattice space. The inverse problem of scattering is then reduced to the inverse boundary value problem for vertex Schr{\"o}dinger operator in a bounded domain. Applying ideas  developed for the inverse boundary value problems for the continuous model as well as the discrete model, from the knowledge of the D-N map, one can recover the Dirichlet eigenvalues for Schr{\"o}dinger operators on each edge. 
By the classical result of Borg \cite{Borg} (see also \cite{Lev49}), which is the starting point of the 1-dimensional inverse spectral theory, one can recover the edge potentials from the knowledge of the S-matrix. 

In the course of the proof, we also prove that the knowledge of the D-N map for the edge model is equivalent to the knowledge of the D-N map for the vertex model (Lemma \ref{LemmaDNedge=DNvertex}). 
This is significant, because it enables us to study the lattice structure from the S-matrix of the edge model (see Theorem \ref{LemmaAlambdatoBSigmalambda}).
Suppose we are given an edge model on the hexagonal lattice without perturbation. We deform its compact part arbitrarily as a planar graph, and add $L^2$ potentials on edges for this compact part. Take an interior domain which contains all the perturbations. One can then  develop the spectral theory for this perturbed edge model in the same way as in this paper. In particular, the S-matrix of any fixed energy for the whole space determines the D-N map for the finite domain of the graph as a vertex model. 
One can then apply the results in \cite{AnIsoMo17(1)} to this finite part, e.g. 
\begin{itemize}
\item Reconstruction of the potential on vertices.
\item Location of defects of the lattice.
\item Reconstruction as a planar graph, if one knows all the S-matrices of the edge model near the 0- energy.
\end{itemize}

\subsection{Related works}
Because of its theoretical as well as practical  importance, plenty of works have been presented for Schr{\"o}dinger operators on the graph. 
The monographs \cite{CDGT88}, \cite{CDS95}, \cite{Ch97}, \cite{Post12}, \cite{BK13} are expositions of the graph spectra and related problems from algebraic, geometric,  physical  and functional analytic view points with slight different emphasis on them.  
The present situation of the study of quantum graph is well explained in the above mentioned books, especially in Chapter 7 of \cite{BK13} together with an abundance of references therein. We note, in particular,  that the relation between the vertex model and the edge model is a key step to understand the spectral structure of the quantum graph and studied by \cite{Below85}, \cite{C97}, \cite{KuchPost},  \cite{Pank06}, \cite{BPG08}. See also \cite{KoLo07}, \cite{Niikuni16} for more recent results. As for the inverse problem for the quantum graph, the issue has been centered around the compact graph, the tree and the scattering problem by the infinite edges (leads). See e.g. \cite{Below85}, 
\cite{KostSchr99}, \cite{Pivo00}, \cite{GutSmil01}, \cite{Bel04}, \cite{BroWei09}, \cite{Yurk05}, \cite{Kura08}, \cite{ChExTu11}, \cite{AvdBelMat11}, \cite{VisComMirSor11}, \cite{MochiTroosh12} and other interesting papers cited in the above books. We must also mention the works \cite{Col98}, \cite{CurtMor00} on the planar discrete graph, which solved the inverse problem by giving   the characterization of the D-N map of the associated boundary value problem and the reconstruction procedure starting from the D-N map.

Let us mention recent developments in the study of inverse problems for perturbed periodic systems dealing with vertex models and edge models.
In \cite{KoSa15}, stationary problems and the existence and completeness of wave operators for the edge model were  proved. For more detailed spectral properties, see \cite{KoSa15b}, \cite{KoSa15c}. The long-range scattering for the perturbed periodic lattice is discussed in \cite{Nakamura14}, \cite{Tadano16} and \cite{ParRich18}.
As for the inverse problem of the perturbed periodic lattices, 
\cite{IsKo12} showed that for the square lattice, compactly supported potentials are determined by the S-matrix of all energies. 
The case of hexagonal lattice was proved by \cite{Ando12}. Inverse scattering at a fixed energy was studied by \cite{IsMo} for the square lattice case. 
Spectral properties and inverse problems for more general vertex models were studied in \cite{AnIsoMo17} and \cite{AnIsoMo17(1)}, where it is shown that the S-matrix of the vertex model with one fixed energy determines compactly supported potentials as well as the convex hull of  defects of the lattice, in the case of hexagonal lattice. Also, for the general perturbation of the finite part of the graph, the knowledge of the S-matrix with energies near the end points of the spectrum determines the graph structure as a planar graph. 
In \cite{AnIsoMo17(1)}, to prove these facts, the works \cite{Col98}, \cite{CurtMor00} played an essential role. The present paper is based on  the results in \cite{AnIsoMo17} and \cite{AnIsoMo17(1)} as well as \cite{KoSa15}. As regards the quantum graph, in the recent work \cite{EKMN17}, spectral properties of quantum graphs are pursued on the structure without assuming the equal length property on edges, in particular, allowing $\inf_{{\bf e} \in \mathcal E}|{\bf e}| =0$. 
 We restrict ourselves to the case of periodic lattices perturbed by edge potentials and pursue detailed spectral properties.
In \cite{KoSa15}, the eigenvectors and the generalized eigenfunctions for the free Hamiltonian are computed explicitly. In this paper, we adopt a more operator theoretical approach. 
The quantum graph is now a rapidly growing area in mathematical physics and material science. However, compared to the case of trees for instance, not so much is known about the perturbed periodic lattice and the corresponding quantum graph, especially on the inverse scattering. Even restricted to the forward problem, the results in this paper are new. 

\subsection{Plan of the paper}
Spectral properties of the metric graph can be studied in a rather general framework. In \S 2, we introduce the periodic graph and the associated vertex model and edge model as well as Laplacians on them. The fundamental graph $\mathcal E_{\ast}$ is also introduced. The basic assumptions $(P)$, $(E)$,  $(A$-$1)$ $\sim$ $(A$-$4)$ are given there. The main ingredients are the Kirchhoff condition $(K)$, the discrete Fourier transforms $\mathcal U_{\mathcal V}, \mathcal U_{\mathcal E}$, function spaces $\mathcal B$, $\mathcal B^{\ast}$, which are used throughout the subsequent arguments. 
In \S3, we derive  representations of the resolvent of the perturbed edge Hamiltonian and the spectral measure of the free edge Hamiltonian. 
The key formula is (\ref{Resolventformula2}), which describes the resolvent of the edge Laplacian in terms of the resolvent of the vertex Laplacian. We are mainly concerned with the formal relation between the vertex Laplacian and the edge Laplacian in this section. We prove the  detailed estimates  in \S 4. We proceed to derive the spectral representation of the edge Hamiltonian and the S-matrix in \S 5. We discuss the exterior and interior boundary value problems for the edge Laplacian in \S 6, and show that the S-matrix for the whole space determines uniquely the Dirichlet-to-Neumann map for the interior domain. Although our arguments are restricted to the hexagonal lattice in this section, we can actually consider in the general framework for the forward  scattering theory on the metric graph until the end of \S \ref{SectionSmatrixtoDNmap}. In  \S \ref{SectionInversescattering}, we  study the inverse scattering for the hexagonal lattice and prove the main theorems.
In this paper, we tried to develop a general theory, especially for the forward problem, and included basis facts well-known to exparts. 
For a shorter version, restricted to the case of hexagonal lattices, see \cite{AnIsKorMo20}.

\subsection{Notation}
The notation used in this paper is standard. 
Let  ${\bf Z}$ be the set of all integers, and  ${\bf N} = \{1, 2, 3, \cdots\}$  the set of natural numbers.  
For Banach spaces $X$ and $Y$, ${\bf B}(X;Y)$ denotes the set of all bounded operators from $X$ to $Y$. For an operator $A$, $D(A)$ denotes the domain of $A$. 
For a self-adjoint operator $A$, let $\sigma(A)$, $\sigma_d(A)$, $\sigma_e(A)$, $\sigma_p(A)$, $\sigma_{ac}(A)$ and $\rho(A)$ be the spectrum, discrete spectrum, essential spectrum, point spectrum, absolutely continuous spectrum and the resolvent set of $A$, respectively. Moreover, ${\mathcal H}_{ac}(A)$ and $\mathcal H_{pp}(A)$ denote the absolutely continuous subspace for $A$ and the closure of the linear hull of the eigenvectors of $A$, respectively.
For an interval $I \subset {\bf R}$ and a Hilbert space $\bf h$, let $L^2(I,{\bf h};\rho(\lambda)d\lambda)$ be the set of ${\bf h}$-valued $L^2$-functions on $I$ with respect to the measure $\rho(\lambda)d\lambda$. 
For a vertex set $\mathcal V$, let $\ell^2(\mathcal V)$ be the set of all square summable sequences on $\mathcal V$, and $\ell^2_{loc}(\mathcal V)$ the set of all sequences on $\mathcal V$. Likewise, for an edge set $\mathcal E$, let $L^2(\mathcal E)$ be the set of all square integrable functions on $\mathcal E$, and $L^2_{loc}(\mathcal E)$ the set of functions which are square summable on any finite subset in $\mathcal E$. Sobolev spaces $H^m(\mathcal E)$ and $H^m_{loc}(\mathcal E)$ are defined similarly. We often call a function defined on  edges as {\it edge function} and the one defined on  vertices as {\it vertex function}.  Similar expressions are also used when we contrast the vertex model and the edge model. 

\subsection{Acknowledgement}
K. A. is supported by Grant-in-Aid for Scientific Reserach (C) 17K05303, Japan Society for the Promotion of Science (JSPS).
H. I.  is supported by Grant-in-Aid for Scientific Reserach (S) 15H05740, (B) 16H0394, JSPS. 
 E. K. is supported by the RSF grant No. 18-11-00032. 
H. M. is supported by Grant-in-Aid for Young Scientists (B) 16K17630, JSPS. 
The authors express their gratitude to these supports.

\section{Basic theory for the quantum graph}

\subsection{Periodic graph}\label{SubsectionPeriodicgraph}
We consider a {\it periodic graph} $\Gamma = \{\mathcal V, \mathcal E\}$ in ${\bf R}^d$, where $d \geq 2$ is an integer, and  $\mathcal V$, $\mathcal E$ are a vertex set and an edge set defined as follows. Let 
$\mathcal L$ be a lattice of rank $d$ in ${\bf R}^d$ with basis ${\bf v}_j, j= 1,\cdots,d$, i.e. 
\begin{equation}
\mathcal L = \big\{{\bf v}(n)\, ; \, n \in 
{\bf Z}^d\big\}, \quad
{\bf v}(n) = \sum_{j=1}^dn_j{\bf v}_j, \quad n =(n_1,\cdots,n_d) \in {\bf Z}^d,
\label{S2DefineMathcalL}
\end{equation}
and  put 
\begin{equation}
\Gamma_{\ast} = \{{\bf v}(x)\, ; \, x \in [0,1)^d\} , \quad {\bf v}(x) = \sum_{j=1}^d x_j{\bf v}_j . 
\label{S2DefineGammaastbyv}
\end{equation}
Given points $p^{(j)} \in {\bf R}^d$, $j = 1,\cdots,s$, we define the vertex set 
$\mathcal V$ by
\begin{equation}
\mathcal V = {\mathop\cup_{j=1}^s}\big(p^{(j)} + \mathcal L\big).
\label{V=cupj=1sp(j)+mathcalL}
\end{equation} 
Here for $s = 1$, we assume that  $p^{(1)} = 0$, and for $s \geq 2$,

\medskip
\noindent
${\bf ( P)}$ \ \ \ \
$p^{(j)} \in \Gamma_{\ast}, \quad
p^{(i)}- p^{(j)} \not\in \mathcal L, \quad {\rm if}\quad i\neq j,$

\medskip
\noindent
There exists a bijection $\mathcal V\ni v \to (j(v),n(v)) 
\in \{1,\cdots,s\}\times{\bf Z}^d$ such that
\begin{equation}
v = p^{(j(v))} + {\bf v}(n(v)).
\label{S1vandn}
\end{equation}
The group ${\bf Z}^d$ acts on ${\mathcal V}$ as follows :
\begin{equation}
{\bf Z}^d\times{\mathcal V} \ni (m,v) \to m\cdot v := p^{(j(v))}+{\bf v}(m+n(v)) \in {\mathcal V}.
\label{DefineZdaction}
\end{equation}
An edge ${\bf e}$ is a segment in ${\bf R}^d$ whose two end points are in ${\mathcal V}$. The edge set $\mathcal E$ is the set of all ends.
For $v, w\in \mathcal V$, $v \sim w$ means that they are the mutually opposite end points of a same edge. 
They are assumed to satisfy
\begin{equation}
v \sim w \Longrightarrow  
m\cdot v \sim m\cdot w, 
\quad \forall m \in {\bf Z}^d.
\nonumber
\end{equation}
For $v \in \mathcal V$, the degree of $v$ is the number of edges whose one end point is $v$, and is denoted by ${\rm deg}(v)$. 
Then ${\rm deg}\,(p^{(j)} + {\bf v}(n))$ depends only on $j$, and is denoted by ${\rm deg}\, (j)$.
In this paper, we deal with the case in which ${\rm deg}\,(j)$ is independent of $j$. This is for the notational simplicity, and is not an essential restricition. In what follows, ${\rm deg}\,(v)$ and ${\rm deg}\,(j)$ are the same constant which we denote by $d_{\mathcal V}$:
\begin{equation}
{\rm deg}\, (v) = {\rm deg}\, (j) = d_{\mathcal V}.
\label{Definedg}
\end{equation}

Let $\Gamma= \{\mathcal V, \mathcal E\}$ be as above. 
We identify each edge 
${\bf e} \in \mathcal E$ with the interval $[0,1]$, which introduces an orientation on the edge set $\mathcal E$. Our argument below, in particular the spectrum of edge Laplacian, does not depend on this orientation. 
We assume that the ${\bf Z}^d$-action (\ref{DefineZdaction}) 
preserves this orientation.  Thus each edge ${\bf e} \in \mathcal E$ is identified with an oriented pair $({\bf e}(0),{\bf e}(1)) \in \mathcal V\times \mathcal V$, and is written as
${\bf e} = ({\bf e}(0),{\bf e}(1))$. 
We call ${\bf e}(0)$ the {\it initial vertex} of ${\bf e}$, and ${\bf e}(1)$ the {\it terminal vertex} of ${\bf e}$.   For a vertex $v \in {\mathcal V}$, we put
\begin{equation}
\mathcal E_v(0) = \{{\bf e} \in \mathcal E \, ; \, {\bf e}(0) = v\}, \quad 
\mathcal E_v(1) = \{{\bf e} \in \mathcal E \, ; \, {\bf e}(1) = v\},
\nonumber
\end{equation}
\begin{equation}
\mathcal E_v = \mathcal E_v(0) \cup \mathcal E_v(1).
\nonumber
\end{equation}
If ${\bf e} \in \mathcal E_v$, we say that the edge $\bf e \in \mathcal E$ is associated with the vertex $v \in \mathcal V$. 

Finally, although not essential, we assume that our periodic graph is connected. 

\subsection{Edge basis}
For an edge ${\bf e}$ and $n \in {\bf Z}^d$, we define its translation  ${\bf e} + [n]$ by
\begin{equation}
{\bf e} + [n] = ({\bf e}(0) + {\bf v}(n),{\bf e}(1) + {\bf v}(n)).
\label{DefineactionofZeone}
\end{equation}
By this ${\bf Z}^d$-action,
${\bf Z}^d\times \mathcal E \ni (n,{\bf e}) \to {\bf e} + [n] \in \mathcal E$, $\mathcal E$ is decomposed into the orbits, i.e. there exist ${\bf e}_1,\cdots,{\bf e}_{\nu} \in \mathcal E$ such that
\begin{equation}
\mathcal E = {\mathop\cup_{\ell=1}^{\nu}}\mathcal E_{\ell}, \quad 
\mathcal E_{\ell} = \{{\bf e}_{\ell} + [n]\, ; \,  n \in {\bf Z}^d\}.
\label{E=cupEell}
\end{equation}
We call the set
\begin{equation}
[\mathcal E] = \{{\bf e}_1,\cdots,{\bf e}_{\nu}\}
\label{Defineedgebasis}
\end{equation}
the {\it edge basis}. 
Note that the edge basis is not defined uniquely. Later, we will fix it in applying the Floquet theory on $ \Gamma $.

\subsection{Index}
Decompose the end points of an edge ${\bf e}$ into 
\begin{equation}
{\bf e}(0) = {\bf e}_{\ast}(0) + {\bf v}(m^{(0)}), \quad 
{\bf e}(1) = {\bf e}_{\ast}(1) + {\bf v}(m^{(1)}),
\nonumber
\end{equation}
where ${\bf e}_{\ast}(0), {\bf e}_{\ast}(1)$, $m^{(0)}, m^{(1)}$ are determined uniquely by the condition
\begin{equation}
{\bf e}_{\ast}(0), {\bf e}_{\ast}(1) \in \Gamma_{\ast}, \quad 
m^{(0)}, m^{(1)} \in {\bf Z}^d.
\nonumber
\end{equation}
The {\it index} of ${\bf e}$ is introduced by \cite{KoSa14} and  defined as 
\begin{equation}
{\rm Ind}\,({\bf e}) = m^{(1)} - m^{(0)}.
\nonumber
\end{equation}
Its meaning is as follows. 
For $m = (m_1,\cdots,m_d) \in {\bf Z}^d$, let 
$$
\Gamma_m = \{{\bf v}(x)\, ; \, x \in [m_1,m_1+1)\times\cdots\times[m_d,m_d+1)\}.
$$
Then, there exist $m^{(0)}, m^{(1)} \in {\bf Z}^d$ such that ${\bf e}(0) \in \Gamma_{m^{(0)}}, {\bf e}(1) \in \Gamma_{m^{(1)}}$. Let ${\bf e}'(0) = {\bf e}(0) + {\bf v}(m^{(1)} - m^{(0)}) \in \Gamma_{m^{(1)}}$. Then, in ${\bf R}^d$, the edge ${\bf e} = ({\bf e}(0), {\bf e}(1))$ is homotopic to  a curve from ${\bf e}(0)$ to ${\bf e}'(0)$ plus $({\bf e}'(0), {\bf e}(1))$, which is a curve in $\Gamma_{m^{(1)}}$. 
Therefore, its projection to $\Gamma_{\ast}$ is homotopic to a curve
\begin{equation}
({\bf e}_{\ast}(0),{\bf e}_{\ast}(1)) + n_1[C_1] + \cdots + n_d[C_d],
\label{HomotopyIndex}
\end{equation}
where ${\bf e}_{\ast}(0)$, ${\bf e}_{\ast}(1)$ are projections of ${\bf e}(0), {\bf e}(1)$ to $\Gamma_{\ast}$, and $\{[C_1],\cdots,[C_d]\}$ is a standard basis of the homotopy group $\pi_1({\bf T}^d)$.
Then, $n = (n_1,\cdots,n_d)$ in (\ref{HomotopyIndex})  is  the index of ${\bf e}$.  

\subsection{Fundamental graph}



Let ${\bf e}_{\ell}$ be one of the edge basis. 
The projection of ${\bf e}_{\ell}$ to $\Gamma_{\ast}$ is naturally identified with  an edge  on ${\bf T}^d$, which is denoted by ${{\bf e}_{\ell}}_{\ast}$.
The assumption (E) implies that $\{{{\bf e}_{1}}_\ast,\cdots,{{\bf e}_{\nu}}_{\ast}\}$ forms a connected graph on ${\bf T}^d$, whose edge set is denoted by ${\mathcal E}_{\ast}$:
\begin{equation}
{\mathcal E}_{\ast} = \{{{\bf e}_{1}}_{\ast},\cdots,{{\bf e}_{\nu}}_{\ast}\}.
\nonumber
\end{equation}

\begin{definition}
\label{S2DefineEast}
We call ${\mathcal E}_{\ast}$ the {\it fundamental graph} of $\Gamma$.
\end{definition}


  Let ${{\bf e}_{\ell}}_{\ast}(0)$, ${{\bf e}_{\ell}}_{\ast}(1)$ be the projections of ${\bf e}_{\ell}(0)$, ${\bf e}_{\ell}(1)$ onto $\Gamma_{\ast}$, and put 
\begin{equation}
\mathcal V_{\ast} = \{{{\bf e}_{\ell}}_{\ast}(0), {{\bf e}_{\ell}}_{\ast}(1)\,; \, 
\ell = 1, \cdots,\nu\}.
\label{DefinemathcalVast}
\end{equation}
Then the assumption (P) implies
\begin{equation}
\mathcal V_{\ast} = \Gamma_{\ast}\cap\mathcal V = \{p^{(1)},\cdots,p^{(s)}\}.
\nonumber
\end{equation}
 Note that $\nu$ and $s$ are  the number of edges and vertices of the fundamental graph. It says that any $p^{(i)}$ is an end point of some ${{\bf e}_{\ell}}_{\ast}$, hence $\nu \geq s$. 
 In fact, we have $\nu \geq s + d - 1$ for connected periodic graphs (\cite{KoSa15a}). 
The end points of ${{\bf e}_{\ell}}_{\ast}$ may coincide,  however, the edge ${{\bf e}_{\ell}}_{\ast}$ is not reduced to a point. The fundamental graph is thus a graph on ${\bf T}^d$ with the vertex set $\mathcal V_{\ast}$ and the edge set $\mathcal E_{\ast}$.

\begin{figure}[hbtp]
\centering
\includegraphics[width=9.5cm, bb=0 0 619 517]{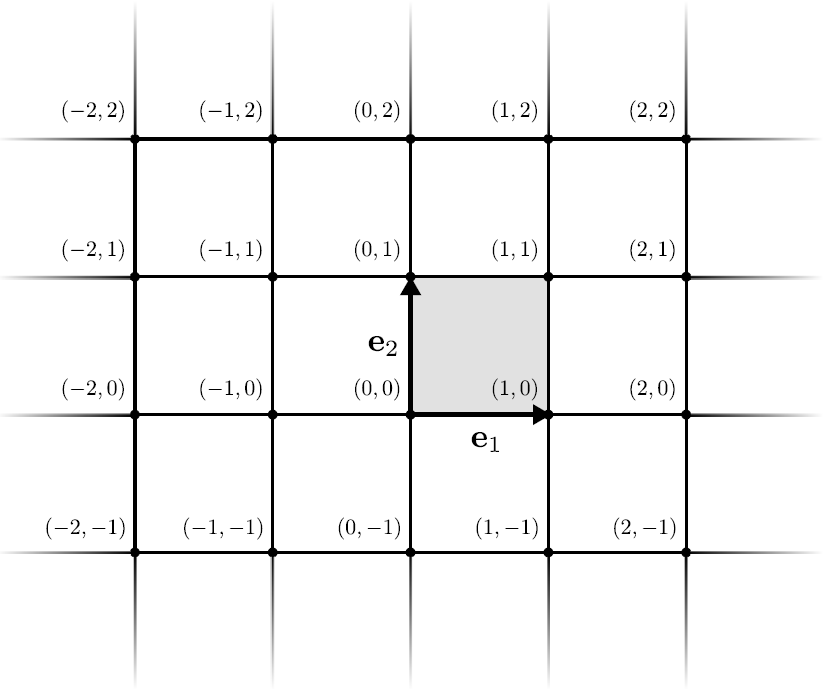}
\caption{Square lattice}
\label{C1SquareLattice}
\end{figure}
\begin{figure}[hbtp]
\centering
\includegraphics[width=5cm, bb=0 0 402 349]{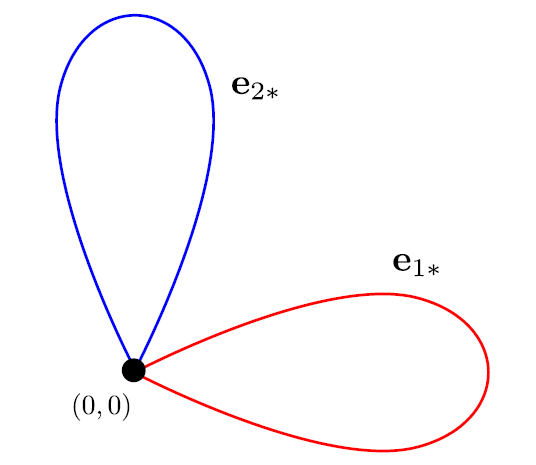}
\caption{Fundamental graph for the square lattice}
\label{Fundamental grapf for square lattice}
\end{figure}

\subsection{Example (1) - $d$-dimensional Square lattice}
 Let $s=1$, $p^{(1)}=0$, and 
$$
{\bf n}_1 = (1,0,\cdots,0),  \cdots,  {\bf n}_d = (0,\cdots,0,1)
$$
be the standard basis of ${\bf R}^d$. Then, letting ${\bf v}_i = {\bf n}_i$, we have
$$
\mathcal V = {\bf Z}^d,  \quad 
\Gamma_{\ast} = [0,1)^d,  
\quad \mathcal V_{\ast} = \{0\},
$$
$$
[{\mathcal E}] = \{{\bf e}_1,\cdots,{\bf e}_d\}, \quad 
{\mathcal E}_{\ast} = \{{{\bf e}_1}_{\ast},\cdots,{{\bf e}_d}_{\ast}\},
$$
$$ 
{\bf e}_{\ell\ast}(0) = {\bf e}_{\ell\ast}(1) = 0, \quad {\rm Ind}({\bf e}_{\ell})= {\bf n}_{\ell}.
$$

\subsection{Example (2) - Hexagonal lattice}
\label{ExampleHexalattice}
 Let $d=s = 2$, and
\begin{equation}
{\bf v}_1 = \Big(\frac{3}{2},-\frac{\sqrt3}{2}\Big), \quad
{\bf v}_2 = \Big(\frac{3}{2},\frac{\sqrt3}{2}\Big), 
\label{S2v1andv2}
\end{equation}
\begin{equation}
p^{(1)} = (1,0), \quad p^{(2)}= (2,0),
\label{S2p1andp2}
\end{equation}
\begin{equation}
\mathcal V = {\mathop\cup_{j=1}^2}\Big(p^{(j)} + \mathcal L\Big),
 \quad \mathcal L = \{{\bf v}(n)\, ; \, n \in {\bf Z}^2\}.
\label{S2DefinemathcalV}
\end{equation}
Then, we have
\begin{equation}
\Gamma_{\ast} = \Big\{(x_1,x_2) \, ; \, -\frac{x_1}{\sqrt3} \leq x_2 \leq \frac{x_1}{\sqrt3}, \ -\sqrt3 + \frac{x_1}{\sqrt3} < x_2 < \sqrt3 - \frac{x_1}{\sqrt3}\Big\},
\nonumber
\end{equation}
which is the shaded region in Figure \ref{Fundamental grapf for square lattice}, and
\begin{equation}
\mathcal V_{\ast} = \{p^{(1)}, p^{(2)}\},
\nonumber
\end{equation}
\begin{equation}
[{\mathcal E}] = \{{\bf e}_1, {\bf e}_2, {\bf e}_3\}, \quad
{\mathcal E}_{\ast} =  \{{{\bf e}_1}_{\ast}, {{\bf e}_2}_{\ast}, {{\bf e}_3}_{\ast}\},
\nonumber
\end{equation}
\begin{equation}
{\bf e}_1 = \Big( p^{(2)} , (\frac{5}{2}, \frac{\sqrt3}{2})  \Big), \quad
{\bf e}_2 = \Big(p^{(1)},p^{(2)}\Big), \quad
{\bf e}_3 = \Big(p^{(1)}, (\frac{1}{2},\frac{\sqrt3}{2})\Big),
\nonumber
\end{equation} 
\begin{equation}
{{\bf e}_{1}}_{\ast}(0) = p^{(2)}, \quad {{\bf e}_{1}}_{\ast}(1) = p^{(1)}, \quad
{\rm Ind}({\bf e}_1) = (0,1),
\nonumber
\end{equation}
\begin{equation}
{{\bf e}_{2}}_{\ast}(0) = p^{(1)}, \quad 
{{\bf e}_{2}}_{\ast}(1) = p^{(2)}, \quad {\rm Ind}({\bf e}_2) = (0,0), 
\nonumber
\end{equation}
\begin{equation}
{{\bf e}_{3}}_{\ast}(0) = p^{(1)}, \quad
{{\bf e}_{3}}_{\ast}(1) = p^{(2)}, \quad {\rm Ind}({\bf e}_3) = (-1,0).
\nonumber
\end{equation}
In Figure \ref{Fundamental grapf for square lattice}, the edges ${\bf e}_1$ and $({\bf e}_1)$,  ${\bf e}_3$ and $({\bf e}_3)$ are identified in $\Gamma_{\ast}$.

Let us recall here the definition of fundamental domain. 
Given a lattice $\bf L$ generated by a discrete group $G$ acting on ${\bf R}^d$, a set $F \subset {\bf R}^d$ is said to be a fundamental domain of ${\bf L}$ (or $G$) if it has the property
\begin{equation}
{\bf R}^d = {\mathop\cup_{g\in G}}g(F), \quad
g(F)\cap g'(F) = \emptyset \quad {\rm if} \quad g \neq g'.
\nonumber
\end{equation}
The fundamental domain is not defined uniquely. An often used choice is 
\begin{equation}
{\mathop\cap_{v\in {\bf L}\setminus\{0\}}}\Big\{x \in {\bf R}^d\, ; \, x\cdot \frac{v}{|v|} \leq \frac{|v|}{2}\Big\},
\nonumber
\end{equation}
the physical terminology of which is the Wigner-Seitz cell. 

\begin{figure}[hbtp]
\centering
\centering
\includegraphics[width=9cm, bb=0 0 605 495]{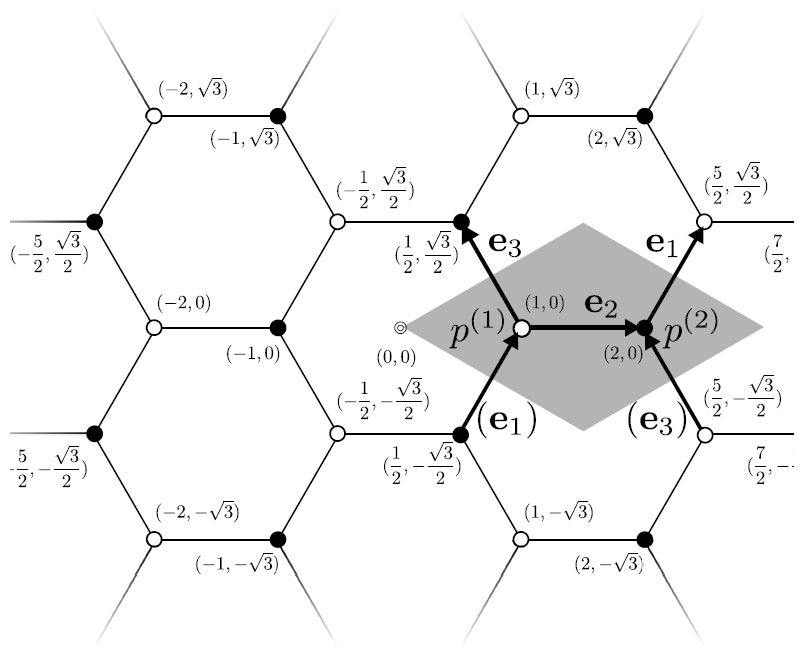}
\caption{Hexagonal lattice}
\label{Fundamental grapf for square lattice}
\end{figure}

\begin{figure}[hbtp]
\centering
\centering
\includegraphics[width=7cm, bb=0 0 496 369]{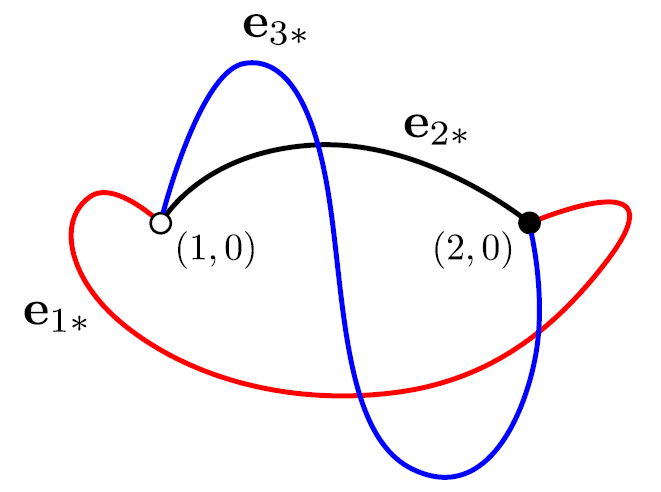}
\caption{Fundamental graph for the hexagonal lattice}
\label{Fundamental graph for the hexagonal lattice}
\end{figure}

Let $\mathcal L$ be the lattice in ${\bf R}^2$ generated by (\ref{S2v1andv2}). Then $\Gamma_{\ast}$ is a fundamental domain of $\mathcal L$. Letting $p^{(1)}, p^{(2)}$ be as in (\ref{S2p1andp2}), put $\mathcal V$ as in (\ref{S2DefinemathcalV}), which we regard as lattice.  
By the choice of $p^{(1)}, p^{(2)}$, the neighboring points of any $v \in \mathcal V$ have the same distance from $v$.
 To fix the idea, we have defined the fundamental domain of $\mathcal L$ of (\ref{S2DefineMathcalL})  by (\ref{S2DefineGammaastbyv}). However, any rotation and translation of $\Gamma_{\ast}$ is again a fundamental domain of $\mathcal L$. 
In \cite{AnIsoMo17}, we chose ${\bf v}_1, {\bf v}_2$ and $p^{(1)}, p^{(2)}$ to be
\begin{equation}
{\bf v}_1 = (\frac{3}{2}, \frac{\sqrt3}{2}), \quad {\bf v}_2 = (0, \sqrt{3}), \quad 
p^{(1)} = (\frac{1}{2}, - \frac{\sqrt3}{2}), \quad p^{(2)} = (1,0),
\label{S2vandpinthepreviouspaper}
\end{equation}
which are obtained from ${\bf v}_1, {\bf v}_2$ of (\ref{S2v1andv2}), and $p^{(1)}, p^{(2)}$ of (\ref{S2p1andp2}) by rotation of angle $\pi/3$ followed by the translation $x \to x + (0,-\sqrt3)$.

Taking the dual basis of ${\bf v}_1, {\bf v}_2$, $({\bf v}_i^{\ast}\cdot{\bf v}_j = 2\pi\delta_{ij})$, we define the dual lattice of $\mathcal V$. The Wigner-Seitz cell of this dual lattice is called the Brioullin zone. As is seen in Figure \ref{Fundamental grapf for square lattice}, the Wigner-Seitz cell for the hexagonal lattice is a hexagon centered at the origin.

\subsection{Fourier series}
Let $\ell^2(\mathcal V)$ be the set of ${\bf C}$-valued functions  $\widehat f = \big\{\widehat f(v)\big\}_{v\in\mathcal V}$ on $\mathcal V$ satisfying
\begin{equation}
\|\widehat f\|^2 := {\mathop\sum_{v\in\mathcal V}}|\widehat f(v)|^2 \,d_{\mathcal V}< \infty,
\nonumber
\end{equation}
$d_{\mathcal V}$ being defined by (\ref{Definedg}), 
which is a Hilbert space equipped with the inner product
\begin{equation}
(\widehat f,\widehat g) = {\mathop\sum_{v\in\mathcal V}}\widehat f(v)\overline{\widehat g(v)}\, d_{\mathcal V}.
\label{Vinnerproduct}
\end{equation}
Recalling that the vertex set $\mathcal V$ is written as a disjoint union $\mathcal V = \cup_{j=1}^s\big(p^{(j)} + \mathcal L\big)$, for any ${\bf C}$-valued function $\widehat f(v)$ on $\mathcal V$, we associate a function on ${\bf Z}^d$ by
\begin{equation}
\widehat f(p^{(j)} + {\bf v}(n)) = \widehat f_j(n), \quad n \in {\bf Z}^d.
\nonumber
\end{equation}
Using the correspondence
\begin{equation}
\widehat f(v) \longleftrightarrow (\widehat f_1(n), \cdots, \widehat f_s(n)),
\label{S2Identificationf(s)withfj(s)}
\end{equation}
we identify $\widehat f(v)$ with $(\widehat f_1(n), \cdots, \widehat f_s(n))$, and write
\begin{equation}
\widehat f(n) = \big(\widehat f_1(n), \cdots,\widehat f_s(n)\big).
\nonumber
\end{equation}
This induces a natural identification:
\begin{equation}
\ell^2(\mathcal V) = \big(\ell^2({\bf Z}^d)\big)^s.
\nonumber
\end{equation}
We then define a unitary operator $\mathcal U_{\mathcal V} : \ell^2(\mathcal V) \to L^2({\bf T}^d)^s$ by
\begin{equation}
\big(\mathcal U_{\mathcal V}\widehat f\big)(x) = (2\pi)^{-d/2}
\sqrt{d_{\mathcal V}}\sum_{n\in {\bf Z}^d}\widehat f(n)e^{in\cdot x},
\label{S1UdDefine}
\end{equation}
where $L^2({\bf T}^d)^s$ is equipped with the inner product
\begin{equation}
(f,g)_{L^2({\bf T}^d)^s} = \sum_{j=1}^s\int_{{\bf T}^d}f_j(x)\overline{g_j(x)}dx.
\label{S2L2bfTdinnerproduct}
\end{equation}

\subsection{Vertex Laplacian}
The {\it vertex Laplacian} $\widehat \Delta_{\mathcal V}$ on  $\mathcal V$  is defined by the following formula
\begin{equation}
\begin{split}
 & (\widehat\Delta_{\mathcal V} \widehat f)(n)  = (\widehat g_1(n),\cdots,\widehat g_s(n)),\\
\widehat g_i(n) & = \frac{1}{d_{\mathcal V}}\sum_{b\sim p^{(i)}+ {\bf v}(n)}\widehat f_{j(b)}(n(b)),\quad b = p^{(j(b))} + {\bf v}(n(b)).  
\end{split}
\label{S2DefineLaplacian}
\end{equation}
Passing to the Fourier series, we rewrite it into the following form :
\begin{equation}
\big(\mathcal U_{\mathcal V}(- \widehat\Delta_{\mathcal V}) {\mathcal U_{\mathcal V}}^{-1} f\big)(x) = H_0(x)f(x), \quad 
f \in L^2({\bf T}^d)^s,
\nonumber
\end{equation}
where $H_0(x)$ is an $s\times s$ Hermitian matrix whose entries are trigonometric functions. Let $D$ be the $s\times s$ diagonal matrix whose $(j,j)$ entry is $\sqrt{{\rm deg}(j)}$. Then 
\begin{equation}
H_0(x) = DH_0^0(x)D^{-1}, \quad H_0^0(x) = \mathcal U_{\mathcal V}(- \widehat\Delta_{\mathcal V}){\mathcal U_{\mathcal V}}^{-1}.
\label{S2H0(x)rewritten}
\end{equation}
Let $\lambda_j(x)$, $j = 1, \cdots,s$, be the eigenvalues of $H_0(x)$:
\begin{equation}
\lambda_1(x) \leq \lambda_2(x) \leq \cdots \leq \lambda_s(x).
\nonumber
\end{equation}
Then, we have
\begin{equation}
\sigma(-\widehat\Delta_{\mathcal V}) = {\mathop \cup_{j=1}^s}\lambda_j({\bf T}^d).
\nonumber
\end{equation}

In the following exmaples, the sum in (\ref{S2DefineLaplacian}) is ranging over all nearest neighboring vertices of 
$p^{(i)} + {\bf v}(n)$. 

\bigskip
\noindent
{\bf Example (1) - $d$-dimensional Square lattice.}  
Here, $\widehat u(n)$ is a ${\bf C}$-valued function on ${\bf Z}^d$ and the Laplacian is
\begin{equation}
\big(\widehat{\Delta}_{\mathcal V}\widehat u\big)(n) = \frac{1}{2d}\sum_{|n'-n|=1}\widehat u(n').
\nonumber
\end{equation}
Hence
\begin{equation}
H_0(x) = -\frac{1}{d}\left(\cos x_1 + \cdots + \cos x_d\right).
\nonumber
\end{equation}
{\bf Example (2) - Hexagonal lattice.} Here, $\widehat u(n)$ has two components $\widehat u(n) = (\widehat u_1(n),\widehat u_2(n))$, and the Laplacian is defined by
$$
\big(\widehat{\Delta}_{\mathcal V}\widehat u\big)(n) = \big(\widehat w_1(n),\widehat w_2(n)\big),
$$
\begin{equation}
\widehat w_1(n) = \frac{1}{3}\sum_{|{\bf v}(n')-{\bf v}(n)|=1}\widehat u_2(n'),
\nonumber
\end{equation}
\begin{equation}
\widehat w_2(n) = \frac{1}{3}\sum_{|{\bf v}(n')-{\bf v}(n)|=1}\widehat u_1(n').
\nonumber
\end{equation}
Hence
\begin{equation}
H_0(x) = -\frac{1}{3}\left(
\begin{array}{cc}
 0 & 1 + e^{-ix_1} + e^{-ix_2} \\
1 + e^{ix_1} + e^{ix_2} & 0
\end{array}
\right).
\nonumber
\end{equation}

\subsection{Kirchhoff condition}
As is noted above, we employ the identification: 
\begin{equation}
\mathcal E \ni ({\bf e}(0),{\bf e}(1))  \longleftrightarrow (0,1) \subset {\bf R}.
\nonumber
\end{equation}
Therefore, given ${\bf e} \in \mathcal E$, we use the abbreviation
\begin{equation}
\widehat u_{\bf e}(z) = \widehat u_{\bf e}((1-z){\bf e}(0) + z{\bf e}(1)).
\label{ue(z)=ue(1-z)e(0)+ze(1))}
\end{equation}
Hence on each edge ${\bf e} \in \mathcal E$, we consider the Hilbert space
$L^2_{\bf e} = L^2(0,1)$, and define the Hilbert space $L^2(\mathcal E)$ of $L^2$-functions $\widehat f = \big\{\widehat f_{\bf e}\big\}_{{\bf e} \in \mathcal E}$ on the edge set $\mathcal E$
\begin{equation}
L^2(\mathcal E) = {\mathop\oplus_{\bf e \in \mathcal E}}L^2_{\bf e}
\nonumber
\end{equation}
equipped with the inner product
\begin{equation}
(\widehat f,\widehat g)_{L^2({\mathcal E})}  = \sum_{{\bf e}\in {\mathcal E}}(\widehat f_{\bf e},\widehat g_{\bf e})_{L^2(0,1)}.
\label{L2mathcalE}
\end{equation}

\begin{definition}
\label{KirchhoffCond}
A function $\widehat u = \{\widehat u_{\bf e}\}_{{\bf e} \in \mathcal E}$ defined on $\mathcal E$ is said to satisfy the {\it Kirchhoff condition} if

\smallskip
\noindent
{\bf (K-1)} \noindent $\ \  \widehat u$ {\it is continuous on} $\mathcal E$, 

\smallskip
\noindent
{\bf (K-2)} \ \ {\it  $\widehat u_{\bf e}  \in C^1([0,1])$ on each edge ${\bf e} \in \mathcal E$, and }
$$
\sum_{{\bf e} \in \mathcal E_v(1)} \widehat u'_{\bf e}(1) - \sum_{{\bf e} \in \mathcal E_v(0)} \widehat u'_{\bf e}(0)  = 0.
$$ 

In (K-1), we regard $\mathcal E$ as a closed subset of ${\bf R}^d$ by the induced topology. Therefore,  $\widehat u_{\bf e}({\bf e}(p)) = \widehat u_{{\bf e}'}({\bf e}'(q))$ if ${\bf e}(p) = {\bf e}'(q)$ for $p, q = 0$ or 1.
\end{definition}

\subsection{Edge Laplacian}
On 
$L^2_{\bf e} = L^2(0,1)$, we consider the 1-dimensional Schr{\"o}dinger operator  
\begin{equation}
h^{(0)}_{\bf e} = - d^2/dz^2, \quad h_{\bf e} = h^{(0)}_{\bf e} + q_{\bf e}(z).
\nonumber
\end{equation} 
Assume that $q_{\bf e}$ satisfies (Q-1) and (Q-2). 
Define the Hamiltonian
\begin{equation}
\widehat H_{\mathcal E} : \widehat u = \left\{{\widehat u}_{\bf e}\right\}_{{\bf e}\in \mathcal E} \to 
\left\{h_{{\bf e}}{\widehat u}_{\bf e}\right\}_{{\bf e}\in \mathcal E}
\label{S2HamitonianhatH}
\end{equation}
with domain $D(\widehat H_{\mathcal E})$ consisting of  $\widehat u_{\bf e} \in H^2(0,1)$ satisfying the Kirchhoff condition (K-1), (K-2)  
 and $\sum_{{\bf e}\in\mathcal E}\|- d^2 {\widehat u}_{\bf e}/dz^2 + q_{\bf e}{\widehat u}_{\bf e}\|_{L^2(0,1)}^2 < \infty$. 
By integration by parts, one can show that $\widehat H_{\mathcal E}$ 
is self-adjoint in $L^2(\mathcal E)$.

When $q_{\bf e} = 0$, $\widehat H_{\mathcal E}$ is denoted by $\widehat H_{\mathcal E}^{(0)}$ or 
$- \widehat\Delta_{\mathcal E}$, i.e. 
\begin{equation}
\big(- \widehat\Delta_{\mathcal E}\widehat u\big)_{\bf e}(z) = - \frac{d^2}{dz^2}\widehat u_{\bf e}(z), \quad {\bf e} \in \mathcal E.
\nonumber
\end{equation}
 We call it the {\it edge Laplacian}.
 Let $q_{\mathcal E}$ be the multiplication operator defined by
\begin{equation}
\big(q_{\mathcal E}\widehat f\big)_{\bf e}(z) = q_{\bf e}(z)\widehat f_{\bf e}(z), \quad {\bf e} \in \mathcal E.
\nonumber
\end{equation}
Then
$$
\widehat H_{\mathcal E} = \widehat H^{(0)}_{\mathcal E} + q_{\mathcal E}.
$$

\subsection{Floquet theory}
The Floquet (or Floquet-Bloch) theory transforms  operators periodic on ${\bf Z}^d$ into the one on ${\bf T}^d\times M$, where $M$ is a suitable manifold. 
Recall that, by fixing a set of edge basis $[\mathcal{E} ] =\{ {\bf e}_1 , \ldots , {\bf e}_{\nu} \} $ satisfying the assumption  (E), 
any ${\bf e} \in \mathcal E$ is uniquely written as 
\begin{equation}
{\bf e} = {\bf e}_{\ell} + [n], \quad 1 \leq \ell \leq \nu, \quad n \in {\bf Z}^d.
\nonumber
\end{equation}
Using the action of ${\bf Z}^d$ on $\mathcal E$:
$n \to {\bf e} + [n]$, (see (\ref{DefineactionofZeone})), 
we  define a unitary operator $\mathcal U_{\mathcal E} : L^2(\mathcal E) \to \big(L^2({\bf T}^d\times(0,1))\big)^{\nu}$ by
 \begin{equation}
\mathcal U_{\mathcal E} = 
(\mathcal U_{\mathcal E,1},\cdots,\mathcal U_{\mathcal E,\nu}),
\label{DefinemathcalU}
\end{equation}
\begin{equation}
\begin{split}
\big(\mathcal U_{{\mathcal E,\ell}}\widehat f\big)(x,z)  = (2\pi)^{-d/2}\sum_{n\in{\bf Z}^d}
e^{in\cdot x}\widehat f_{{\bf e}_{\ell}+[n]}(z) , \quad (x,z)\in {\bf T}^d \times [0,1].
\end{split}
\label{mathcalUEdefine}
\end{equation}
The adjoint operator of $\mathcal U_{\mathcal E,\ell}$ is
\begin{equation}
\big(\mathcal U_{\mathcal E,\ell}^{\ast}g\big)(n,z) = (2\pi)^{-d/2}\int_{{\bf T}^d}
e^{-in\cdot x}g(x,z)dx.
\label{mathcalUEastdefine}
\end{equation}
The Fourier series gives rise to a unitary operator from $\ell^2({\bf Z}^d;L^2(0,1))$, the set of the $L^2(0,1)$-valued $\ell^2$-functions on ${\bf Z}^d$, to $L^2({\bf T}^d\times(0,1))$. 
By Definition \ref{S2DefineEast} and the above formulas (\ref{mathcalUEdefine}) and (\ref{mathcalUEastdefine}), $\mathcal U_{\mathcal E}$ is unitary:
$$
\mathcal U_{\mathcal E} : L^2(\mathcal E) \to L^2({\bf T}^d\times \mathcal E_{\ast}).
$$
 We are thus led to study the operators on ${\bf T}^d\times {\mathcal E}_{\ast}$. 

We transfer the edge Laplacian $\widehat\Delta_{\mathcal E}$ to ${\bf T}^d \times \mathcal E_{\ast}$. 
Put
\begin{equation}
\partial_{\ast}({\bf e}_{\ell}) = \{{{\bf e}_{\ell}}_{\ast}(0),{{\bf e}_{\ell}}_{\ast}(1)\}.
\nonumber
\end{equation}
Because of the ${\bf Z}^d$-action, the orbits of ${\bf e}_{\ell}$ and ${\bf e}_k$ have common end points, i.e. ${\bf e}_{\ell} + [n']$ and ${\bf e}_k + [n'']$ have common end points for some $n', n'' \in {\bf Z}^d$,  if and only if so do the end points of ${{\bf e}_{\ell}}_{\ast} = ({{\bf e}_{\ell}}_{\ast}(0),{{\bf e}_{\ell}}_{\ast}(1))$ and ${{\bf e}_{k}}_{\ast} = ({{\bf e}_{k}}_{\ast}(0),{{\bf e}_{k}}_{\ast}(1))$, i.e. 
\begin{equation}
\partial_{\ast}({\bf e}_{\ell})\cap\partial_{\ast}({\bf e}_{k}) \neq \emptyset.
\nonumber
\end{equation}
Let $\mathcal V_{\ast}$ be as in (\ref{DefinemathcalVast}), and for $v \in \mathcal V_{\ast}$, put
\begin{equation}
L_{\ast}(v) = \{\ell\, ; \, {\bf e}_{\ell\ast}(0)=v \ \ {\rm or} \ \ {\bf e}_{\ell\ast}(1)=v\},
\nonumber
\end{equation}
and for $\ell \in L_{\ast}(v)$
\begin{equation}
 \delta_{\ell}(v) = \left\{
\begin{split}
& 1 \quad {\rm if} \quad v = {{\bf e}_{\ell}}_{\ast}(1),\\
& 0 \quad {\rm if} \quad v = {{\bf e}_{\ell}}_{\ast} (0).
\end{split}
\right.
\label{Definedeltaellv}
\end{equation}

The following Lemma \ref{Lemma4.2}  is proved in \cite{KoSa15}, Theorem 1.1.

\begin{lemma}
\label{Lemma4.2}
Let $\widehat f \in H^2(\mathcal E)$. Then, $\widehat f$ satisfies the Kirchhoff condition if and only if $f = (f_1,\cdots,f_{\nu}) =  \mathcal U_{\mathcal E}\widehat f$ satisfies
\begin{equation}
e^{i\delta_{k} (v) {\rm Ind}({\bf e}_{k})\cdot x}f_{k}(x,\delta_{k} (v) ) = e^{i\delta_{\ell} (v) {\rm Ind}({\bf e}_{\ell })\cdot x}f_{\ell}(x,\delta_{\ell} (v)) ,
\label{E-ideltanxfell=e-ideltaf'} 
\end{equation}
provided  $v\in \partial_{\ast}({\bf e}_{k})\cap\partial_{\ast}({\bf e}_{\ell}) $, and for $v\in \mathcal{V}_* $
\begin{equation}
\sum_{\ell \in L_{\ast}(v)} (-1)^{\delta_{\ell} (v)} e^{i\delta_{\ell} (v) {\rm Ind}({\bf e}_{\ell})\cdot x} f'_{\ell}(x,\delta_{\ell} (v)) = 0,
\label{sum(-1)deltaUfprime=0}
\end{equation}
where $f'_{\ell}(x,z) = \frac{\partial}{\partial z}f_{\ell}(x,z)$. 
\end{lemma}

The {\it Floquet operator} $-\Delta_{\mathcal E_{\ast}}(x)  $ is a differential operator on $L^2(\mathcal E_{\ast})$ with parameter $x \in {\bf T}^d$ defined by
\begin{equation}
( - \Delta_{\mathcal E_{\ast}}(x)u)_{{{\bf e_{\ell}}_{\ast}}}(x,z) = -u'' _{{\bf e} _{\ell *} }(x,z) ,\quad  {} '' =  \frac{\partial^2}{\partial z^2} , 
\nonumber
\end{equation}
where $u_{{{\bf e_{\ell}}_{\ast}}}(x,\cdot) \in H^2(0,1)$ for each $x \in {\bf T}^d$ and satisfies the following {\it quasi-periodic condition} at any $v \in \mathcal V_{\ast}$
\begin{equation}
e^{i\delta_{k} (v) {\rm Ind}({\bf e}_{k})\cdot x} u_{{\bf e} _{k*}}(x,\delta_{k} (v) ) = e^{i\delta_{\ell} (v) {\rm Ind}({\bf e}_{\ell })\cdot x} u_{{\bf e} _{\ell *}}(x,\delta_{\ell} (v)) , 
\nonumber
\end{equation}
where $ {\bf e}_{k*} , {\bf e}_{\ell *} $ satisfy $v\in \partial _* ({\bf e}_k  ) \cap \partial_* ({\bf e}_{\ell} )$, and 
\begin{equation}
\sum_{\ell \in L_{\ast}(v)} (-1)^{\delta_{\ell} (v)} e^{i\delta_{\ell} (v) {\rm Ind}({\bf e}_{\ell})\cdot x} u' _{{\bf e} _{\ell *} } (x,\delta_{\ell} (v)) = 0 .
\nonumber
\end{equation}
For $ \mathcal U_{\mathcal E}^{\ast} u \in D(- \widehat\Delta_{\mathcal E})$
\begin{equation}
- \big(\mathcal U_{\mathcal E} \widehat\Delta_{\mathcal E}\mathcal U_{\mathcal E}^{\ast} u\big)(x,z) = - \Delta_{\mathcal E_{\ast}}(x)u(x,z)
\nonumber
\end{equation}
holds, 
and $-\Delta_{\mathcal E_{\ast}}(x)$ is self-adjoint in $L^2(\mathcal E_{\ast})$.

\subsection{Fermi surface}
Spectral properties of the vertex Laplacian depend on its characteristic surface, i.e. the Fermi surface. To describe it, we use the following notations. 

Let $\lambda_1(x) \leq \lambda_2(x) \leq \cdots \leq \lambda_s(x)$ be the eigenvalues of $H_0(x)$, and 
\begin{equation}
M_{\mathcal V, \lambda, j} = \{x \in {\bf T}^d\, ; \, \lambda_j(x) = \lambda\},
\label{C4S1MlambdajDefine}
\end{equation}
\begin{equation}
p(x,\lambda) = \det\big(H_0(x)-\lambda\big) = \prod_{j=1}^s(\lambda_j(x) - \lambda),
\label{C4S1pxlambda=prodpj}
\end{equation}
\begin{equation}
M_{\mathcal V, \lambda}= \{x \in {\bf T}^d\, ; \, p(x,\lambda)=0\} = {\mathop\cup_{j=1}^s}M_{\mathcal V, \lambda,j},
\label{C4S1Mlambda=cupMlambdaj}
\end{equation}
\begin{equation}
{\bf T}^d_{\bf C} = {\bf C}^d/(2\pi{\bf Z})^d, \quad
M_{\mathcal V,\lambda}^{\bf C} = \{z \in {\bf T}^d_{\bf C}\, ; \, p(z,\lambda)=0\},
\nonumber
\end{equation}
\begin{equation}
M^{\bf C}_{\mathcal V,\lambda,reg} = \{z \in M_{\mathcal V,\lambda}^{ \bf C}\, ; \, \nabla_z p(z,\lambda) \neq 0\},
\nonumber
\end{equation}
\begin{equation}
M^{\bf C}_{\mathcal V, \lambda,sng} = \{z \in M_{\mathcal V,\lambda}^{\bf C}\, ; \, \nabla_z p(z,\lambda) = 0\}.
\nonumber
\end{equation}

Let $H_0$ be the self-adjoint operator of multiplication by $H_0(x)$ on ${\bf T}^d$. As in \cite{AnIsoMo17}, we impose the following assumptions on the free system.

\medskip
\noindent
{\bf(A-1)}  {\it There exists a subset $\mathcal T_1 \subset \sigma(H_0)$ such that for} $\lambda \in \sigma(H_0)\setminus \mathcal T_1$, 

\smallskip
{\bf(A-1-1)} $M_{\mathcal V,\lambda,sng}^{\bf C}$ {\it is discrete}.

\smallskip
{\bf(A-1-2)} {\it Each connected component of $M_{\mathcal V,\lambda,reg}^{\bf C}$ intersects with ${\bf T}^d$ and the intersecton is a $(d-1)$-dimensional real analytic submanifold of ${\bf T}^d$.}

\medskip
\noindent
{\bf(A-2)} {\it  There exists a finite set $\mathcal T_0 \subset \sigma (H_0)$ such that}
$$
M_{\mathcal V,\lambda,i}\cap M_{\mathcal V,\lambda,j} = \emptyset, \ \ if \ \ i \neq j, \ \ 
\lambda \in \sigma(H_0)\setminus \mathcal T_0.
$$

\medskip
\noindent
{\bf(A-3)} \ {\it $\nabla_xp(x,\lambda) \neq 0$ , \ on \ $M_{\mathcal V,\lambda}$, \ $\lambda \in \sigma(H_0)\setminus \mathcal T_0$.}

\medskip
\noindent
{\bf(A-4)}  {\it The unique continuation property holds for $-\widehat \Delta_{\mathcal V}$ in $\mathcal V$, i.e. if there exist $\widehat u$ and a constant $\lambda$ such that $(- \widehat\Delta_{\mathcal V}-\lambda)\widehat u = 0$ holds on $\mathcal V$, and $\widehat u = 0$ except for a finite number of vertices, then $\widehat u = 0$ on $\mathcal V$.}
\footnote{This assumption is incorrectly stated in 
\cite{AnIsoMo17(1)}. Here, we give a correct assumption. See \cite{AnIsoMo17(2)}.}

\medskip
We put 
\begin{equation}
\mathcal T = \mathcal T_0 \cup \mathcal T_1 \cup \sigma_p(- \widehat\Delta_{\mathcal V}).
\label{S2DefineMathcalT}
\end{equation}

\medskip
For the square, triangular, hexagonal, Kagome, diamond lattices and the subdivision of square lattice, $\mathcal T_1$ is a finite set. However, for the  ladder and graphite, $\mathcal T_1$ fills closed intervals.  See \cite{AnIsoMo17}, \S 5. 

\subsection{Function spaces}
We use the following function spaces on the edge set $\mathcal E$:
\begin{equation}
{\widehat {\ell}}^{\; 2,\sigma}(\mathcal E) \ni \widehat f\Longleftrightarrow 
\sum_{\ell=1}^{\nu}\sum_{n\in {\bf Z}^d}
(1 + |n|^2)^{\sigma}\|\widehat f_{{\bf e}_{\ell} +[n]}\|_{L^2(0,1)}^2  < \infty,
\nonumber
\end{equation}
where $\sigma \in {\bf R}$ is a parameter, 
\begin{equation}
\widehat{\mathcal B}({\mathcal E}) \ni \widehat f 
\Longleftrightarrow
\|\widehat f\|_{\widehat{\mathcal B}(\mathcal E)} 
= \sum_{\ell=1}^{\nu}\sum_{j=0}^{\infty}r_j^{1/2}
\Big(\sum_{r_{j-1}\leq |n| < r_j}\|\widehat f_{{\bf e}_{\ell} + [n]}\|^2_{L^2(0,1)}\Big)^{1/2} < \infty,
\nonumber
\end{equation}
where $r_{-1}=0$, $r_j = 2^j$ ($j \geq 0$), 
\begin{equation}
\widehat{\mathcal B}^{\ast}({\mathcal E}) \ni \widehat f 
\Longleftrightarrow
\|\widehat f\|^2_{\widehat{\mathcal B}^{\ast}(\mathcal E)} = \sup_{R>1}\frac{1}{R}\sum_{\ell=1}^{\nu}\sum_{|n| < R}\|\widehat f_{{\bf e}_{\ell} +[n]}\|_{L^2(0,1)}^2  < \infty.
\nonumber
\end{equation}
\begin{equation}
\widehat{\mathcal B}^{\ast}_{0}(\mathcal E)\ni \widehat f 
\Longleftrightarrow
\lim_{R\to\infty}\frac{1}{R}\sum_{\ell=1}^{\nu}\sum_{|n| < R}\|\widehat f_{{\bf e}_{\ell} +[n]}\|_{L^2(0,1)}^2  =0.
\nonumber
\end{equation}
 Their counter parts on the vertex set $\mathcal V$ are:

\begin{equation}
\widehat {\ell}^{\; 2,\sigma}(\mathcal V) \ni \widehat f\Longleftrightarrow 
\sum_{n\in {\bf Z}^d}
(1 + |n|^2)^{\sigma/2}|\widehat f(n)|^2  < \infty,
\nonumber
\end{equation}
\begin{equation}
\widehat{\mathcal B}({\mathcal V}) \ni \widehat f 
\Longleftrightarrow
\|\widehat f\|_{\widehat{\mathcal B}(\mathcal V)} 
= \sum_{j=0}^{\infty}r_j^{1/2}
\Big(\sum_{r_{j-1}\leq |n| < r_j}|\widehat f(n)|^2\Big)^{1/2} < \infty,
\nonumber
\end{equation}
\begin{equation}
\widehat{\mathcal B}^{\ast}({\mathcal V}) \ni \widehat f 
\Longleftrightarrow
\|\widehat f\|^2_{\widehat{\mathcal V}^{\ast}(\mathcal E)} = \sup_{R>1}\frac{1}{R}\sum_{|n| < R}|\widehat f(n)|^2  < \infty.
\nonumber
\end{equation}
\begin{equation}
\widehat{\mathcal B}^{\ast}_{0}(\mathcal V)\ni \widehat f 
\Longleftrightarrow
\lim_{R\to\infty}\frac{1}{R}\sum_{|n| < R}|\widehat f(n)|^2  =0.
\nonumber
\end{equation}

We also introduce their counter parts on the torus: Letting 
$f = \mathcal U_{\mathcal E}\widehat f$,
\begin{equation}
H^{\sigma}({\bf T}^d\times\mathcal E_{\ast}) \ni f \Longleftrightarrow \widehat f \in \widehat\ell^{\; 2,\sigma}(\mathcal E),
\nonumber
\end{equation}
\begin{equation}
\mathcal B({\bf T}^d\times \mathcal E_{\ast}) \ni f 
\Longleftrightarrow \widehat f \in \widehat{\mathcal B}(\mathcal E),
\nonumber
\end{equation}
\begin{equation}
\begin{split}
\mathcal B^{\ast}({\bf T}^d \times {\mathcal E}_{\ast}) \ni f \Longleftrightarrow 
\widehat f \in \widehat{\mathcal B}^{\ast}(\mathcal E),
\end{split}
\nonumber
\end{equation}
\begin{equation}
\begin{split}
\mathcal B^{\ast}_0({\bf T}^d \times {\mathcal E}_{\ast}) \ni f \Longleftrightarrow 
\widehat f \in \widehat{\mathcal B}^{\ast}_0(\mathcal E).
\end{split}
\nonumber
\end{equation}
Their counter parts for the vertex set $\mathcal V$ are:
\begin{equation}
H^{\sigma}({\bf T}^d) \ni f \Longleftrightarrow \widehat f \in \widehat\ell^{\;2,\sigma}(\mathcal V),
\nonumber
\end{equation}
\begin{equation}
\mathcal B({\bf T}^d) \ni f 
\Longleftrightarrow \widehat f \in \widehat{\mathcal B}(\mathcal V),
\nonumber
\end{equation}
\begin{equation}
\begin{split}
\mathcal B^{\ast}({\bf T}^d) \ni f \Longleftrightarrow 
\widehat f \in \widehat{\mathcal B}^{\ast}(\mathcal V),
\nonumber
\end{split}
\end{equation}
\begin{equation}
\begin{split}
\mathcal B^{\ast}_0({\bf T}^d) \ni f \Longleftrightarrow 
\widehat f \in \widehat{\mathcal B}^{\ast}_0(\mathcal V).
\end{split}
\nonumber
\end{equation}
Here, the discrete Fourier transform $\widehat f \to f$ is the usual Fourier series defined on the vertex set $\mathcal V$.

Note that letting $H^{\sigma}({\bf T}^d)$, $ \mathcal B_{1/2}({\bf T}^d)$ and $\mathcal B_{-1/2}({\bf T}^d)$ be the Sobolev and Besov spaces on ${\bf T}^d$ (see \cite{AgHo76}), we have
\begin{equation}
H^{\sigma}({\bf T}^d\times\mathcal E_{\ast})  = H^{\sigma}({\bf T}^d)\otimes L^2(\mathcal E_{\ast}),
\nonumber
\end{equation}
\begin{equation}
\mathcal B({\bf T}^d\times \mathcal E_{\ast}) = \mathcal B_{1/2}({\bf T}^d)\otimes L^2(\mathcal E_{\ast}),
\nonumber
\end{equation}
\begin{equation}
\mathcal B^{\ast}({\bf T}^d\times \mathcal E_{\ast}) = \mathcal B_{-1/2}({\bf T}^d)\otimes L^2(\mathcal E_{\ast}).
\nonumber
\end{equation}

\begin{definition}
For $f, g \in \mathcal B^{\ast}({\bf T}^d\times{\mathcal E}_{\ast})$, we use the notation $f \simeq g$ in the following sense:
$$
f \simeq g \Longleftrightarrow 
f - g \in \mathcal B^{\ast}_0({\bf T}^d\times
{\mathcal E}_{\ast}).
$$
We  use the same notation $\simeq$ for $\widehat{\mathcal B}^{\ast}(\mathcal E)$, $\widehat{\mathcal B}^{\ast}(\mathcal V)$ and $\widehat{\mathcal B}^{\ast}({\bf T}^d)$.
\end{definition}

\section{Resolvent formulae}
The purpose of this section is to represent the resolvents of the Schr{\"o}dinger operators
\begin{equation}
 \widehat R^{(0)}_{\mathcal E}(\lambda) = (\widehat H^{(0)}_{\mathcal E} - \lambda)^{-1}, \quad 
 \widehat R_{\mathcal E}(\lambda) = (\widehat H_{\mathcal E} - \lambda)^{-1}
\label{S3ResolventsRandR0}
\end{equation}
 in terms of those of the vertex Laplacian and the 1-dimensional Schr{\"o}dinger operator on each edge.  We give a formal derivation here, and justify them in the next section.

\subsection{Green operator on the edge}
Let 
$ - (d^2/dz^2)_D$
be the Laplacian on $(0,1)$ with boundary condition $u(0) = u(1)=0$. By the assumption (Q-1),  the operator $ - (d^2/dz^2)_D + q_{\bf e}(z)$ equipped with domain $H^2(0,1)\cap H^1_0(0,1)$ is self-adjoint. Let $r_{\bf e}(\lambda)$ be the resolvent:
\begin{equation}
r_{{\bf e}}(\lambda) = \big(-(d^2/dz^2)_D + q_{\bf e}(z) - \lambda\big)^{-1}.
\nonumber
\end{equation}
Let $\phi_{{\bf e}0}(z,\lambda), \phi_{{\bf e}1}(z,\lambda)$ be the solutions of 
$$
\big(- d^2/dz^2 + q_{\bf e}(z) - \lambda\big)\phi = 0
$$
 with initial data
\begin{equation}
\left\{
\begin{split}
&\phi_{{\bf e}0}(0,\lambda) = 0, \\
& \phi'_{{\bf e}0}(0,\lambda)=1, 
\end{split}
\right.
\qquad
\left\{
\begin{split}
&\phi_{{\bf e}1}(1,\lambda) = 0, \\
& \phi'_{{\bf e}1}(1,\lambda)=-1. 
\end{split}
\right.
\nonumber
\end{equation}
Note that $\phi_{{\bf e}0}(1,\lambda) = 0$ or  $\phi_{{\bf e}1}(0,\lambda) = 0$ if and only if $\lambda$ is an eigenvalue of $- (d^2/dz^2)_D + q_{\bf e}(z)$. 
In the following, we assume that
\begin{equation}
\lambda \not\in {\mathop\cup_{{\bf e} \in \mathcal E}}\sigma\left(-(d^2/dz^2)_D + q_{\bf e}(z)\right).
\nonumber
\end{equation}
The resolvent $r_{\bf e}(\lambda)$ is then written as
\begin{equation}
\begin{split}
\big(r_{\bf e}(\lambda)\widehat f\big)(z)&  = \frac{1}{C(\lambda)}\int_0^z\phi_{{\bf e}1}(z,\lambda)\phi_{{\bf e}0}(t,\lambda)\widehat f(t)dt \\
& + \frac{1}{C(\lambda)}\int_z^1\phi_{{\bf e}0}(z,\lambda)\phi_{{\bf e}1}(t,\lambda)\widehat f(t)dt,
\end{split}
\label{GreenOperartorgeneraledge}
\end{equation}
\begin{equation}
C(\lambda) = \phi_{{\bf e}0}(1,\lambda) = \phi_{{\bf e}1}(0,\lambda),
\nonumber
\end{equation}
the last line of which is proven by computing the Wronskian. 
Then, we have
\begin{equation}
\begin{split}
\frac{d}{dz}r_{{\bf e}}(\lambda)\widehat f(z)\big|_{z=0}& = 
\frac{1}{C(\lambda)}\int_0^1\phi_{{\bf e}1}(t,\lambda)\widehat f(t)dt, \\
\frac{d}{dz}r_{{\bf e}}(\lambda)\widehat f(z)\big|_{z=1} & = -
\frac{1}{C(\lambda)}\int_0^1\phi_{{\bf e}0}(t,\lambda)\widehat f(t)dt.
\end{split}
\label{ddzrelambdaf=int}
\end{equation}
We define  operators $\Phi_{{\bf e}0}(\lambda),  \Phi_{{\bf e}1}(\lambda): L^2(0,1) \to {\bf C}$ by
\begin{equation}
\Phi_{{\bf e}1}(\lambda)\widehat f = 
\int_0^1\frac{\phi_{{\bf e}0}(t,\lambda)}{\phi_{{\bf e}0}(1,\lambda)}\widehat f(t)dt,
\quad
\Phi_{{\bf e}0}(\lambda)\widehat f = 
\int_0^1\frac{\phi_{{\bf e}1}(t,\lambda)}{\phi_{{\bf e}1}(0,\lambda)}\widehat f(t)dt.
\label{Te0Te1formula}
\end{equation}
Their adjoints : ${\bf C} \to L^2(0,1)$ are defined for $c \in {\bf C}$ by
\begin{equation}
\Phi_{{\bf e}1}(\lambda)^{\ast}c = c\frac{\phi_{{\bf e}0}(z,\overline{\lambda})}{\phi_{{\bf e}0}(1,\overline{\lambda})}, \quad 
\Phi_{{\bf e}0}(\lambda)^{\ast}c = c\frac{\phi_{{\bf e}1}(z,\overline{\lambda})}{\phi_{{\bf e}1}(0,\overline{\lambda})}.
\nonumber
\end{equation}

\subsection{Kirchhoff condition and vertex Laplacian}
We put
\begin{equation}
\widehat u_{{\bf e}}(z,\lambda) = \Phi_{{\bf e}1}(\overline{\lambda})^{\ast}c_{{\bf e}}(1,\lambda) + 
\Phi_{{\bf e}0}(\overline{\lambda})^{\ast}c_{{\bf e}}(0,\lambda) + r_{{\bf e}}(\lambda)\widehat f_{\bf e},
\label{S3ue(z)=Ce1+Ce0}
\end{equation}
where $c_{\bf e}(0,\lambda), c_{\bf e}(1,\lambda)$ are to be specified later. 
It satisfies
\begin{equation}
\left\{
\begin{split}
&\Big(- \frac{d^2}{dz^2} + q_{\bf e} - \lambda\Big)\widehat u_{\bf e} = \widehat f_{\bf e}, \quad {\rm on} \quad (0,1), \\
& \widehat u_{\bf e}(0,\lambda) = c_{\bf e}(0,\lambda), \quad 
\widehat u_{\bf e}(1,\lambda) = c_{\bf e}(1,\lambda).
\end{split}
\right.
\nonumber
\end{equation}
Here we have used (\ref{ue(z)=ue(1-z)e(0)+ze(1))}). 
The following lemma reduces the edge Laplacian to the vertex Laplacian.

\begin{lemma}
\label{LemmaKirchhoffandvertexLaplacian}
(1) For $\widehat u_{\bf e}(z,\lambda)$ defined by (\ref{S3ue(z)=Ce1+Ce0}),
the Kirchhoff condition (K-1) is fulfilled  if and only if for two edges ${\bf e}, {\bf e}' \in \mathcal E$ and $p = 0, 1$, $q = 0, 1$,
$$
{\bf e}(p) = {\bf e}'(q) \Longrightarrow 
c_{\bf e}(p,\lambda) = c_{\bf e'}(q,\lambda).
$$ 
(2) 
The condition (K-2) is satisfied if and only if
\begin{equation}
\begin{split}
&- \sum_{{\bf e}\in\mathcal E_v(0)}\frac{1}{\phi_{{\bf e}0}(1,\lambda)}c_{\bf e}(1,\lambda) - \sum_{{\bf e}\in\mathcal E_v(1)}\frac{1}{\phi_{{\bf e}1}(0,\lambda)}c_{\bf e}(0,\lambda)\\
& - \sum_{{\bf e}\in\mathcal E_v(0)}\frac{\phi'_{{\bf e}1}(0,\lambda)}{\phi_{{\bf e}1}(0,\lambda)}c_{\bf e}(0,\lambda) + \sum_{{\bf e}\in\mathcal E_v(1)}\frac{\phi'_{{\bf e}0}(1,\lambda)}{\phi_{{\bf e}0}(1,\lambda)}c_{\bf e}(1,\lambda)\\
&=  \sum_{{\bf e}\in\mathcal E_v(1)} \Phi_{{\bf e}1}(\lambda)\widehat f_{\bf e} + \sum_{{\bf e}\in\mathcal E_v(0)} \Phi_{{\bf e}0}(\lambda)\widehat f_{\bf e}
\end{split}
\label{KirchhoffCond3}
\end{equation}
holds at any $v \in \mathcal V$.
\end{lemma}

Proof. Take $v \in \mathcal V$ and ${\bf e} \in \mathcal E_v$. 
Then ${\bf e}(p) = v$ for $p = 0$ or $1$ and $\widehat u_{\bf e}({\bf e}(p),\lambda) = c_{\bf e}(p,\lambda)$. 
The condition (K-1) means that the value of $\widehat{u} = \{ \widehat{u} _{{\bf e}} \} _{{\bf e} \in \mathcal{E}}$ at $v$ depends only on $v$ and is independent of the edge $\bf e$. This proves (1). 

By (\ref{S3ue(z)=Ce1+Ce0}), for $p = 0, 1$, 
\begin{equation}
\widehat u_{\bf e}'(p,\lambda) = c_{\bf e}(1,\lambda)\frac{\phi_{{\bf e}0}'(p,\lambda)}{\phi_{{\bf e}0}(1,\lambda)} +  c_{\bf e}(0,\lambda)\frac{\phi_{{\bf e}1}'(p,\lambda)}{\phi_{{\bf e}1}(0,\lambda)} + \frac{d}{dz}r_{{\bf e}}(\lambda)\widehat f_{\bf e}\Big|_{z=p},
\label{u'e=CeandC1}
\end{equation}
where $' = d/dz$. In (K-2), we replace $\widehat u_{\bf e}$ and $\widehat u'_{\bf e}$ by (\ref{S3ue(z)=Ce1+Ce0}) and (\ref{u'e=CeandC1}). 
Using (\ref{ddzrelambdaf=int}) and (\ref{Te0Te1formula}), we obtain (\ref{KirchhoffCond3}). 
\qed

\medskip
We rewrite (\ref{KirchhoffCond3}). If $\widehat u_{\bf e}(z,\lambda)$ satisfies the Kirchhoff condition, by virtue of Lemma \ref{LemmaKirchhoffandvertexLaplacian} (1), $c_{\bf e}(p,\lambda)$ depends only on the end point ${\bf e}(p)$ and is independent of the edge ${\bf e}$. Therefore, we denote 
$c_{\bf e}(p,\lambda) =  c(v,\lambda)$ if ${\bf e}(p) = v$. 

For $v, w \in \mathcal V$ such that $v \sim w$, there is a unique ${\bf e} \in \mathcal E$ whose end points are $v$ and $w$. We define a new edge ${\bf e}_{wv}$ by 
$$
{\bf e}_{wv} = \left\{
\begin{split}
&{\bf e} \quad {\rm if} \quad {\bf e}(0) = v, \\
& {\bf e}^{-1} \quad {\rm if} \quad {\bf e}(0) = w,
\end{split}
\right.
$$
where ${\bf e}^{-1}$ means ${\bf e}$ with reverse direction. Therefore, ${\bf e}_{wv}$ is a directed edge with initial vertex $v$ and terminal vertex $w$. We also define 
a function $\psi_{wv}(z,\lambda)$ on the edge ${\bf e}_{wv}$ by
\begin{equation}
\psi_{wv}(z,\lambda) = 
\left\{
\begin{split}
& \phi_{{\bf e}0}(z,\lambda) \quad {\rm if} \quad {\bf e}(0) = v, \\
& \phi_{{\bf e}1}(1-z,\lambda) \quad {\rm if} \quad {\bf e}(1) = v.
\end{split}
\right.
\label{Definephiwvzlambda}
\end{equation}
This means that $\psi_{wv}(z) = \phi_{{\bf e}0}(z,\lambda)$ if ${\bf e}$ and ${\bf e}_{wv}$ have the same direction, and 
$\psi_{wv}(z) = \phi_{{\bf e}1}(1-z,\lambda)$ if ${\bf e}$ and ${\bf e}_{wv}$ have the opposite direction.

Now, the assumption (Q-3) plays an important role.  We put $q_{wv}(z) = q_{\bf e}(z)$, which is well-defined by virtue of (Q-3). 
Moreover, $\psi_{wv}(z,\lambda)$ satisfies
\begin{equation}
\left\{
\begin{split}
& \big(- \frac{d^2}{dz^2} + q_{wv}(z) - \lambda\big)\psi_{wv}(z,\lambda) = 0, \quad 0 < z < 1, \\
& \psi_{wv}(0,\lambda) = 0, \quad \psi_{wv}'(0,\lambda) = 1.
\end{split}
\right.
\nonumber
\end{equation}
The equality $\phi_{{\bf e}0}(z,\lambda) = \phi_{{\bf e}1}(1 - z,\lambda)$ implies
\begin{equation}
\phi_{{\bf e}0}(1,\lambda) = \phi_{{\bf e}1}(0,\lambda), \quad \phi_{{\bf e}0}'(1,\lambda) = - \phi'_{{\bf e}1}(0,\lambda).
\nonumber
\end{equation}
Therefore, by (\ref{Definephiwvzlambda}), we have
\begin{equation}
\psi_{wv}(1,\lambda) = \left\{
\begin{split}
& \phi_{{\bf e}0}(1,\lambda) \quad {\rm if} \quad {\bf e}(0) = v, \\
& \phi_{{\bf e}1}(0,\lambda) \quad {\rm if} \quad {\bf e}(1) = v, 
\end{split}
\right.
\label{S3.2psiwvandphie0e1}
\end{equation}
\begin{equation}
\psi_{wv}'(1,\lambda) = \left\{
\begin{split}
& -\phi_{{\bf e}1}'(0,\lambda) \quad {\rm if} \quad {\bf e}(0) = v, \\
& \phi_{{\bf e}0}'(1,\lambda) \quad {\rm if} \quad {\bf e}(1) = v. 
\end{split}
\right.
\label{S3.2psi'wzandphi'e01}
\end{equation}
\begin{definition}
We define the perturbed {\it vertex Laplacian on $\mathcal V$} by
\begin{equation}
\big(\widehat\Delta_{\mathcal V, \lambda}\widehat u\big)(v) = \frac{1}{d_{\mathcal V}}
\sum_{w\sim v}\frac{1}{\psi_{wv}(1,\lambda)}\widehat u(w), \quad v \in \mathcal V
\label{S3DefineVertexLaplacian}
\end{equation}
for  $\widehat u \in \ell^2_{loc}(\mathcal V)$. 
\end{definition}

What we have defined in (\ref{S3DefineVertexLaplacian}) is the so-called normalized discrete Laplacian. It is not the standard discrete Laplacian. 
By (\ref{S3.2psiwvandphie0e1}), the first and the second terms of  the left-hand side of (\ref{KirchhoffCond3}) divided by $d_{\mathcal V}$ are summarized into $-\big(\widehat\Delta_{\mathcal V, \lambda}\widehat u\big)(v) $. By (\ref{S3.2psi'wzandphi'e01}),
the third and the 4th terms are summarized into a scalar multiplication operator:
\begin{equation}
\big(\widehat Q_{\mathcal V,\lambda}\widehat u\big)(v) = \widehat Q_{v,\lambda}(v)\widehat u(v), 
\nonumber
\end{equation}
where
\begin{equation}
\widehat Q_{v,\lambda}(v) =  \frac{1}{d_{\mathcal V}}\sum_{w\sim v}\frac{\psi'_{wv}(1,\lambda)}{\psi_{wv}(1,\lambda)}.
\label{S3DefinewidehatQVlambda}
\end{equation}

We define an operator $\widehat T_{\mathcal V}(\lambda) : L^2_{loc}(\mathcal E) \to L^2_{loc}(\mathcal V)$ by
\begin{equation}
\begin{split}
\big(\widehat T_{{\mathcal V}}(\lambda)\widehat f\big)(v)  = \frac{1}{d_{\mathcal V}}\Big(\sum_{{\bf e}\in\mathcal E_v(1)} \Phi_{{\bf e}1}(\lambda)\widehat f_{\bf e} + \sum_{{\bf e} \in \mathcal E_v(0)}\Phi_{{\bf e}0}(\lambda)\widehat f_{\bf e}\Big), \quad 
v \in \mathcal V.
\label{DefineFe(lambda)}
\end{split}
\end{equation}
Lemma  \ref{LemmaKirchhoffandvertexLaplacian} (2) is now rewritten as follows.

\begin{lemma}
\label{Lemma3.2}
For $\widehat u$ defined by (\ref{S3ue(z)=Ce1+Ce0}), let $\widehat u\big|_{\mathcal V}$ be the restriction of $\widehat u$ on $\mathcal V$. Then
the  condition (\ref{KirchhoffCond3}) is rewritten as
\begin{equation}
\begin{split}
& \left(- \widehat\Delta_{\mathcal V,\lambda}+  \widehat{Q}_{\mathcal V,\lambda}\right)\widehat u\big|_{\mathcal V} = {\widehat T}_{\mathcal V}(\lambda)\widehat f.
\end{split}
\label{DeltaVlambdawidehataU+lambdaQVU=f}
\end{equation}
\end{lemma}

Therefore, $\widehat u\big|_{\mathcal V}$ should be written  as
\begin{equation}
\widehat u\big|_{\mathcal V} =  \left(- \widehat\Delta_{\mathcal V,\lambda}+  \widehat{Q}_{\mathcal V,\lambda}\right)^{-1}\widehat T_{\mathcal V}(\lambda)\widehat f.
\label{widehatu|mathcalVformal}
\end{equation}
Here, we must be careful about the operator $\big(- \widehat\Delta_{\mathcal V,\lambda}+  \widehat{Q}_{\mathcal V,\lambda}\big)^{-1}$.
For $\lambda \not\in {\bf R}$, the operator $-\widehat \Delta_{\mathcal V,\lambda} + \widehat Q_{\mathcal V,\lambda}$  has complex coefficients, hence is not self-adjoint. Therefore, the existence of its inverse is not obvious.  We discuss the validity of this formula  in Subsection \ref{SubsectionLAP}.  For the moment, we admit (\ref{widehatu|mathcalVformal}) as a formal formula. 

We define an operator $\widehat T_{\mathcal E}(\lambda) : L^2_{loc}(\mathcal V) \to L^2_{loc}(\mathcal E)$ by
\begin{equation}
\big(\widehat T_{\mathcal E}(\lambda)\widehat u\big)_{\bf e}(z) = 
\Phi_{{\bf e}1}(\overline{\lambda})^{\ast}\widehat u({\bf e}(1)) + 
\Phi_{{\bf e}0}(\overline{\lambda})^{\ast}\widehat u({\bf e}(0)) .
\label{DefineTmathcalElambda}
\end{equation}
We also define $r_{\mathcal E}(\lambda) : L^2_{loc}(\mathcal E) \to L^2_{loc}(\mathcal E)$ by
\begin{equation}
\big( r_{\mathcal E}(\lambda)\widehat f\big)_{\bf e} = r_{\bf e}(\lambda)\widehat f_{\bf e}, \quad {\bf e} \in \mathcal E.
\label{DefinermathcalElambda}
\end{equation}
Lemma \ref{Lemma3.2} and (\ref{S3ue(z)=Ce1+Ce0}) yield the following lemma.

\begin{lemma}
\label{Lemma3.3}
The resolvent of $\widehat H_{\mathcal E}$ is  written as
\begin{equation}
\widehat R_{\mathcal E}(\lambda)= 
\widehat T_{\mathcal E}(\lambda)\left(- \widehat\Delta_{\mathcal V,\lambda} + \widehat Q_{\mathcal V,\lambda}\right)^{-1}\widehat T_{\mathcal V}(\lambda) + r_{\mathcal E}(\lambda).
\label{Resolventformalformula1}
\end{equation}
In other words, $\widehat u = \widehat R_{\mathcal E}(\lambda)\widehat f$ is given by $\widehat u = \{\widehat u_{\bf e}\}_{{\bf e}\in \mathcal E}$ with $\widehat u_{{\bf e}}(z,\lambda)$ defined by (\ref{S3ue(z)=Ce1+Ce0}), and 
\begin{equation}
c_{{\bf e}}(p,\lambda) =  
\big(\widehat R_{\mathcal V}(\lambda)\widehat T_{\mathcal V}(\lambda)\widehat f\big)({\bf e}(p)), \quad p = 0, 1,
\end{equation}
\begin{equation}
\widehat R_{\mathcal V}(\lambda) = 
\left(- \widehat\Delta_{\mathcal V,\lambda} + \widehat Q_{\mathcal V,\lambda}\right)^{-1}.
\label{RVlambdaFormula}
\end{equation}
We also have a symmetric form:
\begin{equation}
\widehat R_{\mathcal E}(\lambda)= 
\widehat T_{\mathcal V}(\overline\lambda)^{\ast}\left(- \widehat\Delta_{\mathcal V,\lambda} + \widehat Q_{\mathcal V,\lambda}\right)^{-1}\widehat T_{\mathcal V}(\lambda) + r_{\mathcal E}(\lambda).
\label{Resolventformalformula2}
\end{equation}
\end{lemma}

 The formula (\ref{Resolventformalformula2}) follows from the following lemma. 

\begin{lemma}
\label{TmathcalVlambdaadjoint}
$$
\widehat T_{\mathcal V}(\lambda)^{\ast} = \widehat T_{\mathcal E}(\overline{\lambda}).
$$
\end{lemma}

Proof. Suppose $\widehat f \in L^2_{loc}(\mathcal E)$ and $\widehat u \in L^2_{loc}(\mathcal V)$ are compactly supported. Then, recalling that the inner product of $\ell^2(\mathcal V)$ contains $d_{\mathcal V}$  (see  (\ref{Vinnerproduct})), 
\begin{equation}
\begin{split}
& \big(\widehat T_{\mathcal V}(\lambda)\widehat f,\widehat u\big) \\
&= 
\frac{1}{d_{\mathcal V}}\sum_{v \in \mathcal V}\Big(\sum_{{\bf e}\in \mathcal E_v(1)}(\Phi_{{\bf e}1}(\lambda)\widehat{f}_{\bf e})\overline{\widehat u({\bf e}(1))} + 
\sum_{{\bf e}\in \mathcal E_v(0)}(\Phi_{{\bf e}0}(\lambda)\widehat{f}_{\bf e})\overline{\widehat u({\bf e}(0))}\Big)d_{\mathcal V}\\
&= \sum_{{\bf e} \in \mathcal E}
\int_0^1
\widehat{f}_{\bf e}(\lambda,z)\Big(\overline{\Phi_{{\bf e}1}(\overline\lambda)^{\ast}\widehat u({\bf e}(1))} + 
\overline{\Phi_{{\bf e}0}(\overline\lambda)^{\ast}\widehat u({\bf e}(0))}\Big)(z)dz \\
&= \big(\widehat f,\widehat T_{\mathcal E}(\overline{\lambda})\widehat u),
\end{split}
\nonumber
\end{equation} 
which proves the lemma. \qed


\subsection{Unperturbed resolvent}
We rewrite the above formula for the unperturbed resolvent $\widehat R^{(0)}_{\mathcal E}(\lambda)$ to make it more explicit.  We put the superscript $\ ^{(0)}$ for every term. For $\lambda = re^{i\theta} \in {\bf C}$ with $0 \leq \theta < 2\pi$, we define 
$$
\sqrt{\lambda} = \sqrt{r}e^{i\theta/2} 
$$
so that ${\rm Im}\sqrt{\lambda} \geq 0$ and $\sqrt{r \pm i0} = \pm \sqrt{r}$ for $r > 0$. When $q_{\bf e}=0$, we have
\begin{equation}
\phi_{{\bf e}0}^{(0)}(z) = \frac{\sin\sqrt{\lambda}z}{\sqrt{\lambda}}, \quad \phi_{{\bf e}1}^{(0)}(z) = \frac{\sin\sqrt{\lambda}(1-z)}{\sqrt{\lambda}}, \quad 
C^{(0)}(\lambda) = \frac{\sin\sqrt{\lambda}}{\sqrt{\lambda}}.
\label{phie10unperturbed}
\end{equation}
Then, in view of (\ref{S3DefineVertexLaplacian}), we have
\begin{equation}
\Big(\widehat\Delta^{(0)}_{\mathcal V,\lambda}\widehat u\Big)(v) = \frac{\sqrt{\lambda}}{\sin\sqrt{\lambda}}\frac{1}{d_{\mathcal V}}\sum_{w\sim v}\widehat u(w) = \frac{\sqrt{\lambda}}{\sin\sqrt{\lambda}}\big(\widehat\Delta_{\mathcal V}\widehat u\big)(v).
\label{DefineDeltaVlambda(0)}
\end{equation}
\begin{equation}
\widehat Q^{(0)}_{v,\lambda} = \frac{\sqrt{\lambda}}{\sin{\sqrt{\lambda}}}\cos\sqrt{\lambda},
\label{DefinewidehatQ0}
\end{equation}
We can then factor out the term $\sqrt{\lambda}/\sin\sqrt{\lambda}$
in the formula for $\widehat R_{\mathcal V}^{(0)}(\lambda)$:
\begin{equation}
\widehat R_{\mathcal V}^{(0)}(\lambda)
= \frac{\sin\sqrt{\lambda}}{\sqrt{\lambda}}
\Big(- \widehat\Delta_{\mathcal V} + \cos\sqrt{\lambda}\Big)^{-1},
\label{S3DefineRV0lambda}
\end{equation}
which gives a definite meaning to $\widehat R_{\mathcal V}^{(0)}(\lambda)$ since $\cos\sqrt{\lambda} \not\in {\bf R}$ for $\lambda \not\in {\bf R}$. By virtue of (\ref{GreenOperartorgeneraledge}), the Green operator of  $- (d^2/dz^2)_D$ is written as
\begin{equation}
\begin{split}
\big(r^{(0)}_{{\bf e}}(\lambda)\widehat f_{\bf e}\big)(z) = & \frac{\sqrt{\lambda}}{\sin\sqrt{\lambda}}\Big(
\int_0^z\frac{\sin\sqrt{\lambda}(1-z)}{\sqrt{\lambda}}\frac{\sin\sqrt{\lambda}w}{\sqrt{\lambda}}\, \widehat f_{\bf e}(w)dw \\
& \ \ \ \ \ \ \ \ \ \ \ \ \ + \int_z^1\frac{\sin\sqrt{\lambda}z}{\sqrt{\lambda}}\frac{\sin\sqrt{\lambda}(1-w)}{\sqrt{\lambda}}\, \widehat f_{\bf e}(w)dw\Big).
\end{split}
\nonumber
\end{equation}
This implies, by (\ref{Te0Te1formula}) and (\ref{DefineFe(lambda)}),
\begin{equation}
\begin{split}
\Phi^{(0)}_{{\bf e}1}(\lambda)f &=  \frac{\sqrt{\lambda}}{\sin\sqrt{\lambda}}\int_0^1\frac{\sin\sqrt{\lambda}z}{\sqrt{\lambda}} f(z)dz, 
\nonumber
\\
\Phi^{(0)}_{{\bf e}0}(\lambda)f &=  \frac{\sqrt{\lambda}}{\sin\sqrt{\lambda}}\int_0^1\frac{\sin\sqrt{\lambda}(1-z)}{\sqrt{\lambda}} f(z)dz, \\
\big(\widehat T^{(0)}_{{\mathcal V}}(\lambda)\widehat f\big)(v)  &=  \frac{\sqrt{\lambda}}{\sin\sqrt{\lambda}}\frac{1}{d_{\mathcal V}}\Big(\sum_{{\bf e}\in\mathcal E_v(1)}\int_0^1\frac{\sin\sqrt{\lambda}z}{\sqrt{\lambda}}\widehat f_{\bf e}(z)dz,
\nonumber
\\
&\ \ \ \ \ \ \ \ \ \ \ \ \ \ \ \  + 
\sum_{{\bf e}\in\mathcal E_v(0)}\int_0^1\frac{\sin\sqrt{\lambda}(1-z)}{\sqrt{\lambda}}\widehat f_{\bf e}(z)dz\Big).
\end{split}
\nonumber
\end{equation}
In view of Lemmas \ref{Lemma3.2}, \ref{TmathcalVlambdaadjoint} and (\ref{DefineTmathcalElambda}), we have proven the following lemma.

\begin{lemma}
\label{UnperturbedResolERxpression}
For $\lambda \not\in {\bf R}$,
the resolvent of $\widehat H_{\mathcal E}^{(0)}$ is represented as
\begin{equation}
\begin{split}
\widehat R_{\mathcal E}^{(0)}(\lambda) 
&= \widehat T^{(0)}_{\mathcal E}(\lambda)(- \widehat\Delta^{(0)}_{\mathcal V,\lambda} + \widehat Q^{(0)}_{\mathcal V,\lambda})^{-1}\widehat T^{(0)}_{\mathcal V}(\lambda) + r^{(0)}_{\mathcal E}(\lambda) \\
&= \widehat T^{(0)}_{\mathcal V}(\overline\lambda)^{\ast}\frac{\sin{\sqrt{\lambda}}}{\sqrt{\lambda}}(- \widehat\Delta_{\mathcal V} + \cos\sqrt{\lambda})^{-1}\widehat T^{(0)}_{\mathcal V}(\lambda) + r^{(0)}_{\mathcal E}(\lambda).
\label{Resolventformula2}
\end{split}
\end{equation}
More explicitly, $\widehat u^{(0)} = (\widehat H^{(0)}_{\mathcal E} - \lambda)^{-1}\widehat f = \widehat R^{(0)}_{\mathcal E} (\lambda ) \widehat f$ is written as 
\begin{equation}
{\widehat u}^{(0)}_{{\bf e}}(z,\lambda) =
\frac{\sqrt{\lambda}}{\sin\sqrt{\lambda}}\left( c^{(0)}_{\bf e}(1,\lambda)
	\frac{\sin\sqrt{\lambda}z}{\sqrt{\lambda}} +  c^{(0)}_{\bf e}(0,\lambda)\frac{\sin\sqrt{\lambda}(1-z)}{\sqrt{\lambda}}\right) + r^{(0)}_{{\bf e}}(\lambda){\widehat f}_{{\bf e}}, 
\label{Formulaue(z)unperturbed}
\end{equation}
\begin{equation}
c_{\bf e}^{(0)}(p,\lambda) =  \frac{\sin\sqrt{\lambda}}{\sqrt{\lambda}}\left((- \widehat\Delta_{\mathcal V} + \cos\sqrt{\lambda})^{-1}\widehat T^{(0)}_{\mathcal V}(\lambda)\widehat f\right)({\bf e}(p)), \quad p =  0, 1.
\end{equation}
\end{lemma}

Recall that $\mathcal T$ is defined by (\ref{S2DefineMathcalT}). We put
\begin{equation}
\sigma^{(0)}_{\mathcal T} = \left\{
\lambda \in {\rm Int}\big(\,\sigma_e(\widehat H^{(0)}_{\mathcal E})\big)\, ; \, - \cos\sqrt{\lambda} \in \mathcal T
\right\},
\label{Definesigma0tau}
\end{equation}
\begin{equation}
\sigma^{(0)}_{\mathcal V} = \{(\pi j)^2\, ; \, j = 1, 2, \cdots\},
\label{DefinesigmaV0}
\end{equation}
\begin{equation}
\sigma^{(0)}_{\mathcal E} = \{\lambda \, ; \, - \cos\sqrt{\lambda} \in 
\sigma(- \widehat{\Delta}_{\mathcal V})\}.
\label{DefinesigmaE0}
\end{equation}

As will be proven in the next section, for $\lambda \in \big({\rm Int}\,\sigma_e(\widehat H_{\mathcal E})\big)\setminus\big(\sigma_{\mathcal T}^{(0)}\cup\sigma_{\mathcal V}^{(0)}\big)$, there exists a limit $\big(- \widehat\Delta_{\mathcal V} + \cos
\sqrt{\lambda \pm i0}\big)^{-1} = \lim_{\epsilon \downarrow 0}\big(- \widehat\Delta_{\mathcal V} + \cos
\sqrt{\lambda \pm i\epsilon}\big)^{-1} $ in a suitable topolgy. 
Here, for a subset $A \subset {\bf R}$, ${\rm Int}\, A$ means its interior.
Hence, we have the following expression for $\widehat R_{\mathcal E}^{(0)}(\lambda \pm i0)$.

\begin{lemma} 
\label{S3RElambda+i0Formula1}
For $\lambda \in \big({\rm Int}\,\sigma_e(\widehat H_{\mathcal E})\big)\setminus\big(\sigma_{\mathcal T}^{(0)}\cup\sigma_{\mathcal V}^{(0)}\big)$, the following formula holds 
\begin{equation}
\widehat R_{\mathcal E}^{(0)}(\lambda\pm i0) = 
\widehat T^{(0)}_{\mathcal V}(\lambda)^{\ast}\frac{\sin{\sqrt{\lambda}}}{\sqrt{\lambda}}(- \widehat\Delta_{\mathcal V} + \cos\sqrt{\lambda \pm i0})^{-1}\widehat T^{(0)}_{\mathcal V}(\lambda) + r^{(0)}_{\mathcal E}(\lambda).
\nonumber
\end{equation}
\end{lemma}

We put for $\lambda > 0$
\begin{equation}
\sigma(\lambda) = {\rm sgn}(\sin\sqrt{\lambda}) 
= \left\{
\begin{split}
& 1 \quad {\rm if} \quad \sin{\sqrt{\lambda}} > 0, \\
& -1 \quad {\rm if} \quad \sin{\sqrt{\lambda}} < 0,
\end{split}
\right.
\end{equation}
and
$$
\sigma(\lambda) = 0 \quad {\rm for} \quad \lambda \leq 0.
$$
Using
\begin{equation}
\cos\sqrt{\lambda \pm i0} = 
\left\{
\begin{split}
& \cos\sqrt{\lambda} \mp i\sigma(\lambda)0, \quad \lambda > 0, \\
& \cosh\sqrt{|\lambda|}, \quad \lambda \leq 0,
\end{split}
\right.
\label{Cossqrtlambda}
\end{equation}
we  have 
\begin{equation}
\begin{split}
&  \big(- \widehat\Delta_{\mathcal V} + \cos
\sqrt{\lambda + i0}\big)^{-1} - \big(- \widehat\Delta_{\mathcal V} + \cos
\sqrt{\lambda - i0}\big)^{-1} \\
& =  \sigma(\lambda)\left(
\big(- \widehat\Delta_{\mathcal V} + \cos
\sqrt{\lambda}-i0\big)^{-1} - \big(- \widehat\Delta_{\mathcal V} + \cos
\sqrt{\lambda} + i0\big)^{-1}\right).
\end{split}
\nonumber
\end{equation}
By well-known Stone's formula,  for any self-adjoint operator $A$, 
$$
\frac{1}{2\pi i} \int_I\Big(\big((A - \lambda - i0)^{-1} - (A - \lambda + i0)^{-1}\big)f,g\Big) = 
\int_Id(E_A(\lambda)f,g)
$$
holds,
where $E_A(\lambda)$ is the spectral decomposition of $A$, and $I$ is any compact interval in $\sigma_{ac}(A)\setminus \sigma_p(A)$. Therefore, it is natural to define 
$E_A'(\lambda) = \frac{d}{d\lambda}E_A(\lambda)$ by
$$
E_A'(\lambda) = \frac{1}{2\pi i}\big((A - \lambda - i0)^{-1} - (A - \lambda + i0)^{-1}\big).
$$
We then have
\begin{equation}
 \frac{d}{d\lambda} E_{\widehat H^{(0)}_{\mathcal E}}(\lambda) 
= \sigma(\lambda) \widehat T^{(0)}_{\mathcal V}(\lambda)^{\ast}\frac{\sin\sqrt{\lambda}}{\sqrt{\lambda}}E'_{-\widehat{\Delta}_{\mathcal V}}(-\cos\sqrt{\lambda})\widehat T^{(0)}_{\mathcal V}(\lambda).
\nonumber
\end{equation}
Integrating this equality, we obtain the following formula: 

\begin{lemma} Letting $\lambda = (\arccos(-k))^2$, we have
\begin{equation}
E_{{\widehat H}^{(0)}_{\mathcal E}}(I) = 2\int_J
\widehat T^{(0)}_{\mathcal V}(\lambda)^{\ast}E'_{-\widehat{\Delta}_{\mathcal V}}(k)\widehat T^{(0)}_{\mathcal V}(\lambda)dk,
\nonumber
\end{equation}
where $I$ is any interval in $\big({\rm Int}\,\sigma_e(\widehat H_{\mathcal E})\big)\setminus\big(\sigma_{\mathcal T}^{(0)}\cup\sigma_{\mathcal V}^{(0)}\big)$ and $J$ is the image of $I$ by the mapping $\lambda \to k$ .
\end{lemma}

Here, we define $\sqrt{\lambda} = \arccos(-k)$ to be first the principal branch and next its analytic continuation.
The multi-valuedness of the mapping $k \to \lambda$ then gives  rise to the band structure of the spectrum of $\widehat H^{(0)}_{\mathcal E}$, which has already been studied by many authors. 


\subsection{Spectra}

Let $\{\lambda_j(x)\}_{j=1}^s$ be the eigenvalues of $H_0(x)$. We arrange them in such a way that 
there exists $1 \leq s_0 \leq s$ such that $\lambda_j(x)$ is non-constant for $1 \leq j \leq s_0$, but  constant for $s_0 + 1 \leq j \leq s$. Put
\begin{equation}
E^{(0)}_{j,k} = \left\{
\begin{split}
&\big({\rm arccos} \,(-\lambda_j) + \pi k\big)^2, \quad {\rm when} \ k \ {\rm is} \ {\rm even},\\
&\big(- {\rm arccos}\,(-\lambda_j) + \pi(k+1)\big)^2, \quad {\rm when} \ k \ {\rm is} \ {\rm odd},
\end{split}
\right.
\end{equation}
for $s_0 \leq j \leq s$, 
where ${\rm arccos}$ is the principal value, i.e. ${\rm arccos}\,(\lambda) \in [0,\pi]$ for $- 1 \leq \lambda \leq 1$. We put 
\begin{equation}
\sigma_{fb}(\widehat H^{(0)}_{\mathcal E}) = \sigma^{(0)}_{\mathcal V} \cup 
\big\{E_{j,k}^{(0)} \, ; \, s_0 + 1 \leq j \leq s, \ \  k = 0, 1, 2, \cdots\big\},
\end{equation}
and call it  the {\it flat band} of the spectum of $\widehat H^{(0)}_{\mathcal E}$.
The following theorem is  proved in \cite{KoSa15}.
\begin{theorem}
\begin{equation}
\sigma(\widehat H^{(0)}_{\mathcal E}) = \sigma^{(0)}_{\mathcal V} \cup \sigma^{(0)}_{\mathcal E}.
\end{equation}
\begin{equation}
\sigma_p(\widehat H^{(0)}_{\mathcal E}) = \sigma_{fb}(\widehat H^{(0)}_{\mathcal E}).
\end{equation}
Moreover, $ \sigma_{fb}(\widehat H^{(0)}_{\mathcal E})$ consists of the  eigenvalues of $\widehat H^{(0)}_{\mathcal E}$ with infinite multiplicities.
\end{theorem}
In the sequel, we consider the case in which $s_0 = s$ for the sake of simplicity (mainly for the simplicity of notation). 
The results below are also extended to the case where $s_0 < s$.


\section{Resolvent estimates}

We give a definite meaning to the expressions of the resolvents (\ref{S3ResolventsRandR0}).

\subsection{Rellich type theorem}
We put 
\begin{equation}
\sigma(q_{\mathcal E}) = {\mathop\cup_{{\bf e}\in\mathcal E}}\sigma( - (d^2/dz^2)_D + q_{{\bf e}}).
\end{equation}
Note that  by (Q-1) and (Q-2), $\sigma(q_{\mathcal E})$ is a discrete subset of ${\bf R}$, furthermore $\sigma_{\mathcal V}^{(0)} \subset \sigma (q_{\mathcal E})$. For $R > 0$,  the exterior domain $\mathcal E_{ext,R}$ and the interior domain $\mathcal E_{int,R}$ are defined to be the set of edges such that
\begin{equation}
\mathcal E_{ext,R} \ni {\bf e} \Longleftrightarrow 
|{\bf e}(0)| \geq R \quad {\rm and} \quad |{\bf e}(1)| \geq R, 
\end{equation}
\begin{equation}
\mathcal E_{int,R} \ni {\bf e} \Longleftrightarrow 
|{\bf e}(0)| < R \quad {\rm or} \quad |{\bf e}(1)| < R. 
\end{equation}
\begin{theorem}
\label{RelichTypeTheorem}
Let $\lambda \in \big({\rm Int}\,\sigma_{e}(\widehat H^{(0)}_{\mathcal E})\big)\setminus \big(\sigma^{(0)}_{\mathcal V}\cup\sigma^{(0)}_{\mathcal T}\big)$, and suppose  $\widehat u \in \widehat {\mathcal B}^{\ast}_0(\mathcal E)$ satisfies 
$$
\widehat H^{(0)}_{\mathcal E}\widehat u = \lambda\widehat u \quad {\rm in} \quad \mathcal E_{ext,R},
$$
and the Kirchhoff condition 
for some $R > 0$. Then $\widehat u = 0$ on $\mathcal E_{ext,R_1}$ for some $R_1 > 0$.
 \end{theorem}

Proof. We can assume that $\widehat u$ is real-valued. Take $R > 0$ large enough so that ${\rm supp}\, q_{\mathcal E} \subset \mathcal E_{int,R}$. On ${\bf e} \in \mathcal E_{ext,R}$, $\widehat u$ is written as 
$$
\widehat u(z) = c_{\bf e}(1)\frac{\sin\sqrt{\lambda}z}{\sin\sqrt{\lambda}} + c_{\bf e}(0)\frac{\sin\sqrt{\lambda}(1-z)}{\sin\sqrt{\lambda}}.
$$
Then, we have

\begin{equation}
\begin{split}
\int_0^1|\widehat u(z)|^2dz  = & \frac{1}{2\sin^2\sqrt{\lambda}}\Big(1 - \frac{\sin\sqrt{\lambda}\cos\sqrt{\lambda}}{\sqrt{\lambda}}\Big)\Big(c_{\bf e}(0)^2 + c_{\bf e}(1)^2\Big) \\
&+ \frac{1}{\sin^2\sqrt{\lambda}}\Big(\frac{\sin\sqrt{\lambda}}{\sqrt{\lambda}} - \cos\sqrt{\lambda}\Big)c_{\bf e}(0)c_{\bf e}(1).
\end{split}
\nonumber 
\end{equation}
Note that $\sin\sqrt{\lambda} < \sqrt{\lambda}$ and $|\cos\sqrt{\lambda}| < 1$ for $\lambda > 0$ and  $\sqrt{\lambda}/\pi \not\in {\bf Z}$. Also $|a-b| < 1 - ab$ for $|a| <1$ and $|b| < 1$. Then, letting $a= \frac{\sin{\sqrt{\lambda}}}{\sqrt{\lambda}}$ and $b = \cos\sqrt{\lambda}$, we  have
$$
\Big|\frac{\sin\sqrt{\lambda}}{\sqrt{\lambda}} - \cos\sqrt{\lambda}\Big| < 1 - \frac{\sin\sqrt{\lambda}\cos\sqrt{\lambda}}{\sqrt{\lambda}}.
$$
Hence there exists a constant $C(\lambda) > 0$ such that 
\begin{equation}
C(\lambda)^{-1}\big(|c_{\bf e}(0)| + |c_{\bf e}(1)|\big)  \leq \|{\widehat u}_{\bf e}\|_{L^2_{\bf e}} \leq C(\lambda)\big(|c_{\bf e}(0)| + |c_{\bf e}(1)|\big)
\label{ce(0)ce(1)equivwidehatuL2norm}
\end{equation}
for all ${\bf e} \in \mathcal E_{ext,R}$.  For $v \in \mathcal V$, we put
$\widehat w(v) = \widehat u(v)$. 
Then, taking account of Lemma \ref{Lemma3.2}, we have
$$
\big(- \widehat\Delta_{\mathcal V} + \cos\sqrt{\lambda}\big)\widehat w = 0, \quad {\rm on}
\quad \mathcal E_{ext, R}.
$$
Since $\widehat u \in \widehat {\mathcal B}^{\ast}_0(\mathcal E)$, the inequality (\ref{ce(0)ce(1)equivwidehatuL2norm}) implies that $\widehat w \in 
\mathcal B^{\ast}_0(\mathcal V)$. By the Rellich type theorem for lattice Schr{\"o}dinger operators (see Theorem 5.1 in \cite{AnIsoMo17}), $\widehat w$ vanishes for $|v| > R'$ for some $R'>0$. This proves the theorem. \qed

\medskip
We say that the operator $\widehat H_{\mathcal E}-\lambda$ has the {\it unique continuation property} on $\mathcal E$  if the following assertion holds: If $\widehat u$ satisfies $(\widehat H_{\mathcal E} - \lambda)\widehat u = 0$ on $\mathcal E$ and $\widehat u = 0$ on $\widehat E_{ext,R}$ for some $R > 0$, then $\widehat u = 0$ on $\mathcal E$.

We introduce a new assumption.

\medskip
\noindent
{\bf (UC)}  {\it For any $\lambda \in {\rm Int}\,\sigma_{e}(\widehat H^{(0)}_{\mathcal E})\setminus  \sigma^{(0)}_{\mathcal T}$, $\widehat H_{\mathcal E} - \lambda$ has the unique continuation property on $\mathcal E$.}

\begin{theorem} Assume (UC). Then,
$$
\sigma_p(\widehat H_{\mathcal E}) \cap \sigma_e(\widehat H_{\mathcal E}) \subset \sigma^{(0)}_{\mathcal V}\cup\sigma^{(0)}_{\mathcal T}.
$$
\end{theorem}

Proof. Let $\lambda \in \sigma_p(\widehat H_{\mathcal E}) \cap \sigma_e(\widehat H_{\mathcal E})$, and $\widehat u$ be the associated eigenfunction.  If $\lambda \not\in \sigma^{(0)}_{\mathcal V} \cup \sigma^{(0)}_{\mathcal T}$, $\widehat u$ is compactly supported by Theorem \ref{RelichTypeTheorem}. By the unique continuation property, $\widehat  u$ vanishes identically. This is a contradiction.
\qed

\medskip
Let us check that the square lattice and the hexagonal lattice satisfy (UC).

\begin{lemma}
For the square lattice in ${\bf R}^d$ with $d \geq 2$,  (UC) holds.
\end{lemma}

Proof. Suppose $(\widehat H_{\mathcal E} - \lambda)\widehat u=0$ on $\mathcal E$, and $\widehat u = 0$ on any edge in the region $\{x_n > k\}$, where $k$ is an integer. Then, $\widehat u=0$ on any vertex  on the plane $\{x_n = k\}$. Since $\lambda$ is not a Dirichlet eigenvalue on any edge, $\widehat u = 0$ on any edge in the plane $\{x_n = k\}$. Due to the Kirchhoff condition, this implies that $\widehat u'(z)=0$ on any vertex on the plane $\{x_n=k\}$. By the uniqueness of the initial value problem for $- d^2/dz^2 + q_{\bf e}(z)-\lambda$, $\widehat u = 0$ on any edge in the region $\{x_n > k-1\}$. By induction, we have seen that $\widehat u = 0$ on $\mathcal E$. \qed

\begin{lemma}
For the hexagonal lattice in ${\bf R}^2$, (UC) holds.
\end{lemma}

Proof.  Instead of the region $\{x_n > k\}$, consider the region $\{x_2 > k\sqrt3/2\}$ and argue as above. \qed

\subsection{Radiation condition}
Let $\lambda_j(x)$, $j = 1, 2, \cdots, s$,  be the eigenvalues of $H_0(x)$ and $P_j(x)$ the associated eigenprojections. 
Let $H_0$ be the operator of multiplication by $H_0(x)$ on $\big(L^2({\bf T}^d)\big)^s$. 
In \cite{AnIsoMo17}, Lemma 4.7, we have proven that 
if $\rho \not\in \big({\rm Int}\, \sigma(H_0)\big)\setminus \mathcal T$, the operator
$$
f \to \frac{f(x)}{\lambda_j(x) - \rho \mp i0}
$$
is bounded from $\mathcal B({\bf T}^d)$ to $\mathcal B^{\ast}({\bf T}^d)$.

For a distribution $u \in \mathcal D'({\bf T}^d)$, its wave front set $WF^{\ast}(u)$ is defined as follows: For $(x_0,\omega) \in {\bf R}^d\times S^{d-1}$,
$(x_0,\omega) \not\in WF^{\ast}(u)$  if and only if there exist $0 < \delta < 1$ and $\chi(x) \in C_0^{\infty}({\bf R}^d)$ such that $\chi(x_0) =1$ and 
\begin{equation}
\mathop{\lim_{R\to\infty}}\frac{1}{R}\int_{|\xi|<R}|C_{\omega,\delta}(\xi)
(\widetilde{\chi u})(\xi)|^2d\xi = 0,
\end{equation}
where $\widetilde{\chi u}$ is the Fourier transform of $\chi u$ and $C_{\omega,\delta}(\xi)$  is the characteristic function of the cone 
$\{\xi \in {\bf R}^d\, ; \, \omega\cdot\xi > \delta|\xi|\}$.

In \cite{AnIsoMo17}, Theorem 6.1, we have shown that for any $f \in \mathcal B({\bf T}^d)$,  $1 \leq j \leq s$ and $\rho \in \sigma(H_0)\setminus \mathcal T$, it holds that

\medskip
\noindent
$(RC)_+$  : \ \  \ 
$\displaystyle{WF^{\ast}\big(\frac{P_jf}{\lambda_j(x) - \rho - i0}\big) \subset \{(x,\omega_x) \, ; \, x \in M_{\rho,j}\},}$

\medskip
\noindent
$(RC)_-$ : \ \  \ 
$\displaystyle{WF^{\ast}\big(\frac{P_jf}{\lambda_j(x) - \rho + i0}\big) \subset \{(x,-\omega_x) \, ; \, x \in M_{\rho,j}\}}$,

\medskip
\noindent
where $\omega_x \in S^{d-1}\cap T_x(M_{\lambda,j})^{\perp}$, $\omega(x)\cdot \nabla \lambda_j(x) < 0$. Moreover, for $f \in \mathcal B({\bf T}^d)$, $u = (H_0(x) - \lambda \mp i0)^{-1}f \in \mathcal B^{\ast}({\bf T}^d)$ is a unique solution to the equation $(H_0(x) - \rho)u = f$ satisfying $(RC)_+$, or $(RC)_-$. 

These facts are also extended to the case with compactly supported perturbations.

\medskip
In the following, if we say that $\widehat u \in \widehat{\mathcal B}^{\ast}(\mathcal E)$ is a solution to  the equation $(- \widehat\Delta_{\mathcal E} + q_{\mathcal E}- \lambda)\widehat u = \widehat f$, $\widehat u$ is assumed to satisfy the Kirchhoff condition. Taking account of (\ref{Cossqrtlambda}), we define the radiation condition as follws.

\begin{definition} Let $\lambda > 0$. 
We say that  a solution $\widehat u \in \widehat{\mathcal B}^{\ast}(\mathcal E)$ of the equation $(- \widehat\Delta_{\mathcal E} + q_{\mathcal E}- \lambda)\widehat u = \widehat f$  satisfies the {\it outgoing radiation condition} if either (i) or (ii) holds:

(i)  $\sin\sqrt{\lambda} > 0$, and $u = \mathcal U_{\mathcal E}\widehat u$ satisfies $(RC)_+$ with $\rho = - \cos\sqrt{\lambda}$,

(ii)  $\sin\sqrt{\lambda} < 0$, and $u = \mathcal U_{\mathcal E}\widehat u$ satisfies $(RC)_-$  with $\rho = - \cos\sqrt{\lambda}$.

\noindent
Similarly, we say that  a solution $\widehat u \in \widehat{\mathcal B}^{\ast}(\mathcal E)$ of the equation $(- \widehat\Delta_{\mathcal E} + q_{\mathcal E} - \lambda)\widehat u = \widehat f$  satisfies the {\it  incoming radiation condition} if  either (i) or (ii) holds:

(i)  $\sin\sqrt{\lambda} > 0$, and $u = \mathcal U_{\mathcal E}\widehat u$ satisfies $(RC)_-$  with $\rho = - \cos\sqrt{\lambda}$,  

(ii) $\sin\sqrt{\lambda} < 0$, and $u = \mathcal U_{\mathcal E}\widehat u$ satisfies $(RC)_+$  with $\rho = - \cos\sqrt{\lambda}$.

\noindent
If $\widehat u$ satisfies the outgoing or incoming radiation condition, we say that $\widehat u$ satisfies the radiation condition.
\end{definition}

\begin{lemma}
\label{LemmaRadCondUnique}
Let $\lambda \in \big({\rm Int}\,\sigma_e(\widehat H_{\mathcal E})\big) \setminus \big(\sigma^{0)}_{\mathcal V} \cup\sigma^{(0)}_{\mathcal T}\big)$. Then, the solution $\widehat u \in \widehat{\mathcal B}^{\ast}({\mathcal E})$ of the equation $(- \widehat\Delta_{\mathcal E} + q_{\mathcal E} - \lambda)\widehat u = \widehat f$ satisfying the radiation condition is unique. 
\end{lemma}

Proof. 
We show that if $\widehat u \in \widehat{\mathcal B}^{\ast}(\mathcal E)$ satisfies $(- \widehat\Delta_{\mathcal E} + q_{\mathcal E}- \lambda)\widehat u = 0$ and the radiation condition, then $\widehat u =0$. On each edge ${\bf e}$, $\widehat u(z)$ is rewritten as (see (\ref{S3ue(z)=Ce1+Ce0}))
$$
\widehat u_{\bf e}(z) =  c_{\bf e}(1)\frac{\phi_{{\bf e}0}(z,\lambda)}{\phi_{{\bf e}0}(1,\lambda)} + c_{\bf e}(0)\frac{\phi_{{\bf e}1}(z,\lambda)}{\phi_{{\bf e}1}(1,\lambda)}.
$$
Then, letting $\widehat w(v) = c_{\bf e}(1)$ or $c_{\bf e}(0)$, if ${\bf e}(0) = v$ or ${\bf e}'(1)=v$, we see that $(- \widehat\Delta_{\mathcal V,\lambda} + \widehat Q_{\mathcal V,\lambda})\widehat w = 0$ 
(see (\ref{DeltaVlambdawidehataU+lambdaQVU=f})) and $\widehat w$ satisfies the radiation condition. This implies $\widehat w = 0$ by Lemma 7.6 in \cite{AnIsoMo17}. \qed 

\medskip
In our previous work \cite{AnIsoMo17}, the radiation condition was also introduced for the vertex Laplacian (see Lemmas 4.8 and 6.2 in \cite{AnIsoMo17}). 
Let $\widehat f \in \mathcal{B} (\mathcal E)$. For a solution $\widehat u$ of the edge Schr{\"o}dinger equation $(- \widehat\Delta_{\mathcal E}+ q_{\mathcal E} - \lambda)\widehat u = \widehat f$, let $\widehat u\big|_{\mathcal V}$ be its
 restriction on  $\mathcal V$. 
 Then $\widehat u\big|_{\mathcal V}$ satisfies the vertex Schr{\"o}dinger equation 
\begin{equation}
( - \widehat\Delta_{\mathcal V} + \cos\sqrt{\lambda})\widehat u\big|_{\mathcal V} = \widehat g,
\end{equation}
where $\widehat g \in \mathcal{B} (\mathcal V)$. 
Comparing these two definitions of radiation condition, one can show the following lemma. 

\begin{lemma} 
\label{RadCondequivlatticeedge} A solution
$\widehat u$ of the edge Schr{\"o}dinger equation satisfies the radiation condition if and only if the solution $\widehat u\big|_{\mathcal V}$ of the vertex Schr{\"o}dinger equation satisfies the radiation condition.
\end{lemma}


\subsection{Limiting absorption principle}
\label{SubsectionLAP}
\begin{theorem}
\label{LAPwholespace}
Let $I$ be a compact interval in $\big({\rm Int}\,\sigma_e(\widehat H_{\mathcal E})\big) \setminus \big( \sigma^{(0)}_{\mathcal V} \cup \sigma^{(0)}_{\mathcal T}$\big). \\
\noindent
(1) There exists a constant $C > 0$ such that
\begin{equation}
\|( \widehat H_{\mathcal E} - \lambda \mp i\epsilon)^{-1}\|_{{\bf B}(\mathcal B(\mathcal E);{\mathcal B}^{\ast}(\mathcal E))} \leq C
\end{equation}
for any $\lambda \in I$ and $\epsilon > 0$. \\
\noindent
(2) For any $\lambda \in I$ and $\sigma > 1/2$, there exists a strong limit 
\begin{equation}
{\mathop{\rm s-lim}_{\epsilon\downarrow 0}}( \widehat H_{\mathcal E} - \lambda \mp i\epsilon)^{-1} := (\widehat H_{\mathcal E} - \lambda \mp i0)^{-1} \in 
{\bf B}\big(\ell^{2,\sigma}(\mathcal E);\ell^{2,-\sigma}(\mathcal E)\big).
\end{equation}
\noindent
(3) For any $\widehat f \in \widehat\ell^{2,\sigma}(\mathcal E)$, $(\widehat H_{\mathcal E} - \lambda \mp i0)^{-1}\widehat f$ is an $\widehat\ell^{2,-\sigma}(\mathcal E)$-valued strongly continuous function of $\lambda \in I$.\\
\noindent
(4) For any $\widehat f, \widehat g \in \widehat{\mathcal B}(\mathcal E)$, there exists a limit
\begin{equation}
{\mathop{\lim}_{\epsilon\downarrow 0}}\big((\widehat H_{\mathcal E} - \lambda \mp i\epsilon)^{-1}\widehat f,\widehat g\big) := \big(( \widehat H_{\mathcal E} - \lambda \mp i0)^{-1}\widehat f,\widehat g\big),
\end{equation}
and $ \big((\widehat H_{\mathcal E} - \lambda \mp i0)^{-1}\widehat f,\widehat g\big)$ is a continuous function of $\lambda \in I$.

\noindent
(5) For any $\widehat f \in \widehat{\mathcal B}(\mathcal E)$, $( \widehat H_{\mathcal E} - \lambda - i0)^{-1}\widehat f$ satisfies the outgoing radiation condition, and $(\widehat H_{\mathcal E} - \lambda + i0)^{-1}\widehat f$ satisfies the incoming radiation condition.
\end{theorem}

Proof. The limiting absorption principle for $(-\widehat{\Delta}_{\mathcal V} + \cos\sqrt{\lambda} \mp i0)^{-1}$ is proved in \cite{AnIsoMo17}, Theorem 6.1. 
Taking account of this fact and the formula (\ref{Resolventformula2}), 
one can prove this theorem for the case of $\widehat{H}_{\mathcal E}^{(0)}$. Using the fact that the multiplication operator $q_{\mathcal E}$ is relatively compact, one can prove the theorem for the case of $\widehat H_{\mathcal E}$ utilizing Lemmas \ref{LemmaRadCondUnique} and \ref{RadCondequivlatticeedge} by the perturbation argument.  Since the similar argument is already given in the proof of Theorem 7.7 in \cite{AnIsoMo17}, we do not repeat it. \qed

\medskip
Let us return to the problem of $\big(- \widehat\Delta_{\mathcal V,\lambda}+  \widehat{Q}_{\mathcal V,\lambda}\big)^{-1}$ we have encountered in \S 3. 
First we consider the case $q_{\mathcal E} = 0$. 
If $\lambda \in {\bf R}$, $- \widehat\Delta_{\mathcal V,\lambda}+  \widehat{Q}_{\mathcal V,\lambda}$ is self-adjoint by modifying the inner product. 
Arguing as in the proof of Theorem \ref{LAPwholespace}, one can prove the existence of $\big(- \widehat\Delta_{\mathcal V,\lambda}+  \widehat{Q}_{\mathcal V,\lambda \pm i0} \big)^{-1}$ as a bounded operator in ${\bf B}(\mathcal B(\mathcal V) ; {\mathcal B}^{\ast}(\mathcal V))$.  
Using this fact, one can see that when $\epsilon \to 0$,
$\big(- \widehat\Delta_{\mathcal V,\lambda\pm i\epsilon}+  \widehat{Q}_{\mathcal V,\lambda\pm i\epsilon}\big)^{-1}$
is uniformly bounded as an operator from $\mathcal B(\mathcal V)$ to $\mathcal B^{\ast}(\mathcal V)$. The same fact holds true if we add $q_{\mathcal E}$ and use the resolvent equation. Moreover, the existence of the limit $\lim_{\epsilon \to 0}\big(- \widehat\Delta_{\mathcal V,\lambda\pm i\epsilon}+  \widehat{Q}_{\mathcal V,\lambda\pm i\epsilon}\big)^{-1}$ is guaranteed.  
The arguments in \S 3 are then justified if we consider
all operators in $\mathcal B(\mathcal V)$ or $\mathcal B^{\ast}(\mathcal V)$.

\subsection{Analytic continuation of the resolvent}
It is well-known that for the continuous model, the resolvent of the Schr{\"o}dinger operator $- \Delta + V(x)$, where $V(x)$ has compact support, the boundary value of the resolvent $( - \Delta + V(x) - \lambda - i0)^{-1}$ has a meromorphic continuation into the lower half plane $\{{{\rm Re}\, \lambda > 0, \ \rm Im}\, \lambda < 0\}$ as an operator from the space of compactly supported $L^2({\bf R}^n)$ functions to $L^2_{loc}({\bf R}^n)$. This is proven by considering the free case, i.e.  the operator
$$
\int_{{\bf R}^n}\frac{e^{ix\cdot\xi}\widetilde f(\xi)}{|\xi|^2 - \zeta}d\xi = 
\int_0^{\infty}\frac{\int_{S^{n-1}}e^{ir\omega\cdot x}\widetilde f(r\omega)d\omega}{r^2 - \zeta}r^{n-1}dr 
$$
($\widetilde f(\xi)$ being the Fourier transfrom of $f$), 
for ${\rm Im}\,\zeta > 0$, deforming the path of integration into the lower half-plane, and then applying the perturbation theory. This idea also works for the discrete model, and one can show that the boundary value of the resolvent of the vertex Hamiltonian and the edge Hamitonian can be continued meromorphically into the lower half-plane $\{{{\rm Re}\,\lambda > 0, \ \rm Im}\, \lambda < 0\}$ with possible branch points on $\mathcal T$, when the perturbation is compactly supported. 


\section{Spectral representation and  S-matrix}
The spectral representation, also called the generalized Fourier transformation, was introduced by K. O. Friedrichs. Given a selfadjont operator $A$  with absolutely continuous spectrum $I \subset {\bf R}$, one prepares an auxiliary Hilbert space $\bf h$ and a unitray operator $\mathcal F$ from $\mathcal H_{ac}(A)$ to $L^2(I,{\bf h};d\lambda)$ so that $(\mathcal F Au)(\lambda) = \lambda ({\mathcal F}u)(\lambda)
$ holds for any $\lambda \in I$ and $u \in D(A)$. We apply this framework  to scattering theory, where $A$ is a differential operator or a discrete operator, ${\bf h}$ is the $L^2$-space over the characteristic set of $A$, and $\mathcal F$ is constructed by observing the behavior at infinity of the  resolvent of $A$. The boundary values of the resolvent $(A - \lambda \mp i0)^{-1}$ give rise to two spectral representations, and their difference is described by the S-matrix, which is a unitary integral operator on the characteristic set. This is the operator theoretical background of the scattering experiment, which is originally a time-dependent phenomenon. Below, we elucidate this picture for the case of the edge model. 

\subsection{Spectral representation}
We need to introduce a little more notation. Letting $P_{\mathcal V,j}(x)$ be the projection associated with the eigenvalue $\lambda_j(x)$ of $H_0(x)$, we put
\begin{equation}
D^{(0)}(\lambda \pm i0) = \frac{\sin\sqrt{\lambda}}{\sqrt{\lambda}}\sum_{j=1}^s\frac{1}{\lambda_j(x) + \cos\sqrt{\lambda} \mp i\sigma(\lambda)0}P_{\mathcal V,j}(x).
\label{A0lambdapmi02ndFormula}
\end{equation}
Note that by (\ref{S2H0(x)rewritten})
\begin{equation}
D^{(0)}(\lambda \pm i0) = \frac{\sin\sqrt{\lambda}}{\sqrt{\lambda}}\, \mathcal U_{\mathcal V}\big(- \widehat\Delta_{\mathcal V} + \cos\sqrt{\lambda \pm i0}\big)^{-1}{\mathcal U_{\mathcal V}}^{\ast},
\label{S5D0lambdaandresolvent}
\end{equation}
where $\mathcal U_{\mathcal V}$ is the discrete Fourier transformation  defined by (\ref{S1UdDefine}).

Recall that the multiple lattice structure comes from (\ref{V=cupj=1sp(j)+mathcalL}), to which,  by passing to the Fourier series, one associates the function space $\big(L^2({\bf T}^d)\big)^{s}$ on the torus.
Noting that by Lemma \ref{Lemma2.1Vast=p(s)}, the mapping 
$$
\{1,\cdots,\nu\} \ni i \to k \in \{1,\cdots,s\}
$$
defined by
$$
\mathcal V_{\ast} = \{{\bf e}_{\ell\ast}(0), {\bf e}_{\ell\ast}(1) \, ; \, 
1 \leq \ell\leq \nu\} \ni {\bf e}_{\ell\ast}(i) \to p^{(k)} \in 
\{p^{(1)}, \cdots, p^{(s)}\},
$$
where ${\bf e}_{\ell\ast}(i) = p^{(k)}$,
is surjective. We define a projection $P^{\mathcal E}_{{\bf e}_{\ell\ast}(i)}$ by
\begin{equation}
P^{\mathcal E}_{{{\bf e}_{\ell}}_{\ast}(i)} : \big(L^2({\bf T}^d)\big)^{s} \ni 
\left(f_1(x),\cdots,f_s(x)\right) \to f_{k}(x). 
 \end{equation} 
We then define the operators $\Pi_{\ell}(\lambda)$ and $\Pi(\lambda)$  by
\begin{equation}
\Pi_{\ell}(\lambda) = \frac{1}{\sqrt{d_{\mathcal V}}} \frac{\sqrt{\lambda}}{\sin \sqrt{\lambda}} \left( e^{-i{\rm Ind}({\bf e}_{\ell})\cdot x} \frac{\sin\sqrt\lambda z}{\sqrt\lambda}P^{\mathcal E}_{{{\bf e}_{\ell}}_{\ast}(1)} + 
 \frac{\sin\sqrt\lambda (1-z)}{\sqrt\lambda}P^{\mathcal E}_{{{\bf e}_{\ell}}_{\ast}(0)} \right) , 
\nonumber
\end{equation} 
\begin{equation}
\Pi(\lambda) = \big(\Pi_1(\lambda),\cdots,\Pi_{\nu}(\lambda)\big) : 
\big(L^2({\bf T}^d)\big)^s \to L^2({\bf T}^d\times \mathcal E_{\ast}),
\nonumber
\end{equation}
which is naturally extended to spaces of distributions
$$
\Pi(\lambda) :  \big(\mathcal D'({\bf T}^d)\big)^s \to \mathcal D'({\bf T}^d\times \mathcal E_{\ast}).
$$
In view of (\ref{DefineTmathcalElambda}), we have
\begin{equation}
\big(\widehat T^{(0)}_{\mathcal E}(\lambda)\widehat u\big)_{\bf e}(z) =
\frac{\sqrt{\lambda}}{\sin\sqrt{\lambda}}\left( \frac{\sin\sqrt{\lambda}z}{\sqrt{\lambda}}\widehat u({\bf e}(1)) + 
\frac{\sin\sqrt{\lambda}(1-z)}{\sqrt{\lambda}}\widehat u({\bf e}(0))\right).
\label{DefineT0mathcalElambda}
\end{equation}
\begin{lemma}
\label{LemmaUEI0=Blambda}
For $\lambda \in \big({\rm Int}\,\sigma_e(\widehat H^{(0)}_{\mathcal E})\big) \setminus \big(\sigma^{(0)}_{\mathcal T}\cup\sigma^{(0)}_{\mathcal V}\big)$, 
\begin{equation}
\mathcal U_{\mathcal E}\widehat T^{(0)}_{\mathcal E}(\lambda) = 
\Pi(\lambda)\,\mathcal U_{\mathcal V}
\nonumber
\end{equation}
\end{lemma}

Proof. Recall that  each edge $\bf e$ is written as 
${\bf e} = {\bf e}_{\ell} + [n]$, where
$$
{\bf e}(0) = {{\bf e}_{\ell}}_{\ast}(0) + {\bf v}(n), \quad {\bf e}(1) = {{\bf e}_{\ell}}_{\ast}(1) + {\bf v}({\rm Ind}\,({\bf  e}_{\ell}) + n).
$$
By the definition (\ref{mathcalUEdefine}) and (\ref{DefineT0mathcalElambda}), we compute
\begin{equation}
\begin{split}
&\mathcal U_{\mathcal E,\ell}\widehat T^{(0)}_{\mathcal E}(\lambda)\widehat f(x,z) \\ 
& = (2\pi)^{-d/2} \frac{\sqrt{\lambda}}{\sin \sqrt{\lambda} }
\sum_{n\in {\bf Z}^d}e^{in\cdot x}
 \widehat f\big({\bf e}_{\ell\ast}(1)+ {\bf v}({\rm Ind}({\bf e}_{\ell}) +n)\big)\frac{\sin\sqrt{\lambda}z}{\sqrt{\lambda}}  \\
& + (2\pi)^{-d/2} \frac{\sqrt{\lambda}}{\sin \sqrt{\lambda} }
\sum_{n\in {\bf Z}^d}e^{in\cdot x}
 \widehat f\big({\bf e}_{\ell\ast}(0)+ {\bf v}( n)\big)\frac{\sin\sqrt{\lambda}(1-z)}{\sqrt{\lambda}} \\
&  = (2\pi)^{-d/2} \frac{\sqrt{\lambda}}{\sin \sqrt{\lambda} }
\sum_{n\in {\bf Z}^d}e^{in\cdot x} e^{-i{\rm Ind}({\bf e}_{\ell})\cdot x}
 \widehat f\big({\bf e}_{\ell\ast}(1)+ {\bf v}( n)\big)\frac{\sin\sqrt{\lambda}z}{\sqrt{\lambda}}  \\
& + (2\pi)^{-d/2} \frac{\sqrt{\lambda}}{\sin \sqrt{\lambda} }
\sum_{n\in {\bf Z}^d}e^{in\cdot x}
 \widehat f\big({\bf e}_{\ell\ast}(0)+ {\bf v}(n)\big)\frac{\sin\sqrt{\lambda}(1-z)}{\sqrt{\lambda}}.
\end{split}
\nonumber
\end{equation} 
The lemma then follows from this. \qed

\medskip

Using the identification (\ref{S2Identificationf(s)withfj(s)}), we  put
\begin{equation}
\Phi^{(0)}(\lambda) = \mathcal U_{\mathcal V}\widehat T^{(0)}_{\mathcal E}(\lambda)^{\ast} = \mathcal U_{\mathcal V}\widehat T^{(0)}_{\mathcal V}(\lambda).
\label{DefinePhi(0)lambdaS5}
\end{equation}

\begin{lemma}
The resolvent $\widehat R^{(0)}_{\mathcal E}(\lambda \pm i0)$ has the following expressions:
\begin{eqnarray}
\widehat R^{(0)}_{\mathcal E}(\lambda \pm i0) &=& 
\Phi^{(0)}(\lambda)^{\ast}D^{(0)}(\lambda \pm i0)\Phi^{(0)}(\lambda) + r^{(0)}_{\mathcal E}(\lambda)
\label{R0Elambda=Phi0AlambdaPhi0} \\
 &=& {\mathcal U_{\mathcal E}}^{\ast}\Pi(\lambda)D^{(0)}(\lambda \pm i0)\Phi^{(0)}(\lambda) + r^{(0)}_{\mathcal E}(\lambda).
\label{R0Elambda=Phi0AlambdaPhi0(1)}
\end{eqnarray}
\end{lemma}

Proof. The formula (\ref{R0Elambda=Phi0AlambdaPhi0}) follows from 
 Lemma \ref{S3RElambda+i0Formula1} and (\ref{S5D0lambdaandresolvent}). 
By Lemma \ref{TmathcalVlambdaadjoint}, we have 
$\widehat T_{\mathcal V}(\lambda)^{\ast} = 
\widehat T_{\mathcal E}(\lambda)$. Therefore  by 
(\ref{DefinePhi(0)lambdaS5}), $\Phi_0(\lambda)^{\ast} = \mathcal U_{\mathcal E}^{\ast}\Pi(\lambda)$, which proves  (\ref {R0Elambda=Phi0AlambdaPhi0(1)}). \qed

\medskip
Using these formulas, we can construct a spectral representation of the absolutely continuous part of $\widehat H^{(0)}_{\mathcal E}$.   We put
\begin{equation}
M_{\mathcal E,\lambda,j} = \{x \in {\bf T}^d\, ; \, \lambda_j(x) + \cos\sqrt{\lambda} = 0\}, 
\nonumber
\end{equation} 
\begin{equation}
(\varphi,\psi)_{\lambda,j} = \int_{M_{\mathcal E,\lambda,j}}\varphi(x)\overline{\psi(x)}dS_j, 
\nonumber
\end{equation}
\begin{equation}
dS_j = \frac{|\sin \sqrt{\lambda}|}{\sqrt{\lambda}} \frac{dM_{\mathcal E,\lambda,j}}{|\nabla_x\lambda_j(x)|}.
\nonumber
\end{equation}
Then, by virtue of (\ref{R0Elambda=Phi0AlambdaPhi0}), for $\lambda \in \big({\rm Int}\,\sigma_e(\widehat H^{(0)}_{\mathcal E})\big) \setminus \big(\sigma^{(0)}_{\mathcal T}\cup\sigma^{(0)}_{\mathcal V}\cup\sigma_p(\widehat H^{(0)}_{\mathcal E})\big)$, 
\begin{equation}
\begin{split}
&  \frac{1}{2\pi i}\Big(\big(\widehat R^{(0)}_{\mathcal E}(\lambda + i0) - \widehat R^{(0)}_{\mathcal E}(\lambda - i0)\big)\widehat f, \widehat g\Big)\\
&  = \sum_{j=1}^s
\big(P_{\mathcal V,j}\Phi^{(0)}(\lambda)\widehat f,P_{\mathcal V,j}\Phi^{(0)}(\lambda)\widehat g\big)_{\lambda,j}, 
\end{split}
\label{Parseval0forR0lambda}
\end{equation}
where we have used 
\begin{equation}
\begin{split}
& \big(\widehat R_{\mathcal V}^{(0)}(-\cos\sqrt{\lambda + i0}) \widehat f-
\widehat R_{\mathcal V}^{(0)}(-\cos\sqrt{\lambda - i0})\widehat f, \widehat g\big) \\
& = 2\pi i \sum_j\int_{M_{\mathcal E, \lambda, j}}
P_{\mathcal V,j}\widehat f \cdot \overline{P_{\mathcal V,j}\widehat g}
\frac{dM_{\mathcal E,\lambda,j}}{|\nabla \lambda_j(x)|},
\end{split}
\nonumber
\end{equation}
(cf. (6.7) of \cite{AnIsoMo17}). 
Let $u_j(x)$ be a normalized eigenvector of $H_0(x)$ associated with the eigenvalue $\lambda_j(x)$:
$$
H_0(x)u_j(x) = \lambda_j(x)u_j(x).
$$
We put
\begin{equation}
\widehat{\mathcal F}_j^{(0)}(\lambda)\widehat f = ( u_j(x) \cdot \Phi^{(0)}(\lambda)\widehat f ) u_j (x), 
\label{S5DefineFj(0)lambda}
\end{equation} 
\begin{equation}
\widehat{\mathcal F}^{(0)}(\lambda) = \big(\widehat{\mathcal F}^{(0)}_{1}(\lambda),\cdots,\widehat{\mathcal F}^{(0)}_{s}(\lambda)\big),
\nonumber
\end{equation} 
\begin{equation}
{\bf h}_{\lambda} = {\mathop\oplus_{j=1}^s}L^2\big(M^{\mathcal E}_{\lambda,j};dS_j\big),
\nonumber
\end{equation}
\begin{equation}
{\bf H} = L^2\big((0,\infty),{\bf h}_{\lambda};d\lambda\big).
\nonumber
\end{equation}
By (\ref{Parseval0forR0lambda})
\begin{equation}
\frac{1}{2\pi i}\Big(\big(\widehat R^{(0)}_{\mathcal E}(\lambda + i0)-\widehat R^{(0)}_{\mathcal E}(\lambda - i0)\big)\widehat f,\widehat g\Big) 
= (\widehat{\mathcal F}^{(0)}(\lambda)\widehat f,\widehat{\mathcal F}^{(0)}(\lambda)\widehat g)_{{\bf h}_{\lambda}}.
\nonumber
\end{equation}
Letting $E^{(0)}(\lambda)$ be the spectral measure for $\widehat H^{(0)}_{\mathcal E}$ and integrating this equality, 
we have
\begin{equation}
(E^{(0)}(I)\widehat f,\widehat g) = \int_I(\widehat{\mathcal F}^{(0)}(\lambda)\widehat f,\widehat{\mathcal F}^{(0)}(\lambda)\widehat g)_{{\bf h}_{\lambda}}d\lambda
\nonumber
\end{equation}
for any interval $I \subset \big({\rm Int}\,\sigma_e(\widehat H^{(0)}_{\mathcal E})\big) \setminus \big(\sigma^{(0)}_{\mathcal T}\cup\sigma^{(0)}_{\mathcal V}\cup\sigma_p(\widehat H^{(0)}_{\mathcal E})\big)$. 
Therefore, $\widehat{\mathcal F}^{(0)}$ is uniquely extended to an isometry from $\mathcal H_{ac}(\widehat H^{(0)}_{\mathcal E})$ to ${\bf H}$. We define
\begin{equation}
\widehat{\mathcal F}^{(0)} = 0, \quad {\rm on} \quad {\mathcal H}_p(\widehat H^{(0)}_{\mathcal E}).
\nonumber
\end{equation}

The generalized Fourier transform for $\widehat H_{\mathcal E}$ is constructed by the perturbation method. Define $\widehat{\mathcal F}^{(\pm)}(\lambda)$ by
\begin{equation}
\widehat{\mathcal F}^{(\pm)}(\lambda) = \widehat{\mathcal F}^{(0)}(\lambda)\Big(1 - q_{\mathcal E}\widehat R_{\mathcal E}(\lambda \pm i0)\Big) \in 
{\bf B}(\mathcal B(\mathcal E)\, ; \,{\bf h}_{\lambda}).
\label{DefineFpmlambdainmathcalE}
\end{equation}
By using the resolvent equation (see Lemma 7. 8 in \cite{AnIsoMo17}), 
we have 
\begin{equation}
\frac{1}{2\pi i}\Big(\big(\widehat R_{\mathcal E}(\lambda + i0)-\widehat R_{\mathcal E}(\lambda - i0)\big)\widehat f,\widehat g\Big) 
= (\widehat{\mathcal F}^{(\pm)}(\lambda)\widehat f,\widehat{\mathcal F}^{(\pm)}(\lambda)\widehat g)_{{\bf h}_{\lambda}}.
\nonumber
\end{equation}
We define the operator $\widehat{\mathcal F}^{(\pm)}$ by 
$(\widehat{\mathcal F}^{(\pm)}\widehat f)(\lambda) = \widehat{\mathcal F}^{(\pm)}(\lambda)\widehat f$. We define also
\begin{equation}
\widehat{\mathcal F}^{(\pm)} = 0,  \quad {\rm on} \quad {\mathcal H}_p(\widehat H_{\mathcal E}).
\nonumber
\end{equation}
Then, it gives a spectral representation for  $\widehat H_{\mathcal E}$ in the following sense.

\begin{theorem}
\label{EigenfunctionExpansionWholespace}
(1) The operator $\widehat{\mathcal F}^{(\pm)}$ is uniquely extended to a unitary operator from ${\mathcal H}_{ac}(\widehat H_{\mathcal E})$ to ${\bf H}$ annihilating ${\mathcal H}_p(\widehat H_{\mathcal E})$. 

\noindent
(2) It diagonalizes $\widehat H_{\mathcal E}$:
$$
\big(\widehat{\mathcal F}^{(\pm)}\widehat H_{\mathcal E}\widehat f\big)(\lambda) = 
\lambda\big(\widehat{\mathcal F}^{(\pm)}\widehat f\big)(\lambda), \quad \forall \widehat f \in D(\widehat H_{\mathcal E}).
$$
(3) The adjoint operator $\widehat{\mathcal F}^{(\pm)}(\lambda)^{\ast} \in {\bf B}({\bf h}_{\lambda};\mathcal B^{\ast}(\mathcal E))$ is an eigenoperator in the sense that
$$
(\widehat H_{\mathcal E} - \lambda)\widehat{\mathcal F}^{(\pm)}(\lambda)^{\ast}\phi = 0, \quad \forall \phi \in {\bf h}_{\lambda}.
$$
(4) For $\widehat f \in \mathcal H_{ac}(\widehat H_{\mathcal E})$, the inversion formula holds:
$$
\widehat f = \int_{\sigma_{ac}(\widehat H_{\mathcal E})}\widehat{\mathcal F}^{(\pm)}(\lambda)^{\ast}\big(\widehat{\mathcal F}^{(\pm)}\widehat f\big)(\lambda)d\lambda.
$$
\end{theorem}

The proof is almost the same as that for Theorem 7.11 in 
\cite{AnIsoMo17}, hence is omitted.

Note that  
the generalized eigenfunctions for the unperturbed operator $\widehat H^{(0)}_{\mathcal E}$ are constructed in 
\cite{KoSa15}. Using this result, it is not difficult to construct a complete family of generalized eigenfunctions for the perturbed operator $\widehat H_{\mathcal E}$, which turns out to be the integral kernel of the above generalized Fourier transformation $\widehat{\mathcal F}^{(\pm)}$.

\subsection{Resolvent expansion}
We observe the behavior at infinity of $\widehat R_{\mathcal E}(\lambda \pm i0)\widehat f$ in the sense of $\widehat{\mathcal B}^{\ast}(\mathcal E)$, which is equivalent to observing its singularities in the sense of $\mathcal B^{\ast}(\mathcal E)$.

\begin{lemma} 
\label{rbfelambdainL2}
For any compact interval $I \subset \big({\rm Int}\,\sigma_e(\widehat H^{(0)}_{\mathcal E})\big) \setminus \big(\sigma^{(0)}_{\mathcal T}\cup\sigma^{(0)}_{\mathcal V}\cup\sigma_p(\widehat H^{(0)}_{\mathcal E})\big)$, there exists a constant $C > 0$ such that
\begin{equation}
\|\{r^{(0)}_{\bf e}(\lambda)\widehat f_{\bf e}\}_{{\bf e}\in\mathcal E}\|_{\ell^2(\mathcal E)} \leq C\|\widehat f\|_{\ell^2(\mathcal E)}
\nonumber
\end{equation}
holds for all $\lambda \in I$ and ${\bf e} \in \mathcal E$.
\end{lemma}

Proof. Since $I$ is in the resolvent set of $- (d/dz)^2_D + q_{\bf e}$, the lemma follows. \qed

\medskip
Therefore, taking account of  (\ref{Resolventformula2}) and Lemma \ref{rbfelambdainL2}, we obtain the following asymptotic expansion.
For $\widehat f, \widehat g \in \mathcal B^{\ast}(\mathcal E)$, we use the following notation
\begin{equation}
\widehat f \simeq \widehat g \Longleftrightarrow 
\widehat f - \widehat g \in \mathcal B^{\ast}_0(\mathcal E).
\nonumber
\end{equation}

\begin{theorem}
\label{ResolventExpansionWholespace}
For any  $\lambda \in \big({\rm Int}\,\sigma_e(\widehat H^{(0)}_{\mathcal E})\big) \setminus \big(\sigma^{(0)}_{\mathcal T}\cup\sigma^{(0)}_{\mathcal V}\cup\sigma_p(\widehat H^{(0)}_{\mathcal E})\big)$ and $\widehat f \in \mathcal B(\mathcal E)$, we have
\begin{equation}
 \mathcal U_{\mathcal E}\widehat R^{(0)}_{\mathcal E}(\lambda \pm i0)\widehat f 
 \simeq \Pi(\lambda)D^{(0)}(\lambda \pm i0)\, \widehat{\mathcal F}_0(\lambda)\widehat f.
\nonumber
\end{equation}
\end{theorem}

Proof. Use (\ref{R0Elambda=Phi0AlambdaPhi0(1)}) and Lemma \ref{rbfelambdainL2}.
\qed

\medskip
By using the resolvent equation
\begin{equation}
\widehat R_{\mathcal E}(\lambda \pm i0) = \widehat R^{(0)}_{\mathcal E}(\lambda \pm i0)\big(1 - q_{\mathcal E}\widehat R_{\mathcal E}(\lambda \pm i0)\big),
\nonumber
\end{equation}
and (\ref{DefineFpmlambdainmathcalE}), 
one can extend Theorem \ref{ResolventExpansionWholespace} to the perturbed case.

\begin{theorem}
\label{ResolventExpansionPerturbedCase}
For any  $\lambda \in \big({\rm Int}\,\sigma_e(\widehat H_{\mathcal E})\big) \setminus \big(\sigma^{(0)}_{\mathcal T}\cup\sigma^{(0)}_{\mathcal V}\cup\sigma_p(\widehat H_{\mathcal E})\big)$ and $\widehat f \in \mathcal B(\mathcal E)$, we have
\begin{equation}
 \mathcal U_{\mathcal E}\widehat R_{\mathcal E}(\lambda \pm i0)\widehat f 
 \simeq \Pi(\lambda)D^{(0)}(\lambda \pm i0)\, \widehat{\mathcal F}^{(\pm)}(\lambda)\widehat f.
\nonumber
\end{equation}
\end{theorem}

This theorem shows that the spectral representation for the edge model arizes from that of the vertex model, and that the theory developped for the vertex model in \cite{AnIsoMo17} works for the edge model as well. In fact, the only difference is that the latter contains the injection operator term $\Pi(\lambda)$. 

\subsection{Helmholtz equation and S-matrix}
Theorem \ref{ResolventExpansionPerturbedCase} enables us to characterize the solution space to the Helmholtz equation.

\begin{lemma}
\label{LemmaRangeFlambdaast=solutionspace}
Let  $\lambda \in \big({\rm Int}\,\sigma_e(\widehat H_{\mathcal E})\big) \setminus \big(\sigma^{(0)}_{\mathcal T}\cup\sigma^{(0)}_{\mathcal V}\cup\sigma_p(\widehat H_{\mathcal E})\big)$ and $\widehat f \in \mathcal B(\mathcal E)$. Then
\begin{equation}
\{\widehat u \in \widehat{\mathcal B}^{\ast}(\mathcal E)\, ; \, 
(\widehat H_{\mathcal E} - \lambda)\widehat u = 0\} 
= \widehat{\mathcal F}^{(-)}(\lambda)^{\ast}{\bf h}_{\lambda}.
\end{equation}
\end{lemma}

One can then obtain the asymptotic expansion of solutions to the Helmholtz equation and derive the S-matrix.

\begin{theorem}
\label{HelmhlotzEqWholespace}
For any incoming data $\phi^{in} \in L^2(\mathcal M_{-\cos\sqrt\lambda})$, there exist a unique solution $\widehat u \in \widehat{\mathcal B}^{\ast}(\mathcal E)$ of the equation 
$$
(\widehat H_{\mathcal E} - \lambda)\widehat u = 0
$$
and an outgoing data $\phi^{out} \in L^2(M_{-\cos\sqrt\lambda})$ satisfying

\begin{equation}
\begin{split}
 \mathcal U_{\mathcal E}\widehat u
 \simeq &  
- \Pi(\lambda)\sum_{j=1}^s\frac{1}{\lambda_j(x) + \cos\sqrt\lambda + i\sigma(\lambda)}P_{\mathcal V,j}(x)\phi^{in} \\
& +  \Pi(\lambda)\sum_{j=1}^s\frac{1}{\lambda_j(x) + \cos\sqrt\lambda - i\sigma(\lambda)}P_{\mathcal V,j}(x)\phi^{out}.
\end{split}
\nonumber
\end{equation}
The mapping
\begin{equation}
S(\lambda) : \phi^{in} \to \phi^{out}
\nonumber
\end{equation}
is the S-matrix, which is unitary on ${\bf h}_{\lambda}$.
\end{theorem}

We omit the proof of Lemma \ref{LemmaRangeFlambdaast=solutionspace} and 
Theorem \ref{HelmhlotzEqWholespace}, since they are almost the same as that of  Theorem 7. 15 of \cite{AnIsoMo17} by the reasoning given above.

As is proven in \cite{KoSa15}, the wave operator
\begin{equation}
\widehat W_{\pm} = {\mathop{\rm s-lim}_{t\to\pm\infty}}\, e^{it\widehat H_{\mathcal E}}e^{-it\widehat H^{(0)}_{\mathcal E}}\widehat P_{ac}(\widehat H^{(0)}_{\mathcal E}),
\nonumber
\end{equation}
where $\widehat P_{ac}(\widehat H^{(0)}_{\mathcal E})$ is the projection onto the absolutely continuous subspace of $\widehat H^{(0)}_{\mathcal E}$, exists and is complete, i.e. ${\rm Ran}\, \widehat W_{\pm} = \mathcal H_{ac}(\widehat H_{\mathcal E})$. 
One can then follow the general scheme of scattering theory.
The scattering operator 
$$
\widehat S = (\widehat W^{(+)})^{\ast}\widehat W^{(-)}
$$
is unitary. Define $S$ by
$$
S =\widehat{\mathcal F}^{(0)}\widehat S (\widehat{\mathcal  F}^{(0)})^{\ast}.
$$
The S-matrix $S(\lambda)$ and the scattering amplitude $A(\lambda)$ are defined by
\begin{equation}
S(\lambda) = 1 - 2\pi iA(\lambda),
\nonumber
\end{equation}
\begin{equation}
A(\lambda) = \widehat{\mathcal F}^{(+)}(\lambda)q_{\mathcal E}\widehat{\mathcal F}^{(0)}(\lambda).
\label{S5FormulaScateringamplitude}
\end{equation}
Then $S(\lambda)$ is unitary on ${\bf h}_{\lambda}$, and for $\lambda \in \big({\rm Int}\,\sigma_e(\widehat H_{\mathcal E})\big) \setminus \big(\sigma^{(0)}_{\mathcal T}\cup\sigma^{(0)}_{\mathcal V}\cup\sigma_p(\widehat H_{\mathcal E})\big)$
\begin{equation}
(Sf)(\lambda) = S(\lambda)f(\lambda), \quad f \in {\bf H}.
\nonumber
\end{equation}
Since the resolvent has a meromorphic extension into the lower half-plane $\{{\rm Re}\, \lambda > 0, \ {\rm Im}\, \lambda < 0\}$ with possible branch points on $\mathcal T$, the formula (\ref{S5FormulaScateringamplitude}) implies that 
the S-matrix $S(\lambda)$ is also meromorphic in the same domain.


\section{From S-matrix to interior D-N map}
\label{SectionSmatrixtoDNmap}
We first recall the framework of boundary value problems for both of the edge model and the vertex model. We then show that the S-matrix and the D-N map for the edge model determine each other. 

\subsection{Boundary value problem}
For a subgraph $\Omega = \{\mathcal V_{\Omega}, \mathcal E_{\Omega}\} \subset \{\mathcal V, \mathcal E\}$ and $v \in \mathcal V$, $v \sim \Omega$ means that there exist a vertex $w \in \mathcal V_{\Omega}$ and an  edge ${\bf e} \in \mathcal E$ such that 
$v \sim w$, ${\bf e}(0) = v$ or ${\bf e}(1) = v$. 
For a connected subgraph $\Omega \subset \{\mathcal V, \mathcal E\}$, we define a subset $\partial\Omega = \{\mathcal V_{\partial\Omega}, \mathcal E_{\partial\Omega}\} \subset \{\mathcal V, \mathcal E\}$ by 
\begin{equation}
\mathcal V_{\partial\Omega} = \{v \not\in \mathcal V_{\Omega}\, ; \, 
v \sim {\Omega}\},
\nonumber
\end{equation}
\begin{equation}
\mathcal E_{\partial\Omega} = \{{\bf e} \in \mathcal E\, ; \, {\bf e}(0) \in \mathcal V_{\partial\Omega} \ {\rm or} \ {\bf e}(1) \in \mathcal V_{\partial\Omega}\}.
\nonumber
\end{equation}
We then put $\overline{\Omega} = \Omega \cup \partial\Omega$ and
\begin{equation}
\stackrel{\circ}{\mathcal V_{\overline{\Omega}}} = {\mathcal V}_{\Omega}, \quad \partial \mathcal V_{\overline{\Omega}} = \mathcal V_{\partial\Omega},
\nonumber
\end{equation}
which are called the set of {\it interior vertices}  and the set of {\it boundary vertices} of $\overline{\Omega}$, respectively. We  put
\begin{equation}
\mathcal V_{\overline{\Omega}} = \stackrel{\circ}{\mathcal V_{\overline{\Omega}}}\cup\, \partial\mathcal V_{\overline{\Omega}}.
\nonumber
\end{equation}
As for the edges, we simply put
\begin{equation}
{\mathcal E}_{\overline{\Omega}} = {\mathcal E}_{\Omega}\cup \mathcal E_{\partial\Omega}.
\nonumber
\end{equation}
We then define the {\it edge Dirichlet Laplacian} $\widehat\Delta_{\mathcal E_{\overline{\Omega}}}$ by
\begin{equation}
\widehat\Delta_{\mathcal E_{\overline{\Omega}}} u_{\bf e}(z) = \frac{d^2}{dz^2}u_{\bf e}(z), \quad 
{\bf e} \in \mathcal E_{\overline{\Omega}}
\nonumber
\end{equation}
whose domain $D(\widehat\Delta_{\mathcal E_{\overline{\Omega}}})$ is the set of all $u = \{u_{\bf e}\}_{{\bf e} \in \mathcal E_{\overline{\Omega}}} \in H^2(\mathcal E_{\overline{\Omega}})$ satisfying $u(v) = 0$ at any boundary vertex $v \in \partial \mathcal V_{\overline{\Omega}}$ and the Kirchhoff condition at any interior vertex $v \in \stackrel{\circ}{\mathcal V}_{\overline{\Omega}}$. By the standard argument, $\widehat\Delta_{\mathcal E_{\overline{\Omega}}}$ is self-adjoint.

The {\it vertex Dirichlet Laplacian on $\mathcal V_{\overline{\Omega}}$}  is defined in the same way as in (\ref{S3DefineVertexLaplacian}) :
\begin{equation}
\big(\widehat\Delta_{\mathcal V_{\overline{\Omega}}, \lambda}\widehat u\big)(v) = \frac{1}{{\rm deg}_{\mathcal V_{\overline{\Omega}}}(v)}
\sum_{w\sim v, w\in \mathcal V_{\overline{\Omega}}}\frac{1}{\psi_{wv}(1,\lambda)}\widehat u(w), \quad 
v \in \mathcal V_{\overline{\Omega}}.
\nonumber
\end{equation}
Recall that for a domain $\mathcal W \subset \mathcal V$, we define
\begin{equation}
\deg_{\mathcal W}(v) = 
\left\{
\begin{split}
& \sharp\, \{w \in \mathcal W \, ; \, w\sim v\}, \quad  v \in \stackrel{\circ}{\mathcal W}, \\
& \sharp\, \{w \in \stackrel{\circ}{\mathcal W} \, ; \, w\sim v\}, \quad  v \in \partial{\mathcal W}.
\end{split}
\right.
\nonumber
\end{equation}
(See (2.6) of \cite{AnIsoMo17(1)}). 
We impose the Dirichlet boundary condition for the domain $D(\widehat\Delta_{\mathcal V_{\overline{\Omega}}, \lambda})$ : 
\begin{equation}
\widehat u \in D(\widehat\Delta_{\mathcal V_{\overline{\Omega}}, \lambda}) \Longleftrightarrow \widehat u \in \ell^2(\mathcal V_
{\overline{\Omega}})\cap\{\widehat u \, ; \, \widehat u(v) = 0, \ v \in \partial \mathcal V_{\overline{\Omega}}\}.
\nonumber
\end{equation}
As in \S 3, we first define the vertex Dirichlet Laplacian for the  case without potential and then add the pontntial $\widehat Q_{\mathcal V,\lambda}$ as a perturbation. By modifying the inner product, $- \widehat\Delta_{\mathcal V_{\overline{\Omega}},\lambda} + \widehat Q_{\mathcal V,\lambda}$ is self-adjoint.
 The normal derivative at the boundary associated with $\widehat\Delta_{\mathcal V_{\overline{\Omega}},\lambda}$ is defined by
\begin{equation}
\big(\partial^{\nu}_{\widehat\Delta_{\mathcal V_{\overline{\Omega}},\lambda}}\widehat u\big)(v) = - \frac{1}{{\rm deg}_{\mathcal V_{\overline{\Omega}}}(v)}\sum_{w\sim v, w \in \stackrel{\circ}{\mathcal V_{\overline{\Omega}}}}\frac{1}{\psi_{wv}(1,\lambda)}
\widehat u(w).
\label{DefinenormalderivativeVDlambda}
\end{equation}
(c.f. (2.7) of \cite{AnIsoMo17(1)}).
Note that in the right-hand side, $w$ is taken only from $\stackrel{\circ}{\mathcal V_{\overline{\Omega}}}$.

\begin{figure}[hbtp]
\centering
\includegraphics[width=9cm, bb=0 0 606 564]{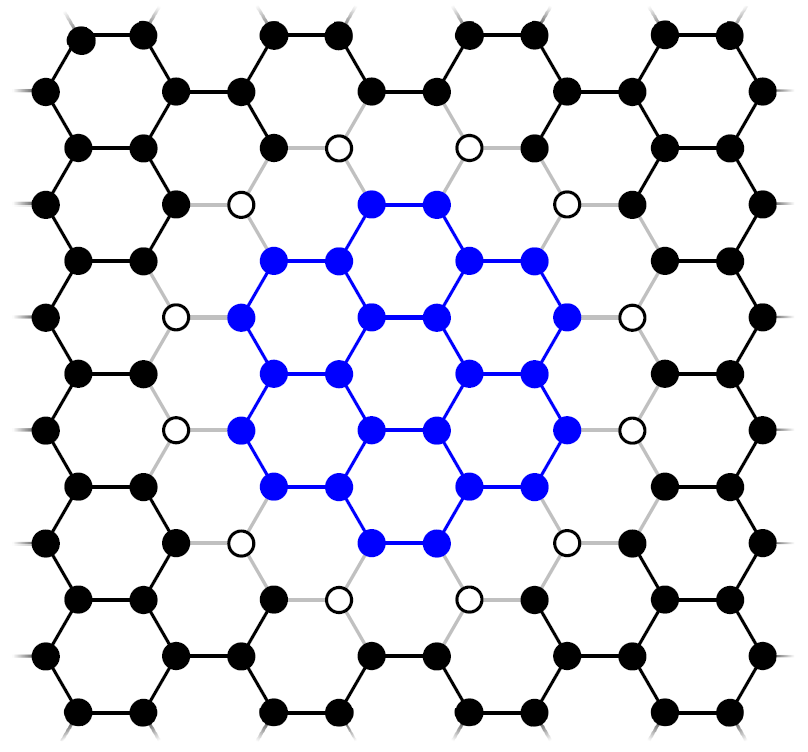}
\caption{Boundary of a domain in the hexagonal lattice}
\label{BoundaryHexagonal}
\end{figure}

Let us give examples of interior and exterior domains as well as their boundaries for the case of hexagonal lattice. In the sequel, all of our arguments are centered around these examples, although formulations and definitions are given for the general case.

We identify ${\bf R}^2$ with ${\bf C}$ and put $\omega = e^{2\pi i/6} = (1 + \sqrt{3}i)/2$. 
Let $\mathcal D$ be the hexagon with center at the origin and vertices $\omega^n, 1 \leq n \leq 6$. Recalling that the basis of the hexagonal lattice are $2 - \omega$ and $1 + \omega$, we put
\begin{equation}
\mathcal D_{k\ell} = \mathcal D + k(2-\omega) + \ell(1 + \omega),
\nonumber
\end{equation}
which denotes the translation of $\mathcal D$ by $k(2-\omega)$ and $\ell(1 + \omega)$.
For an integer $L \geq 1$, let 
\begin{equation}
\mathcal D_L = {\mathop\cup_{|k|\leq L, |\ell| \leq L}}\mathcal D_{k\ell}.
\nonumber
\end{equation}
 As is illustrated in Figure \ref{BoundaryHexagonal}, we take an interior domain $\Omega_{int}$ in such a way that
\begin{equation}
\stackrel{\circ}{\mathcal V_{\Omega_{int}}} = \mathcal V\cap \mathcal D_L, \quad 
\stackrel{\circ}{\mathcal E_{\Omega_{int}}} = \mathcal E\cap \mathcal D_L.
\nonumber
\end{equation}
In Figure \ref{BoundaryHexagonal}, $\partial \mathcal V_{\Omega_{int}}$ is denoted by white dots. The exterior domain $\Omega_{ext}$ is defined similarly. We then put
$$
\mathcal V_{int} = \mathcal V_{\overline{\Omega_{int}}}, \quad \mathcal E_{int} = \mathcal E_{\overline{\Omega_{int}}}, 
$$
$$
\mathcal V_{ext} = \mathcal V_{\overline{\Omega_{ext}}}, \quad \mathcal E_{ext} = \mathcal E_{\overline{\Omega_{ext}}}, 
$$
for the sake of simplicity. 
Note that
\begin{equation}
\mathcal V = \mathcal V_{int} \cup \mathcal V_{ext}, \quad 
\partial\mathcal V_{int} = \partial \mathcal V_{ext},
\nonumber
\end{equation}
\begin{equation}
\mathcal E = \mathcal E_{int}\cup\mathcal E_{ext}, \quad 
\mathcal E_{int}\cap\mathcal E_{ext} = \emptyset.
\nonumber
\end{equation}
We define the edge Dirichlet Laplacians on $\mathcal E_{int}$, $\mathcal E_{ext}$, which are denoted by $\widehat\Delta_{int,\mathcal E}$, $\widehat\Delta_{ext,\mathcal E}$:
$$
\widehat \Delta_{int,\mathcal E} = \widehat \Delta_{\mathcal E_{int}}, \quad \widehat \Delta_{ext,\mathcal E} = \widehat\Delta_{\mathcal E_{ext}}.
$$ 
We assume that the support of the potential lies strictly inside of  $\mathcal E_{int}$. Namely introducing a set:
\begin{equation}
\widetilde{\mathcal E_{int}} = \{{\bf e} \in \mathcal E_{int}\, ; \, {\bf e}(0) \not\in \partial\mathcal V_{int}, \ {\bf e}(1) \not\in \partial\mathcal V_{int}\},
\nonumber
\end{equation}
we assume
\begin{equation}
{\rm supp}\, q_{\mathcal E} \subset \widetilde{\mathcal E_{int}}.\label{SupportCond}
\end{equation}

The formal formulas (\ref{Resolventformalformula1}), (\ref{Resolventformula2})  are also valid for  boundary value problems of  edge Laplacians. For the case of the exterior problem, the resolvent of $- \widehat{\Delta}_{ext,\mathcal E}$ is written by (\ref{Resolventformula2}) with $\widehat H^{(0)}_{\mathcal E}$ replaced by $- \widehat{\Delta}_{ext,\mathcal E}$.
In our previous work \cite{AnIsoMo17(1)}, we studied the spectral properties of the vertex Laplacian in the exterior domain by reducing them to the whole space problem. Therefore, all the results for the edge Laplacian in the previous section also hold in the exterior domain. In particular, we have 
\begin{itemize}
\item Rellich type theorem (Theorem \ref{RelichTypeTheorem}),
\item Limiting absorption principle (Theorem \ref{LAPwholespace}),
\item Spectral representation (Theorem \ref{EigenfunctionExpansionWholespace}),
\item Resolvent expansion (Theorem \ref{ResolventExpansionWholespace}),
\item Exapansion of solutions to the Helmholtz equation  (Theorem \ref{HelmhlotzEqWholespace}), 
\item S-matrix (Theorem \ref{HelmhlotzEqWholespace})
\end{itemize}
in the exterior domain $\mathcal E_{ext}$. In fact, Theorem \ref{RelichTypeTheorem} holds without any change. Using the formula (\ref{S3ue(z)=Ce1+Ce0}) and the limiting absorption principle for $\widehat\Delta_{ext}$ proven in Theorem 7.7 in \cite{AnIsoMo17}, one can extend Theorem \ref{LAPwholespace} for the exterior domain. 
The radiation condition is also extended to the exterior domain. Then, the remaining theorems 
(Theorems \ref{EigenfunctionExpansionWholespace}, \ref{ResolventExpansionWholespace},  \ref{HelmhlotzEqWholespace}) are proven by the same argument.

\subsection{Exterior and interior D-N maps}

We consider the edge model for the exterior problem. 
Let
$\widehat u^{(\pm)} = \{\widehat u^{(\pm)}_{\bf e}\}_{{\bf e}\in \mathcal E_{ext}}$ be the solution to the equation
\begin{equation}
\left\{
\begin{split}
& ( - \widehat{\Delta}_{ext,\mathcal E} - \lambda)\widehat u = 0, \quad 
{\rm in} \quad \stackrel{\circ}{{\mathcal E}_{ext}}, \\
& \widehat u = \widehat f, \quad {\rm on} \quad \partial \mathcal E_{ext},
\end{split}
\right.
\label{EdgeEquationExterior}
\end{equation}
 satisfying the radiation
condition (outgoing for $\widehat u^{(+)}$ and incoming for $\widehat u^{(-)}$).
Then, the extrior D-N map $\Lambda^{(\pm)}_{ext,\mathcal E}(\lambda)$ is defined by
\begin{equation}
\Lambda^{(\pm)}_{ext,\mathcal E}(\lambda)\widehat f(v) = - \frac{d}{dz}\widehat u^{(\pm)}_{\bf e}(v), \quad v \in \partial \mathcal V_{ext}, 
\label{DefineLambdaextE}
\end{equation}
where ${\bf e}$ is the edge having $v$ as its end point.
Here, to compute $\frac{d}{dz}\widehat u_{\bf e}^{(\pm)}(v)$, we negelect the original orientation of ${\bf e}$. Namely, we parametrize ${\bf e}$ by $z \in [0,1]$ so that $v \in \partial\mathcal V$ corresponds to $z= 0$, and define $\frac{d}{dz}\widehat u^{(\pm)}_{\bf e}(v) = \frac{d}{dz}\widehat u^{(\pm)}_{\bf e}(z)\big|_{z=0}$.

 For the case of the interior problem, the Dirichlet boundary value problem for the edge Laplacian 
\begin{equation}
\left\{
\begin{split}
& ( - \widehat{\Delta}_{int,\mathcal E} + q_{{\bf e}} - \lambda)\widehat u = 0, \quad 
{\rm in} \quad \stackrel{\circ}{{\mathcal E}_{int}}, \\
& \widehat u = \widehat f, \quad {\rm on} \quad \partial \mathcal V_{int}
\end{split}
\right.
\label{EdgeEquationInterior}
\end{equation}
is formulated as above.
Note that the spectrum of $- \widehat\Delta_{int,\mathcal E} + q_{\mathcal E}$ is discrete. In the following, we assume that 
\begin{equation}
\lambda \not\in \sigma(- \widehat\Delta_{int,\mathcal E} + q_{\mathcal E}).
\label{lambdanotinsigamDeltaintE}
\end{equation}
The D-N map $\Lambda_{int,\mathcal E}(\lambda)$ is defined by
\begin{equation}
\Lambda_{int,\mathcal E}(\lambda)\widehat f(v) = \frac{d}{dz}\widehat u_{\bf e}(v), \quad v \in \partial \mathcal V_{int}, 
\label{LambdaintE}
\end{equation}
where ${\bf e}$ is the edge having $v$ as its end point and $\widehat u = \{\widehat u^{(\pm)}_{\bf e}\}_{{\bf e}\in \mathcal E_{int}}$ 
is the solution to the equation (\ref{EdgeEquationInterior}).
The same remark as above is applied to $\frac{d}{dz}\widehat u_{\bf e}(v), v \in \partial\mathcal V_{int}$.

The D-N maps are also defined for vertex operators. 
Let us slightly change the  notation. 
For a subset $\mathcal V_D \subset \mathcal V$ and $v \in \mathcal V_D$, let
\begin{equation}
(\widehat\Delta^{(0)}_{\mathcal V}\widehat u)(v) = \frac{1}{d_{\mathcal V}}
\sum_{w \sim v}\widehat u(w),
\nonumber
\end{equation}
\begin{equation}
(\widehat\Delta^{(0)}_{\mathcal V_D}\widehat u)(v) = \frac{1}{{\rm deg}_{\mathcal V_D}(v)}
\sum_{w \sim v, w \in \mathcal V_D}\widehat u(w).
\nonumber
\end{equation}
By this definition, we have (see (\ref{DefineDeltaVlambda(0)}))
\begin{equation}
\widehat\Delta^{(0)}_{\mathcal V,\lambda} = \frac{\sqrt{\lambda}}{\sin\sqrt{\lambda}}\widehat\Delta^{(0)}_{\mathcal V}.
\label{Delta0Vlambda=fracsinlambdalambdaDelta0}
\end{equation}
For the exterior and interior domains $\Omega_{ext}$ and $\Omega_{int}$ defined in the previous section, $\widehat\Delta_{\mathcal V_D}^{(0)}$ is denoted by $\widehat\Delta_{ext,\mathcal V}$ and $\widehat\Delta_{int,\mathcal V}$, respectively:
\begin{equation}
\widehat\Delta_{ext,\mathcal V} = \widehat\Delta^{(0)}_{\mathcal V_{ext}}, \quad 
\widehat\Delta_{int,\mathcal V} = \widehat\Delta^{(0)}_{\mathcal V_{int}}.
\nonumber
\end{equation}

Now, consider the exterior boundary value problem
\begin{equation}
\left\{
\begin{split}
& \big(- \widehat\Delta_{\mathcal V,\lambda} + \widehat Q_{\mathcal V,\lambda}\big)\widehat u = 0, \quad {\rm in} \quad \stackrel{\circ}{\mathcal V_{ext}}, \\
& \widehat u = \widehat f, \quad {\rm on} \quad \partial\mathcal V_{ext}.
\end{split}
\right.
\label{ExteriorBVPVertexmodel}
\end{equation}
Note that by (\ref{Delta0Vlambda=fracsinlambdalambdaDelta0}) and (\ref{DefinewidehatQ0}) this is equivalent to 
\begin{equation}
\left\{
\begin{split}
& (- \widehat\Delta^{(0)}_{\mathcal V} + \cos\sqrt{\lambda})\widehat u = 0, \quad {\rm in} \quad \stackrel{\circ}{\mathcal V_{ext}}, \\
& \widehat u = \widehat f, \quad {\rm on} \quad \partial\mathcal V_{ext}.
\end{split}
\right.
\nonumber
\end{equation}
Let $\widehat u^{(\pm)}_{ext,\mathcal V}$ be the solution of this equation satisfying the radiation condition.
Then, taking account of (\ref{DefinenormalderivativeVDlambda}) and  (\ref{Delta0Vlambda=fracsinlambdalambdaDelta0}), 
we define the exterior D-N map by
\begin{equation}
\begin{split}
\widehat \Lambda^{(\pm)}_{ext,\mathcal V}(\lambda)\widehat f &= 
- \frac{\sin\sqrt{\lambda}}{\sqrt{\lambda}}\partial^{\nu}_{\widehat\Delta_{\mathcal V_{ext},\lambda}}\widehat u^{(\pm)}_{ext,\mathcal V} = \partial^{\nu}_{\widehat\Delta_{ext,\mathcal V}}\widehat u^{(\pm)}_{ext,\mathcal V} \\
& = \frac{1}{{\rm deg}_{\mathcal V_{ext}}(v)}\sum_{w\sim v, w\in \stackrel{\circ}{\mathcal V_{ext}}}\widehat u^{(\pm)}_{ext,\mathcal V}(w).
\end{split}
\label{LambdaextV}
\end{equation}

We also consider the interior boundary value problem
\begin{equation}
\left\{
\begin{split}
& \big(- \widehat\Delta_{\mathcal V,\lambda} + \widehat Q_{\mathcal V,\lambda}\big)\widehat u = 0, \quad {\rm in} \quad \stackrel{\circ}{\mathcal V_{int}}, \\
& \widehat u = \widehat f, \quad {\rm on} \quad \partial\mathcal V_{int}.
\end{split}
\right.
\label{IntBVPVertex}
\end{equation}
Taking account of (\ref{SupportCond}), we define the interior D-N map by
\begin{equation}
\begin{split}
\widehat \Lambda_{int,\mathcal V}(\lambda)\widehat f(v) &= 
 \frac{\sin\sqrt{\lambda}}{\sqrt{\lambda}}\partial^{\nu}_{\widehat\Delta_{\mathcal V_{int},\lambda}}\widehat u_{int,\mathcal V} = - \partial^{\nu}_{\widehat\Delta_{int,\mathcal V}}\widehat u_{int,\mathcal V} \\
& = - \frac{1}{{\rm deg}_{\mathcal V_{int}}(v)}\sum_{w\sim v, w\in \stackrel{\circ}{\mathcal V_{int}}}\widehat u_{int,\mathcal V}(w).
\end{split}
\label{DefineLambdaintV}
\end{equation}

Note that by virtue of Lemma \ref{Lemma3.2}, if $\widehat u$ satisfies the edge Schr{\"o}dinger equation $(\widehat H_{\mathcal E} - \lambda)\widehat u = 0$ and the Kirchhoff condition, $\widehat u\big|_{\mathcal V}$ satisfies the vertex Schr{\"o}dinger equation $\big(- \widehat \Delta_{\mathcal V,\lambda} + \widehat Q_{\mathcal V,\lambda}\big)\widehat u\big|_{\mathcal V} = 0$. 
Therefore, if the exterior boundary value problem (\ref{EdgeEquationExterior}) for the edge model is solvable, so is the exterior boundary value problem  
(\ref{ExteriorBVPVertexmodel}) for the vertex model. The same remark applies to the interior boundary value problem.

If $\varphi(z)$ satisfies $- \varphi''(z) - \lambda\varphi(z) = 0$ in $(0,1)$, we have
$$
\varphi(1) = \varphi(0)\cos\sqrt{\lambda} +  \varphi'(0)\frac{\sin\sqrt{\lambda}}{\sqrt{\lambda}},
$$
Since the D-N map for the vertex model is computed by $\widehat u\big|_{\mathcal V}$, where $\widehat u$ is the solution to the edge Schr{\"o}dinger equation, this implies, by (\ref{DefineLambdaextE}), (\ref{LambdaintE}), (\ref{LambdaextV}) and (\ref{DefineLambdaintV}), the following formulas between the D-N maps of edge-Laplacian and vertex Laplacian.
We put
\begin{equation}
I_{ext} = \big({\rm Int}\,\sigma_e(\widehat H_{\mathcal E})\big)\setminus
\big(\sigma^{(0)}_{\mathcal V}\cup \sigma^{(0)}_{\tau}\big),
\nonumber
\end{equation}
and let $I_{int}$ be the set of $\lambda \in {\bf C}\setminus \sigma(- \widehat{\Delta}_{int,\mathcal E} + q_{\mathcal E})$ for which there exists $(- \widehat{\Delta}_{\mathcal V,\lambda} + \widehat Q_{\mathcal V,\lambda})^{-1}$. Let us note that 
$$
\sigma_e\big(- \widehat{\Delta}_{\mathcal E}\big) = \sigma_e\big(- \widehat{\Delta}_{ext,\mathcal E}\big).
$$

\begin{lemma} 
\label{LemmaDNedge=DNvertex}
The following equalities hold: 
\begin{equation}
\widehat\Lambda^{(\pm)}_{ext,\mathcal V}(\lambda) = \cos\sqrt{\lambda} - \frac{\sin\sqrt{\lambda}}{ \sqrt{\lambda}}\, \Lambda^{(\pm)}_{ext,\mathcal E}(\lambda), \quad 
\lambda \in I_{ext},
\nonumber
\end{equation}
\begin{equation}
\widehat\Lambda_{int,\mathcal V}(\lambda) = - \cos\sqrt{\lambda} -  \frac{\sin\sqrt{\lambda}}{\sqrt{\lambda}}\, \Lambda_{int,\mathcal E}(\lambda), \quad \lambda \in I_{int}.
\nonumber
\end{equation}
\end{lemma}

Therefore, the D-N map for the edge model and the D-N map for the vertex model determine each other.

\medskip
We put
\begin{equation}
\Sigma = \partial\mathcal V_{int} = \partial\mathcal V_{ext},
\nonumber
\end{equation}
and define an operator  $\widehat S_{\Sigma} : \ell^2(\Sigma) \to \ell^2(\Sigma)$ by
\begin{equation}
(\widehat S_{\Sigma}\widehat f)(v) = \frac{1}{{\rm deg}_{\mathcal V}(v)}\sum_{w\sim v, w \in \Sigma}\widehat f(w),
\label{SSigma}
\end{equation}
where
$$
{\rm deg}_{\mathcal V}(v) = {\rm deg}\,(v) = \sharp\{w \in \mathcal V\, ; \, w \sim v\}
$$
is the degree on $\mathcal V$. Let $\chi_{\Sigma}$ be the characterisitic function of $\Sigma$. We use $\chi_{\Sigma}$ to mean both of the operator of restriction
$$
\chi_{\Sigma} : \ell^2_{loc}(\mathcal V) \ni \widehat f \to 
\widehat f\big|_{\Sigma},
$$
and the operator of extension
$$
\chi_{\Sigma} : \ell^2(\Sigma) \ni \widehat f \to 
\left\{
\begin{split}
& \widehat f , \quad {\rm on} \quad \Sigma,\\
& 0, \quad {\rm otherwise}.
\end{split}
\right.
$$
Then, we have for $\widehat f \in \ell^2(\Sigma)$
\begin{equation}
\widehat\Delta_{\mathcal V}\chi_{\Sigma}\widehat f = \widehat S_{\Sigma}\widehat f.
\nonumber
\end{equation}
We also introduce multiplication operators by
\begin{equation}
\big(\mathcal M_{int}\widehat f\big)(v) = \frac{{\rm deg}_{\mathcal V_{int}}(v)}{{\rm deg}_{\mathcal V}(v)}\widehat f(v),
\nonumber
\end{equation}
\begin{equation}
\big(\mathcal M_{ext}\widehat f\big)(v) = \frac{{\rm deg}_{\mathcal V_{ext}}(v)}{{\rm deg}_{\mathcal V}(v)}\widehat f(v).
\nonumber
\end{equation}

Given $\widehat f \in \ell^2(\Sigma)$, 
let $\widehat u^{(\pm)}_{ext,\mathcal E}$  be the solution to the exterior boundary value problem
\begin{equation}
\left\{
\begin{split}
& (- \widehat{\Delta}_{ext,\mathcal E} - \lambda)\widehat u^{(\pm)}_{ext,\mathcal E} = 0, \quad {\rm in} \quad {\mathcal E}_{ext}, \\
& \widehat u^{(\pm)}_{ext,\mathcal E} = \widehat f, \quad {\rm on} \quad 
\partial{\mathcal E}_{ext}
\end{split}
\right.
\label{S6ExteriorBVP}
\end{equation}
satisfying the radiation condition.  Let $\widehat u_{int,\mathcal E}$ be the solution to the interior problem
\begin{equation}
\left\{
\begin{split}
& (- \widehat{\Delta}_{int,\mathcal E} + q_{\mathcal E}- \lambda)\widehat u^{(\pm)}_{int,\mathcal E} = 0, \quad {\rm in} \quad {\mathcal E}_{int}, \\
& \widehat u^{(\pm)}_{int,\mathcal E} = \widehat f, \quad {\rm on} \quad 
\partial{\mathcal E}_{int}.
\end{split}
\right.
\label{S6InteriorBVP}
\end{equation}
We put
\begin{equation}
\widehat u^{(\pm)}_{\mathcal E} = 
\left\{
\begin{split}
& \widehat u^{(\pm)}_{ext,\mathcal E} \quad {\rm on} \quad \mathcal E_{ext}\setminus\Sigma, \\
& \widehat f \quad {\rm on} \quad \Sigma. \\
& \widehat u_{int,\mathcal E} \quad {\rm on} \quad \mathcal E_{int}\setminus\Sigma.
\end{split}
\right.
\nonumber
\end{equation}
Then, $\widehat u^{(\pm)}_{\mathcal E}$ satisfies
\begin{equation}
\widehat u^{(\pm)}_{\mathcal E} \in \mathcal B^{\ast}(\mathcal E) \cap H^1_{loc}(\mathcal E)\cap H^2_{loc}(\mathcal E_{ext})\cap H^2(\mathcal E_{int}), 
\nonumber
\end{equation}
and on any edge ${\bf e} \in \mathcal E$
\begin{equation}
(- \widehat \Delta_{\mathcal E} + q_{\mathcal E} - \lambda)\widehat u^{(\pm)}_{\mathcal E} =  0 
\nonumber
\end{equation}
holds. We define a vertex function on $\mathcal V$ by
\begin{equation}
\widehat f^{(\pm)}(\lambda) = \widehat f^{(\pm)}(\lambda,v) = 
\big(- \widehat\Delta_{\mathcal V,\lambda} + \widehat Q_{\mathcal V,\lambda}\big)\left(\widehat u^{(\pm)}_{\mathcal E}\big|_{\mathcal V}\right), \quad v \in \mathcal V.
\label{widehatflambdavpmdefine}
\end{equation}
Then, since $\widehat u^{(\pm)}_{\mathcal E}$ satisfies the Kirchhoff condition outside $\Sigma$, we have
\begin{equation}
{\rm supp}\, \widehat f^{(\pm)}(\lambda) \subset \Sigma.
\label{suppwidehatfpmlambdainSigma}
\end{equation}
Define a bounded operator $B^{(\pm)}_{\Sigma}(\lambda) : \ell^2(\Sigma) \to \ell^2(\Sigma)$ by
\begin{equation}
B^{(\pm)}_{\Sigma}(\lambda) = \frac{\sqrt{\lambda}}{\sin\sqrt{\lambda}}\left(\mathcal M_{int}\Lambda_{int,\mathcal V}(\lambda) - \mathcal M_{ext}\Lambda^{(\pm)}_{ext,\mathcal V}(\lambda) - \widehat S_{\Sigma} + \cos\sqrt{\lambda}\right).
\label{OperatorBSigmalambda}
\end{equation}
We have, using (\ref{DefineDeltaVlambda(0)}), (\ref{DefinewidehatQ0}) and  (\ref{SSigma}), 
\begin{equation}
\widehat f^{(\pm)}(\lambda) = B^{(\pm)}_{\Sigma}(\lambda)\widehat f, \quad {\rm on} \quad \Sigma.
\label{fpmlambda=BpmSigmaf}
\end{equation}
We put an edge function $\widehat w^{(\pm)}_{\mathcal E} = \{\widehat w^{(\pm)}_{\bf e}\}_{\bf e \in \mathcal E}$ by
\begin{equation}
\widehat w^{(\pm)}_{\bf e}(z) = c^{(\pm)}_{\bf e}(1,\lambda)\frac{\phi_{{\bf e}0}(z,\lambda)}{\phi_{{\bf e}0}(1,\lambda)} +  c^{(\pm)}_{\bf e}(0,\lambda)\frac{\phi_{{\bf e}1}(z,\lambda)}{\phi_{{\bf e}1}(1,\lambda)}, \quad {\bf e} \in \mathcal E,
\nonumber
\end{equation}
where $c_{\bf e}^{(\pm)}(p,\lambda)$ is defined by
\begin{equation}
c_{\bf e}^{(\pm)}(p,\lambda) = \Big(- \widehat\Delta_{\mathcal V,\lambda \pm i0} + \widehat Q_{\mathcal V,\lambda\pm i0}\Big)^{-1}
\widehat f^{(\pm)}(\lambda), \quad p = 0, 1,
\nonumber
\end{equation}
 at the end point of an edge ${\bf e}$, and 
\begin{equation}
\Big(- \widehat\Delta_{\mathcal V,\lambda \pm i0} + \widehat Q_{\mathcal V,\lambda\pm i0}\Big)^{-1} = \lim_{\epsilon \to 0}
\Big(- \widehat\Delta_{\mathcal V,\lambda \pm i\epsilon} + \widehat Q_{\mathcal V,\lambda\pm i\epsilon}\Big)^{-1}.
\nonumber
\end{equation}
We then have
\begin{equation}
\widehat w^{(\pm)}_{\mathcal E} = \widehat T^{(0)}_{\mathcal E}(\lambda)\Big(- \widehat\Delta_{\mathcal V,\lambda \pm i0} + \widehat Q_{\mathcal V,\lambda\pm i0}\Big)^{-1}
\widehat f^{(\pm)}(\lambda).
\label{wpmandfpmlambda}
\end{equation}
Note that outside $\widetilde{\mathcal E}_{int}$
\begin{equation}
\widehat w^{(\pm)}_{\bf e}(z) = c^{(\pm)}_{\bf e}(1,\lambda)\frac{\sin\sqrt{\lambda}z}{\sqrt{\lambda}} +  c^{(\pm)}_{\bf e}(0,\lambda)\frac{\sin\sqrt{\lambda}(1-z)}{\sqrt{\lambda}}.
\end{equation}
Moreover, $\widehat w^{(\pm)}_{\mathcal E}$ satisfies the Kirchhoff condition
on $\mathcal V\setminus\Sigma$.

We prepare one more notation. 
Let the operator of restriction $r\big|_{\mathcal V} : H^1(\mathcal E) \to L^2(\mathcal V)$ be defined by
\begin{equation}
\big(r\big|_{\mathcal V}\widehat u\big)(v) = \widehat u(v), \quad v \in \mathcal V.
\nonumber
\end{equation}
The adjoint : $\big(r\big|_{\mathcal V}\big)^{\ast} : L^2(\mathcal V) \to H^{-1}(\mathcal E)$ is defined by
\begin{equation}
\big(r\big|_{\mathcal V}\widehat u,\widehat v\big)_{L^2(\mathcal V)} = (\widehat u,\big(r\big|_{\mathcal V}\big)^{\ast}\widehat v\big)_{L^2(\mathcal E)}.
\nonumber
\end{equation}
We then have in view of (\ref{Vinnerproduct})
\begin{equation}
\big(r\big|_{\mathcal V}\big)^{\ast} = \frac{1}{d_{\mathcal V}}\sum_{v \in \mathcal V}
\delta_v,
\label{r|Vast=sumdeltafunction}
\end{equation}
where $\delta_v$ denotes the Dirac distribution on $\mathcal E$ supported at $v \in \mathcal V$.

By the elliptic regularity
$$
\widehat R_{\mathcal E}(\lambda \pm i\epsilon) \in {\bf B}(L^2(\mathcal E);H^2(\mathcal E)).
$$
Taking the adjoint 
$$
\widehat R_{\mathcal E}(\lambda \pm i\epsilon) \in {\bf B}(H^{-2}(\mathcal E);L^2(\mathcal E)).
$$
By an interpolation, we then have
\begin{equation}
\widehat R_{\mathcal E}(\lambda \pm i\epsilon) \in {\bf B}(H^{-t}(\mathcal E);H^{2 - t}(\mathcal E)), \quad 0 \leq t \leq 2,
\nonumber
\end{equation}
which implies 
\begin{equation}
\widehat R_{\mathcal E}(\lambda \pm i\epsilon) \in {\bf B}(H^{-t}_0(\mathcal E);H^{2 - t}_{loc}(\mathcal E)), \quad 0 \leq t \leq 2,
\nonumber
\end{equation}
where $H^{-t}_0({\mathcal E})$ is the set of the compactly supported distributions in $H^{-t}(\mathcal E)$. 
Therefore
$$
\widehat R_{\mathcal E}(\lambda \pm i0)\big(r\big|_{\mathcal V}\big)^{\ast}\chi_{\Sigma} \in {\bf B}(L^2(\mathcal V);H^1_{loc}(\mathcal E))
$$
is well-defined.

\begin{lemma}
\label{uE=wEf=TEflambdaonSigma}
Let $\widehat u^{(\pm)}_{\mathcal E}$ be defined by
\begin{equation}
\widehat u^{(\pm)}_{\mathcal E} = 
\left\{
\begin{split}
& \chi_{ext}\widehat u^{(\pm)}_{ext,\mathcal E} + \chi_{int}\widehat u_{int,\mathcal E} \quad {\rm outside} \quad \Sigma, \\
& \widehat f \quad {\rm on} \quad \Sigma,
\end{split}
\right.
\label{Lemma6.3Defineupm}
\end{equation}
where $\widehat f \in \ell^2(\Sigma)$,  $\chi_{ext}$ and $\chi_{int}$ are the characteristic functions of $\mathcal E_{ext}$ and $\mathcal E_{int}$, and $\widehat u^{(\pm)}_{ext,\mathcal E}$, $\widehat u_{int,\mathcal E}$ are the solutions of the boundary value problems (\ref{S6ExteriorBVP}), (\ref{S6InteriorBVP}), respectively. 
Let $\widehat w_{\mathcal E}^{(\pm)}$ be as in (\ref{wpmandfpmlambda}). Then we have
\begin{equation}
\begin{split}
\widehat u^{(\pm)}_{\mathcal E} &=   \widehat w^{(\pm)}_{\mathcal E}= \widehat R_{\mathcal E}(\lambda \pm i0)\big(r\big|_{\mathcal V}\big)^{\ast}\chi_{\Sigma}B^{(\pm)}_{\Sigma}(\lambda)\widehat f \quad {\rm on} \quad \mathcal E.
\end{split}
\label{Lemma6.2upm=wpm}
\end{equation}
In particular, we have
\begin{equation}
\begin{split}
\widehat u^{(\pm)}_{ext,\mathcal E} & =  \widehat R_{\mathcal E}(\lambda \pm i0)\big(r\big|_{\mathcal V}\big)^{\ast}\chi_{\Sigma}B^{(\pm)}_{\Sigma}(\lambda)\widehat f \\
& =  \widehat T^{(0)}_{\mathcal E}(\lambda)\Big(- \widehat\Delta_{\mathcal V,\lambda \pm i0} + \widehat Q_{\mathcal V,\lambda\pm i0}\Big)^{-1}
\chi_{\Sigma}B^{(\pm)}_{\Sigma}(\lambda)\widehat f \quad {\rm on} \quad \mathcal V_{ext}, 
\end{split}
\label{upmext=I0mathcalE()-1widehatfpmlambfaformula}
\end{equation}
\begin{equation}
\widehat f  =  \Big(- \widehat\Delta_{\mathcal V,\lambda \pm i0} + \widehat Q_{\mathcal V,\lambda\pm i0}\Big)^{-1}
\chi_{\Sigma}B^{(\pm)}_{\Sigma}(\lambda)
\widehat f \quad {\rm on} \quad \Sigma.
\label{f=ImathcalElambdaetcfonSigma}
\end{equation}
\end{lemma}

Proof. By  definition, for any edge $\bf e$,
$$
\big(- \frac{d^2}{dz^2} + q_{\bf e}(z) - \lambda\big)\widehat u^{(\pm)}_{\mathcal E, \bf e}(z) = 0, \quad {\rm on} \quad {\bf e}.
$$ 
In view of (\ref{widehatflambdavpmdefine}) and (\ref{wpmandfpmlambda}), we have
$$
\big( - \widehat\Delta_{\mathcal V,\lambda} + \widehat Q_{\mathcal V,\lambda}\big)\left({\widehat w^{(\pm)}}_{\mathcal E}\big|_{\mathcal V}\right) = \widehat f^{(\pm)}(\lambda) = \big( - \widehat\Delta_{\mathcal V,\lambda} + \widehat Q_{\mathcal V,\lambda}\big)\left(\widehat u^{(\pm)}_{\mathcal E}\big|_{\mathcal V}\right),
$$
where we use
\begin{equation}
r\big|_{\mathcal V}\widehat T_{\mathcal E}(\lambda) = 
1_{\mathcal V},
\label{TElambda|V=|V}
\end{equation}
 $1_{\mathcal V}$ being the identity on $\mathcal V$.
These two formulas imply that $u^{(\pm)}_{\mathcal E} - \widehat w^{(\pm)}_{\mathcal E}$ satisfies the equation
$$
\Big(- \frac{d^2}{dz^2} + q_{\bf e}(z) - \lambda\Big)(\widehat u^{(\pm)}_{\mathcal E,\bf e}(z) - \widehat w^{(\pm)}_{\mathcal E,\bf e}(z)) = 0, \quad {\rm on} \quad {\bf e},
$$ 
and the Kirchhoff condition
$$
\big( - \widehat\Delta_{\mathcal V,\lambda} + \widehat Q_{\mathcal V}\big)(\widehat u^{(\pm)}_{\mathcal E}\big|_{\mathcal V} - \widehat w^{(\pm)}_{\mathcal E}\big|_{\mathcal V}) = 0.
$$
Then, $\widehat u^{(\pm)}_{\mathcal E} - \widehat w^{(\pm)}_{\mathcal E} = 0$ since it satisfies the radiation condition. 

Again using (\ref{TElambda|V=|V}) and Lemma  \ref{TmathcalVlambdaadjoint},
\begin{equation}
1_{\mathcal V}  = \widehat T_{\mathcal V}(\lambda)\big(r\big|_{\mathcal V}\big)^{\ast}.  
\nonumber
\end{equation}
Note that 
$$
r\big|_{\mathcal V}r_{\mathcal E}(\lambda) = 0,
$$
hence taking the adjoint,
$$
r_{\mathcal E}(\lambda)\big(r\big|_{\mathcal V}\big)^{\ast} = 0.
$$
In view of (\ref{Resolventformalformula1}), we have
\begin{equation}
\begin{split}
& \widehat R_{\mathcal E}(\lambda + i0)\big(r\big|_{\mathcal V}\big)^{\ast}\chi_{\Sigma}B^{(\pm)}_{\Sigma}(\lambda)\chi_{\Sigma} \\
&= \widehat T_{\mathcal E}(\lambda)\big( - \widehat\Delta_{\mathcal V,\lambda + i0} + \widehat Q_{\mathcal V,\lambda + i0}\big)^{-1}
\widehat T_{\mathcal V}(\lambda)\big(r\big|_{\mathcal V}\big)^{\ast}\chi_{\Sigma}B^{(+)}_{\Sigma}(\lambda)\chi_{\Sigma} \\
&= \widehat T_{\mathcal E}(\lambda)\big( - \widehat\Delta_{\mathcal V,\lambda + i0} + \widehat Q_{\mathcal V,\lambda + i0}\big)^{-1}
\chi_{\Sigma}B^{(+)}_{\Sigma}(\lambda)\chi_{\Sigma}.
\end{split}
\nonumber
\end{equation}
Using (\ref{suppwidehatfpmlambdainSigma}),  (\ref{fpmlambda=BpmSigmaf}) and (\ref{wpmandfpmlambda}), we have
proven (\ref{Lemma6.2upm=wpm}) and (\ref{upmext=I0mathcalE()-1widehatfpmlambfaformula}). 
By (\ref{TElambda|V=|V}), one can prove (\ref{f=ImathcalElambdaetcfonSigma}).
\qed


\subsection{Spectral representation in an exterior domain}
Note  that 
\begin{equation}
\chi_{\Sigma}\, r\big|_{\mathcal V}\widehat R^{(0)}_{\mathcal E}(\lambda \pm i0) \in {\bf B}(\mathcal B(\mathcal E) ; \ell^2(\Sigma)),
\nonumber
\end{equation} 
which yields
\begin{equation}
\widehat R^{(0)}_{\mathcal E}(\lambda \pm i0)\big(r\big|_{\mathcal V})^{\ast}\chi_{\Sigma} \in 
{\bf B}(\ell^2(\Sigma);\mathcal B^{\ast}(\mathcal E)).
\nonumber
\end{equation}
With this in mind, we prove the following lemma.

\begin{lemma}
\label{Lemma6.3}
The following equalities hold 
\begin{equation}
\widehat R_{ext,\mathcal E}(\lambda \pm i0) = \chi_{ext}\Big(I - \widehat R_{\mathcal E}(\lambda \pm i0)\big(r\big|_{\mathcal V}\big)^{\ast}\chi_{\Sigma}B^{(\pm)}_{\Sigma}(\lambda)\chi_{\Sigma}\, r\big|_{\mathcal V}\Big)\widehat R^{(0)}_{\mathcal E}(\lambda \pm i0)\chi_{ext},
\label{widahatRextlambdapmi0formula01}
\end{equation}
\begin{equation}
\widehat R_{ext,\mathcal E}(\lambda \pm i0) = \chi_{ext}\widehat R^{(0)}_{\mathcal E}(\lambda \pm i0)\Big(I -\big(r\big|_{\mathcal V}\big)^{\ast}\chi_{\Sigma}\big(B^{(\mp)}_{\Sigma}(\lambda)\big)^{\ast}\chi_{\Sigma}\, r\big|_{\mathcal V}\widehat R_{\mathcal E}(\lambda\pm i0)\Big)\chi_{ext}.
\label{widahatRextlambdapmi0formula02}
\end{equation}
\end{lemma}

Proof. 
Take $\widehat f \in \mathcal B(\mathcal E)$ arbitrarily, and replace $\widehat f$ in (\ref{S6ExteriorBVP}), (\ref{S6InteriorBVP}) and (\ref{Lemma6.3Defineupm}) 
by $\chi_{\Sigma}\, r\big|_{\mathcal V}\widehat R^{(0)}_{\mathcal E}(\lambda \pm i0)\widehat f$.
Then, letting 
$$
\widehat u_0 = \widehat R_{\mathcal E}(\lambda \pm i0)\big(r\big|_{\mathcal V}\big)^{\ast}\chi_{\Sigma}B_{\Sigma}^{(\pm)}(\lambda)\chi_{\Sigma}\,r\big|_{\mathcal V}\widehat R^{(0)}_{\mathcal E}(\lambda \pm i0)\widehat f
$$
 and using (\ref{Lemma6.2upm=wpm}), we have
\begin{equation}
\begin{split}
\widehat u^{(\pm)}_{\mathcal E} = 
\widehat u_0.
\end{split}
\nonumber
\end{equation}
This implies $\widehat u_0 = \widehat u^{(\pm)}_{ext}$ in $\mathcal E_{ext}$, hence
\begin{equation}
\left\{
\begin{split}
& ( - \widehat\Delta_{\mathcal E} - \lambda)\widehat u_0 = 0, \quad {\rm in} \quad \mathcal E_{ext}, \\
& \widehat u_0 = r\big|_{\mathcal V}\widehat R^{(0)}_{\mathcal E}(\lambda \pm i0)\widehat f, \quad {\rm on} \quad \Sigma.
\end{split}
\right.
\nonumber
\end{equation}
Let $\widehat w = \widehat u_0 - \widehat R^{(0)}_{\mathcal E}(\lambda \pm i0)\widehat f$. Then
\begin{equation}
\left\{
\begin{split}
& ( - \widehat \Delta_{\mathcal E} - \lambda)\widehat w = - \widehat f, \quad {\rm in} \quad \mathcal E_{ext}, \\
& \widehat w = 0, \quad {\rm on} \quad \Sigma.
\end{split}
\right.
\nonumber
\end{equation}
Taking account of the radiation condition, we then have $\widehat w = - \widehat R_{ext}(\lambda \pm i0)\chi_{ext}\widehat f = 
\widehat u_0 - \widehat R^{(0)}_{\mathcal E}(\lambda \pm i0)\widehat f$, which implies 
 (\ref{widahatRextlambdapmi0formula01}). By taking the adjoint, we obtain (\ref{widahatRextlambdapmi0formula02}). 
\qed

\medskip
By virtue of (\ref{DefinePhi(0)lambdaS5}) and (\ref{S5DefineFj(0)lambda}), $\widehat{\mathcal F}^{(0)}(\lambda)$ is extended to a bounded operator from $\ell^2(\Sigma)$ to ${\bf h}_{\lambda}$.
We then define a spectral representation for $\widehat H_{ext} = - \widehat \Delta_{ext, \mathcal E}$ by
\begin{equation}
\widehat{\mathcal F}^{(\pm)}_{ext}(\lambda) = \widehat{\mathcal F}^{(0)}(\lambda)\left(1 - \big(r\big|_{\mathcal V}\big)^{\ast}\chi_{\Sigma}\big(B^{(\mp)}_{\Sigma}(\lambda)\big)^{\ast}\chi_{\Sigma}r\big|_{\mathcal V}\widehat R_{\mathcal E}(\lambda \pm i0)\right)\chi_{ext}.
\nonumber
\end{equation}
Taking the adjoint, we also have
\begin{equation}
\widehat{\mathcal F}^{(\pm)}_{ext}(\lambda)^{\ast} = \chi_{ext}\left(1 -
\widehat R_{\mathcal E}(\lambda \mp i0)\big(r\big|_{\mathcal V}\big)^{\ast}\chi_{\Sigma}B^{(\mp)}_{\Sigma}(\lambda)\chi_{\Sigma}r\big|_{\mathcal V}\right)\widehat{\mathcal F}^{(0)}(\lambda)^{\ast}.
\label{Fastlambdaexpression}
\end{equation}
By (\ref{widahatRextlambdapmi0formula02}), we have
$$
\widehat{\mathcal F}^{(\pm)}_{ext}(\lambda) = \widehat{\mathcal F}^{(0)}(\lambda)(- \widehat\Delta_{\mathcal E} - \lambda)\widehat R_{ext,\mathcal E}(\lambda \pm i0).
$$
Therefore, $\widehat{\mathcal F}^{(\pm)}_{ext}(\lambda)$ is independent of the perturbation $q_{\mathcal E}$. 

\begin{lemma}
\label{Fextatphioutgoingsolofextproblem}
For any $\phi \in {\bf h}_{\lambda}$, $\widehat{\mathcal F}^{(-)}_{ext}(\lambda)^{\ast} \phi$ satisfies the equation
\begin{equation}
\left\{
\begin{split}
& (- \widehat \Delta_{ext,\mathcal E}- \lambda)\widehat{\mathcal F}^{(-)}_{ext}(\lambda)^{\ast} \phi = 0, \quad {\rm in} \quad \mathcal E_{ext}, \\
& \widehat{\mathcal F}^{(-)}_{ext}(\lambda)^{\ast} \phi = 0 \quad {\rm on} \quad \Sigma.
\end{split}
\right.
\nonumber
\end{equation}
Moreover, $\widehat{\mathcal F}^{(-)}_{ext}(\lambda)^{\ast} \phi  - \widehat{\mathcal F}^{(0)}(\lambda)^{\ast}\phi$ is outgoing.
\end{lemma}

Proof. We put $
\widehat v = \widehat R_{\mathcal E}(\lambda + i0)\big(r\big|_{\mathcal V}\big)^{\ast}\chi_{\Sigma}B^{(+)}_{\Sigma}(\lambda)\chi_{\Sigma}r\big|_{\mathcal V}\widehat{\mathcal F}^{(0)}(\lambda)^{\ast}\phi$. 
 Then, 
$$
(- \widehat{\Delta}_{\mathcal E} - \lambda)\widehat v = 0, \quad 
{\rm in} \quad \mathcal E_{ext}.
$$
Letting $\widehat f = \chi_{\Sigma}r\big|_{\mathcal V}\widehat{\mathcal F}^{(0)}(\lambda)^{\ast}\phi$,
and using (\ref{Lemma6.2upm=wpm}), we have $\widehat v = \widehat u^{(+)}_{\mathcal E}$.  Then,  by (\ref{Lemma6.3Defineupm}), $\widehat v = \widehat f = \widehat{\mathcal  F}^{(0)}(\lambda)^{\ast}\phi$ on $\Sigma$. Since $\widehat v$ is outgoing, we obtain the lemma. \qed

\subsection{Imbedding of $\ell^2(\Sigma)$ to ${\bf h}_{\lambda}$} We put
\begin{equation}
\widehat I^{(\pm)}(\lambda) = \widehat{\mathcal F}^{(\pm)}(\lambda)\big(r\big|_{\mathcal V}\big)^{\ast}\chi_{\Sigma}B^{(\pm)}_{\Sigma}(\lambda) : \ell^2(\Sigma) \to {\bf h}_{\lambda}.
\nonumber
\end{equation}
Then, by (\ref{DefineFpmlambdainmathcalE}) and (\ref{Lemma6.2upm=wpm}), 
\begin{equation}
\begin{split}
\widehat{\mathcal F}^{(\pm)}(\lambda)\big(r\big|_{\mathcal V}\big)^{\ast}\chi_{\Sigma}B^{(\pm)}_{\Sigma}(\lambda)\widehat f &= \widehat{\mathcal F}^{(0)}(\lambda)\big(1 - q_{\mathcal E}\widehat R_{\mathcal E}(\lambda \pm i0)\big)\big(r\big|_{\mathcal V}\big)^{\ast}\chi_{\Sigma}B^{(\pm)}_{\Sigma}(\lambda)\widehat f\\
& = \widehat{\mathcal F}^{(0)}(\lambda)\big(- \widehat\Delta_{\mathcal E} - \lambda\big)\widehat u_{\mathcal E}^{(\pm)}.
\end{split}
\nonumber
\end{equation}
This formula shows that $\widehat I^{(\pm)}(\lambda)$
does not depend on the perturbation $q_{\mathcal E}$.

\begin{lemma}
(1) $\ \widehat I^{(\pm)}(\lambda) : \ell^2(\Sigma) \to {\bf h}_{\lambda}$ is 1 to 1. \\
\noindent
(2) $\ \widehat I^{(\pm)}(\lambda)^{\ast} : {\bf h}_{\lambda} \to \ell^2(\Sigma)$ is onto.
\end{lemma}

Proof. Assume $\widehat I^{(\pm)}(\lambda)\widehat f = 0$, and let $\widehat u^{(\pm)}_{ext,\mathcal E}$ be the solution of the exterior problem (\ref{EdgeEquationExterior}). Then, by virtue of (\ref{Lemma6.2upm=wpm})
$$
\widehat u^{(\pm)}_{ext,\mathcal E} =  \widehat R_{\mathcal E}(\lambda \pm i0)\widehat g, \quad
\widehat g = \big(r\big|_{\mathcal V}\big)^{\ast}\chi_{\Sigma}B^{(\pm)}_{\Sigma}(\lambda)\chi_{\Sigma}\widehat f.
$$
Theorem \ref{ResolventExpansionPerturbedCase} yields
$$
\mathcal U_{\mathcal E,\ell}\widehat R_{\mathcal E}(\lambda \pm i0)\widehat g \simeq \Pi_{\ell}(\lambda)D^{(0)}(\lambda \pm i0)\widehat{\mathcal F}^{(\pm)}(\lambda)\widehat g.
$$
Since $\widehat I^{(\pm)}(\lambda)\widehat f = \widehat{\mathcal F}^{(\pm)}(\lambda)\widehat g$, this implies $\widehat u^{(\pm)}_{ext,\mathcal E} \in \widehat{\mathcal B}^{\ast}_0$. Using the Rellich type theorem (Theorem \ref{RelichTypeTheorem}) and the unique continuation property, we obtain $\widehat u^{(\pm)}_{ext,\mathcal E}=0$, which implies $\widehat f = 0$. This proves (1). 
To prove the assertion (2), let us note the following fact:
Let $X, Y$ be Hilbert spaces, and assume that a bounded operator $A : X \to Y$ is injective, and ${\rm dim}\, X < \infty$. Then
\begin{itemize}
\item $A : X \to {\rm Ran}\, (A)$ is bijective, 
\item $A^{\ast} : {\rm Ran}\, (A) \to X$ is surjective.
\end{itemize}
(The first assertion is elementary. The second assertion follows from the fact that $A^{\ast}A : X \to X$ is injective, hence surjective.)
\qed

\subsection{Scattering amplitude in the exterior domain}
We define the scattering amplitude in the exterior domain by
\begin{equation}
A_{ext}(\lambda) = \widehat{\mathcal F}^{(+)}(\lambda)\big(r\big|_{\mathcal V}\big)^{\ast}\chi_{\Sigma}B^{(\pm)}_{\Sigma}(\lambda)\chi_{\Sigma}r\big|_{\mathcal V}\widehat{\mathcal F}^{(0)}(\lambda)^{\ast}.
\nonumber
\end{equation}
Using Theorem \ref{ResolventExpansionPerturbedCase} and (\ref{Fastlambdaexpression}), and extending $\widehat{\mathcal  F}^{(-)}_{ext}(\lambda)^{\ast}\phi - \widehat{\mathcal F}^{(0)}(\lambda)^{\ast}\phi$ to be 0 on $\mathcal E_{int}$, we have
\begin{equation}
\begin{split}
&\mathcal U_{\mathcal E,\ell}\left(\widehat{\mathcal  F}^{(-)}_{ext}(\lambda)^{\ast}\phi - \widehat{\mathcal F}^{(0)}(\lambda)^{\ast}\phi\right) \\
 & = - \mathcal U_{\mathcal E,\ell}
\chi_{ext}\widehat R_{\mathcal E}(\lambda + i0)\big(r\big|_{\mathcal V}\big)^{\ast}\chi_{\Sigma}B^{(+)}_{\Sigma}(\lambda)\chi_{\Sigma}r\big|_{\mathcal V}\widehat{\mathcal F}^{(0)}(\lambda)^{\ast}\phi \\
&\simeq - \Pi_{\ell}(\lambda)D^{(0)}(\lambda + i0)A_{ext}(\lambda)\phi.
\end{split}
\nonumber
\end{equation}
This shows that $A_{ext}(\lambda)$ depends only on $\mathcal E_{ext}$.

\subsection{Single layer and double layer potentials}
The operator 
\begin{equation}
\widehat R_{\mathcal E}(\lambda \pm i0)\big(r\big|_{\mathcal V}\big)^{\ast}\chi_{\Sigma}B^{(\pm)}_{\Sigma}(\lambda)\chi_{\Sigma} : 
\ell^2(\Sigma) \to \mathcal B^{\ast}(\mathcal E)
\nonumber
\end{equation}
is an analogue of the double layer potential.  The operator $M_{\Sigma}^{(\pm)}(\lambda) : \ell^2(\Sigma) \to \ell^2(\Sigma)$  defined by
\begin{equation}
M^{(\pm)}_{\Sigma}(\lambda)\widehat f := \big(\widehat R_{\mathcal E}(\lambda \pm i0)\chi_{\Sigma}\widehat f\big)\big|_{\Sigma} 
\nonumber
\end{equation}
is an analogue of the single layer potential.

\begin{lemma} 
\label{MSigmatimesBsimga=1Lemma}
(1) $M^{(\pm)}_{\Sigma}(\lambda) = \chi_{\Sigma}\big(- \widehat{\Delta}_{\mathcal V, \lambda \pm i0} + 
\widehat Q_{\mathcal V,\lambda \pm i0}\big)^{-1}\chi_{\Sigma}$. \\
\noindent
(2) \ 
$M_{\Sigma}^{(\pm)}(\lambda)B^{(\pm)}_{\Sigma}(\lambda) = 1$ on $\ell^2(\Sigma)$. 
\end{lemma}

Proof. In view of (\ref{Resolventformalformula1}) and (\ref{TElambda|V=|V}), we have
\begin{equation}
\chi_{\Sigma}\widehat R_{\mathcal E}(\lambda \pm i0)\widehat f = \chi_{\Sigma}\big(- \widehat{\Delta}_{\mathcal V,\lambda \pm i0} + \widehat Q_{\mathcal V,\lambda \pm i0}\big)^{-1}\widehat T_{\mathcal V}(\lambda)\chi_{\Sigma}\widehat f.
\nonumber
\end{equation}
This proves (1). 
By (\ref{f=ImathcalElambdaetcfonSigma}), 
\begin{equation}
\widehat f = \big(- \widehat{\Delta}_{\mathcal V,\lambda \pm i0} + \widehat Q_{\mathcal V,\lambda \pm i0}\big)^{-1}B^{(\pm)}_{\Sigma}(\lambda)\widehat f.
\nonumber
\end{equation}
The assertion (2)  then follows from these equalities. 
 \qed

\subsection{S-matrix and interior D-N map}

\begin{theorem}
\label{Aext-A=IBSigma-1ILemma}
The following equality holds:
\begin{equation}
A_{ext}(\lambda) - A(\lambda) = \widehat I^{(+)}(\lambda)\big(r\big|_{\mathcal V}\big)^{\ast}\big(B^{(+)}_{\Sigma}(\lambda)\big)^{-1}r\big|_{\mathcal V}\widehat I^{(-)}(\lambda)^{\ast}.
\nonumber
\end{equation}
\end{theorem}

Proof. For $\phi \in {\bf h}_{\lambda}$, we put
\begin{equation}
\widehat u = \widehat{\mathcal F}^{(-)}(\lambda)^{\ast}\phi - \widehat{\mathcal F}^{(-)}(\lambda)^{\ast}_{ext}\phi.
\nonumber
\end{equation}
By (\ref{DefineFpmlambdainmathcalE}) and (\ref{Fastlambdaexpression}), we have
\begin{equation}
\begin{split}
\widehat u&  = (\chi_{ext}-1)(1 - \widehat R_{\mathcal E}(\lambda + i0)q_{\mathcal E})\widehat{\mathcal F}^{(0)}(\lambda)^{\ast}\phi 
 \\
& + \chi_{ext}\widehat R_{\mathcal E}(\lambda + i0)\Big(\big(r\big|_{\mathcal V}\big)^{\ast}\chi_{\Sigma}B^{(+)}_{\Sigma}(\lambda)\chi_{\Sigma}r\big|_{\mathcal V} - q_{\mathcal E}\Big)\widehat{\mathcal F}^{(0)}(\lambda)^{\ast}\phi.
\end{split}
\nonumber
\end{equation}
The first term of the right-hand side is a smooth function when passed to the Fourier series. 
Theorem \ref{ResolventExpansionPerturbedCase} then implies
\begin{equation}
\mathcal U_{\mathcal E,\ell}\widehat u \simeq 
\Pi_{\ell}(\lambda)D^{(0)}(\lambda + i0)
\widehat{\mathcal F}^{(+)}(\lambda)\Big(\big(r\big|_{\mathcal V}\big)^{\ast}\chi_{\Sigma}B^{(+)}_{\Sigma}(\lambda)\chi_{\Sigma} r\big|_{\mathcal V}- q_{\mathcal E}\Big)\widehat{\mathcal F}^{(0)}(\lambda)^{\ast}\phi.
\nonumber
\end{equation}
By Lemma \ref{Fextatphioutgoingsolofextproblem}, $\widehat u$ is the outgoing solution of the equation
\begin{equation}
(- \widehat{\Delta}_{\mathcal E} - \lambda)\widehat u = 0, \quad {\rm in} \quad \mathcal E_{ext}, \quad 
\widehat u\big|_{\Sigma} = \widehat{\mathcal F}^{(-)}(\lambda)^{\ast}\phi.
\nonumber
\end{equation}
In view of (\ref{upmext=I0mathcalE()-1widehatfpmlambfaformula}), we have
\begin{equation}
\widehat u = \widehat R_{\mathcal E}(\lambda + i0)\big(r\big|_{\mathcal V}\big)^{\ast}\chi_{\Sigma}B^{(\pm)}_{\Sigma}(\lambda)\chi_{\Sigma}\widehat{\mathcal F}^{(-)}(\lambda)^{\ast}\phi.
\nonumber
\end{equation}
Again using Theorem \ref{ResolventExpansionPerturbedCase}, 
\begin{equation}
\mathcal U_{\mathcal E,\ell}\widehat u \simeq 
\Pi_{\ell}(\lambda)D^{(0)}(\lambda + i0)
\widehat{\mathcal F}^{(+)}\big(r\big|_{\mathcal V}\big)^{\ast}\chi_{\Sigma}B^{(\pm)}_{\Sigma}(\lambda)\chi_{\Sigma}\widehat{\mathcal F}^{(-)}(\lambda)^{\ast}\phi..
\nonumber
\end{equation}
This implies
\begin{equation}
A_{ext}(\lambda) - A(\lambda) = \widehat{\mathcal F}^{(+)}(\lambda)\big(r\big|_{\mathcal V}\big)^{\ast}\chi_{\Sigma}B^{(\pm)}_{\Sigma}(\lambda)\chi_{\Sigma}\widehat{\mathcal F}^{(-)}(\lambda)^{\ast}\phi.
\nonumber
\end{equation}
Let us note here that $1 = B^{(-)}_{\Sigma}(\lambda)M^{(-)}_{\Sigma}(\lambda)$ by virtue of Lemma \ref{MSigmatimesBsimga=1Lemma}. Since $M^{(\pm)}_{\Sigma}(\lambda) = 
\chi_{\Sigma}r\big|_{\mathcal V}\widehat R(\lambda \pm i0)\big(r\big|_{\mathcal V}\big)^{\ast}\chi_{\Sigma}$, we have $(M^{(-)}_{\Sigma}(\lambda))^{\ast} = M^{(+)}_{\Sigma}(\lambda)$, which implies 
\begin{equation}
1 = M^{(+)}_{\Sigma}(\lambda)\big(B^{(-)}_{\Sigma}(\lambda)\big)^{\ast}.
\label{1=M+B-ast}
\end{equation}
Inserting (\ref{1=M+B-ast}) 
between  $B^{(+)}_{\Sigma}(\lambda)$ and $\widehat{\mathcal F}^{(-)}(\lambda)^{\ast}\phi$, we obtain
\begin{equation}
\begin{split}
& \widehat{\mathcal F}^{(+)}(\lambda)
\big(r\big|_{\mathcal V}\big)^{\ast}\chi_{\Sigma}B^{(+)}\chi_{\Sigma}r\big|_{\mathcal V}\widehat{\mathcal F}^{(-)}(\lambda)^{\ast} \\
&= \widehat{\mathcal F}^{(+)}(\lambda)\big(r\big|_{\mathcal V}\big)^{\ast}\chi_{\Sigma}B^{(+)}_{\Sigma}(\lambda)M^{(+)}_{\Sigma}(\lambda)\big(B^{(-)}_{\Sigma}(\lambda)\big)^{\ast}\chi_{\Sigma}r\big|_{\mathcal V}\widehat{\mathcal F}^{(-)}(\lambda)^{\ast} \\
&= \widehat I^{(+)}(\lambda)\big(r\big|_{\mathcal V}\big)^{\ast}M^{(+)}_{\Sigma}(\lambda)r\big|_{\mathcal V}\widehat I^{(-)}(\lambda)^{\ast}.
\end{split}
\nonumber
\end{equation}
We have thus proven Theorem \ref{Aext-A=IBSigma-1ILemma}. \qed


\subsection{The operator $\widehat J^{(\pm)}(\lambda)$}
To construct $A(\lambda)$ from $B^{(+)}_{\Sigma}(\lambda)$, we need to invert $\widehat I^{(\pm)}(\lambda)$ and its adjoint.
 To compute them, we first construct a solution $\widehat u^{(\pm)}_{ext}$ to the exterior Dirichlet problem satisfying (\ref{EdgeEquationExterior}) 
and the radiation condition in the form 
$\widehat R^{(0)}_{\mathcal E}(\lambda \pm i0)\widehat \psi$, where $\widehat\psi \in \ell^2(\Sigma)$. Then it is the desired solution if and only if
\begin{equation}
\widehat R^{(0)}_{\mathcal E}(\lambda \pm i0)\widehat \psi = \widehat f \quad {\rm on} \quad \Sigma.
\label{S4BoundaryIntEq}
\end{equation}
Suppose $\widehat R^{(0)}_{\mathcal E}(\lambda \pm i0)\widehat\phi^{(\pm)}=0$ on $\Sigma$. Then, $\widehat v^{(\pm)} = \widehat R^{(0)}_{\mathcal E}(\lambda \pm i0)\widehat\phi^{(\pm)}$ is the solution to the equation (\ref{EdgeEquationExterior})  with 0 boundary data. Since $\widehat v^{(\pm)}$ satisfies the radiation condition, and Lemma \ref{LemmaRadCondUnique} holds also for  the exterior problem, it vanishes identically in ${\mathcal E}_{ext}$ hence on all $\mathcal E$. It then follows that $\widehat\phi^{(\pm)} =0$. Therefore, the equation (\ref{S4BoundaryIntEq}) is uniquely solvable for any $\widehat f \in \ell^2(\Sigma)$. Let $\widehat\psi = \widehat r_{\Sigma}^{(\pm)}(\lambda)\widehat f$ be the solution. 
Then, we have
\begin{equation}
\widehat u^{(\pm)}_{ext} = \widehat R^{(0)}_{\mathcal E}(\lambda \pm i0)\widehat r_{\Sigma}^{(\pm)}(\lambda)\widehat f, 
\label{S4Solformulauext}
\end{equation}
which is a potential theoretic solution to the boundary value problem (\ref{EdgeEquationExterior}).

Let $\widehat g_n, n = 1,\cdots,N$, be a basis of $\ell^2(\Sigma)$ and put
\begin{equation}
\begin{split}
v^{(\pm)}_n & = \widehat I^{(\pm)}(\lambda)\widehat g_n \\
& = 
\mathcal F^{(0)}(\lambda)\mathcal U_{\mathcal E}(\widehat H^{(0)}_{\mathcal E} -\lambda)\widehat P_{ext}\widehat R^{(0)}_{\mathcal E}(\lambda \pm i0)\widehat r_{\Sigma}^{(\pm)}(\lambda)\widehat g_n \in {\bf h}_{\lambda}.
\end{split}
\nonumber
\end{equation}
Let $\mathcal M_{\Sigma}^{(\pm)}$ be the linear hull of $v^{(\pm)}_1,\cdots,v^{(\pm)}_N$. Then, the mapping $\widehat g_n \to v_n^{(\pm)}$ induces a bijection 
$$
\widehat J^{(\pm)}(\lambda) : \ell^2(\Sigma) \ni \sum_{n=1}^Nc_n\widehat g_n \to \sum_{n=1}^Nc_nv^{(\pm)}_n \in \mathcal M_{\Sigma}^{(\pm)}.
$$
In view of  Theorem 
\ref{Aext-A=IBSigma-1ILemma}, we have the following theorem.


\begin{theorem} \label{LemmaAlambdatoBSigmalambda}
The following equality holds:
\begin{equation}
\big(B^{(+)}_{\Sigma}(\lambda)\big)^{-1} = r\big|_{\mathcal V}
\big(\widehat J^{(+)}(\lambda)\big)^{-1}\big( A_{ext}(\lambda) - A(\lambda)\big)
\big(\widehat J^{(-)}(\lambda)^{\ast}\big)^{-1}\big(r\big|_{\mathcal V}\big)^{\ast}.
\nonumber
\end{equation}
\end{theorem}

Theorems \ref{Aext-A=IBSigma-1ILemma} and \ref{LemmaAlambdatoBSigmalambda} imply that the S-matrix and the D-N map determine each other.


\section{Inverse scattering}
\label{SectionInversescattering}
\subsection{Hexagonal parallelogram}
We are now in a position to considering the inverse scattering problem. 
As was discussed in Subsection \ref{ExampleHexalattice},
 the choice of  fundamental domain of the lattice $\mathcal L$ is not unique. However,  different choice gives rise to unitarily equivalent Hamiltonians. 
 In this section, we take  $p^{(1)}, p^{(2)}$ and ${\bf v}_1, {\bf v}_2$ as in (\ref{S2vandpinthepreviouspaper}) in order to make use of our previous results in \cite{AnIsoMo17},  \cite{AnIsoMo17(1)}. 
We identify ${\bf R}^2$ with ${\bf C}$, and put
$$
\omega = e^{\pi i/3}.
$$
 For $n = n_1 + i n_2 \in {\bf Z}[i]= {\bf Z} + i{\bf Z}$, 
let
$$
\mathcal L_0 = \left\{{\bf v}(n)\, ; \, n \in {\bf Z}[i]\right\}, \quad
{\bf v}(n) = n_1{\bf v}_1 + n_2{\bf v}_2,
$$
$$
{\bf v}_1 = 1 + \omega, \quad {\bf v}_2 = \sqrt3 i,
$$
$$
p_1 = \omega^{-1} = \omega^5, \quad p_2 = 1,
$$
and define the vertex set $\mathcal V_0$ by
$$
\mathcal V_0 = \mathcal V_{01} \cup \mathcal V_{02}, \quad \mathcal V_{0i} = p_i + \mathcal L_0.
$$
The adjacent points of $a_1 \in \mathcal V_{01}$ and $a_2 \in \mathcal V_{02}$ are given by
\begin{equation}
\begin{split}
\mathcal N_{a_1} &= \{z \in {\bf C}\, ; \, |a_1 - z|=1\}\cap \mathcal V_{02} \\
&= \left\{a_1 + \omega, a_1 + \omega^{3}, a_1 + \omega^5\right\},
\end{split}
\nonumber
\end{equation}
\begin{equation}
\begin{split}
\mathcal N_{a_2} &= \{z \in {\bf C}\, ; \, |a_2 - z|=1\}\cap \mathcal V_{01} \\
&= \left\{a_2 + 1, a_2 + \omega^{2}, a_2 + \omega^4\right\}.
\end{split}
\nonumber
\end{equation}

By virtue of the formula (\ref{OperatorBSigmalambda}) and Theorem \ref{LemmaAlambdatoBSigmalambda}, given an S-matrix and a bounded domain $\mathcal E_{int}$, we can compute the D-N map 
associated with $\mathcal E_{int}$. Let $\mathcal V_{int}$ be the set of the vertices in $\mathcal E_{int}$. By Lemma \ref{LemmaDNedge=DNvertex}, we can compute the D-N map associated with $\mathcal V_{int}$. The problem is now reduced to the reconstruction of the potentials on the edges from the knowledge of the D-N map for the vertex Schr{\"o}dinger operator.

\begin{figure}[hbtp]
\includegraphics[width=9cm, bb=0 0 664 676]{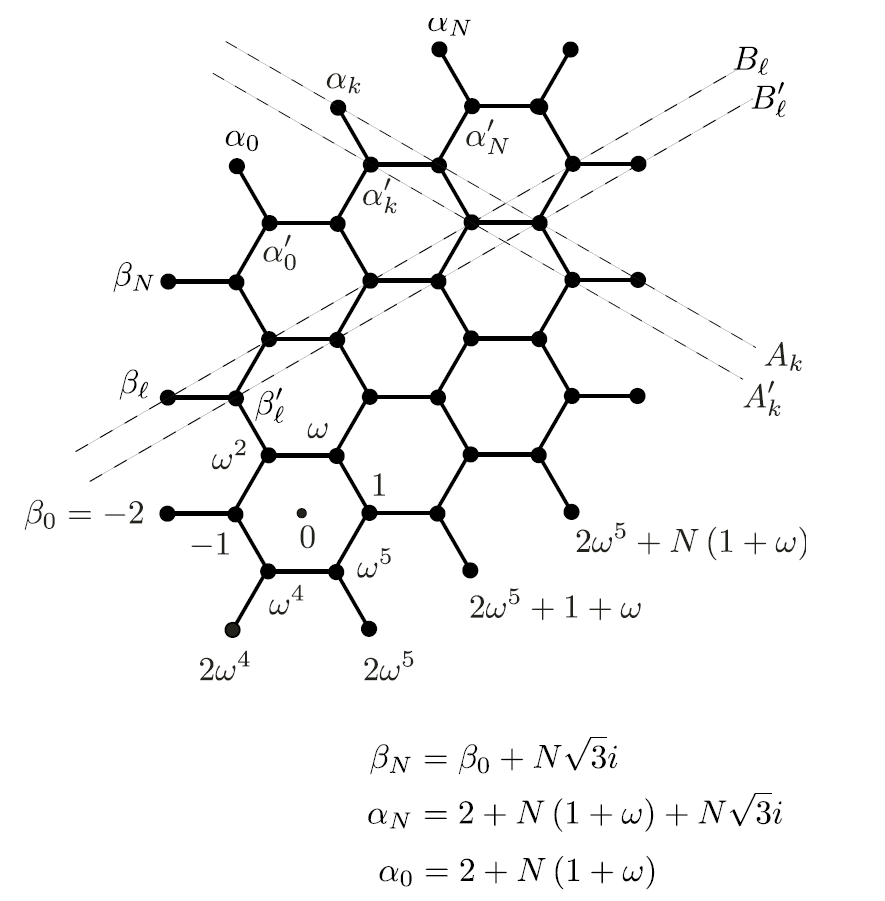}
\caption{Hexagonal parallelogram ($N = 2$)}
\label{S6HexaParallel}
\end{figure}

As $\mathcal V_{int}$, we use the following domain which is different from the one in Figure \ref{BoundaryHexagonal}.
Let $\mathcal D_0$ be the Wigner-Seitz cell of $\mathcal V_0$.
 It is a hexagon
 having 6 vertices 
 $\omega^k,\ 0 \leq k \leq 5$,
 with center at the origin. Take $D_N = \{n \in {\bf Z}[i]\, ; \, 0 \leq n_1 \leq N, \  0 \leq n_2 \leq N\}$, where $N$ is chosen large enough,  and put
 $$
 \mathcal D_N = {\mathop\cup_{n \in D_N}}\Big( \mathcal D_0 + {\bf v}(n)\Big).
 $$
 This is a parallelogram in the hexagonal lattice (see Figure \ref{S6HexaParallel}).
The interior angle of each vertex on the periphery of $\mathcal D_N$ is either $2\pi/3$ or $4\pi/3$. Let $\mathcal A$ be the set of the former, and for  each $z \in \mathcal A$, we assign a new edge $ e_{z,\zeta}$, and a new vertex $\zeta = t(e_{z,\zeta})$ on its terminal point, hence $\zeta$ is in the outside of $\mathcal D_N$. 
Let 
$$
\Omega = \{v \in \mathcal V_0\, ; \, v \in \mathcal D_N\}
$$
 be the set of vertices in the inside of the resulting graph.
The boundary $\partial\Omega = \{t(e_{z,\zeta})\, ; \, z \in \mathcal A\}$ is  divided into 4 parts,   called  top, 
 bottom, right,  left sides, which are denoted by 
$(\partial\Omega)_T, (\partial\Omega)_B, (\partial\Omega)_R, (\partial\Omega)_L$, i.e. 
\begin{gather*}
\begin{split}
 (\partial\Omega)_T =& \{ \alpha_0 , \cdots , \alpha_N \}, \\
 (\partial\Omega)_B = & \{2\omega^5 + k(1 + \omega)\,  ; \, 0 \leq k \leq N\}, \\
 (\partial\Omega)_R = & \{ 2+ N(1+\omega ) + k\sqrt{3} i \, ; \, 1 \leq k \leq N \} \cup \{ 2+N(1+\omega ) +N\sqrt{3} i + 2 \omega ^2  \} , \\
 (\partial\Omega)_L =& \{2\omega^4\}\cup\{\beta_0,\cdots,\beta_N\} , 
\end{split}
\end{gather*}
where $ \alpha_k = \beta _N + 2\omega + k (1+\omega ) $ and $ \beta _k = -2 + k\sqrt{3} i $ for $ 0\leq k \leq N$.

\subsection{Special solutions to the vertex Schr{\"o}dinger equation}
Taking $N$ large enough so that $\mathcal D_N$ contains all the supports of the potentials $q_{\bf e}(z)$ in its interior, 
we consider the following Dirichlet problem for the vertex Schr{\"o}dinger equation
\begin{equation}
\left\{
\begin{split}
& ( - \widehat{\Delta}_{\mathcal V,\lambda} + \widehat Q_{\mathcal V,\lambda})\widehat u = 0, \quad 
{\rm in} \quad \stackrel{\circ}{{\Omega}}, \\
& \widehat u = \widehat f, \quad {\rm on} \quad \partial \Omega.
\end{split}
\right.
\end{equation}
Let  ${\bf\Lambda_{\widehat{\bf Q}}}$ be the associated D-N map. 
The key to the inverse procedure is the following partial data problem.


\begin{lemma}\label{S6partialDNdata}
(1) Given a partial Dirichlet data $\widehat f$ on $\partial\Omega\setminus(\partial \Omega)_R$, and a partial Neumann data $\widehat g$ on $(\partial\Omega)_L$, there is a unique solution $\widehat u$ on $\stackrel{\circ}\Omega \cup (\partial\Omega)_R$ to the equation
\begin{equation}
\left\{
\begin{split}
& (- \widehat\Delta_{\mathcal V,\lambda} + \widehat Q_{\mathcal V,\lambda})\widehat u = 0, \quad {\it in} \quad \stackrel{\circ}\Omega,\\
& \widehat u =\widehat f, \quad {\it on} \quad \partial\Omega\setminus(\partial\Omega)_R, \\
& \partial_{\nu}^{\mathcal D_N}\widehat u = \widehat g, \quad {\it on} \quad (\partial\Omega)_L.
\end{split}
\right.
\label{Lemma61Equation}
\end{equation}
\noindent
(2) Given the D-N map ${\bf\Lambda_{\widehat{\bf Q}}}$, a partial Dirichlet data $\widehat f_2$ on $\partial\Omega\setminus(\partial\Omega)_R$ and a partial Neumann data $\widehat g$ on $(\partial\Omega)_{L}$, there exists a unique $\widehat f$ on $\partial\Omega$ such that $\widehat f = \widehat f_2$ on $\partial\Omega\setminus(\partial\Omega)_R$ and ${\bf\Lambda_{\widehat{\bf Q}}}\widehat f = \widehat g$ on $(\partial\Omega)_{L}$.
\end{lemma}

For the proof, see \cite{AnIsoMo17(1)}, Lemma 6.1.

\begin{figure}[hbtp]
\centering
\includegraphics[width=8cm, bb=0 0 470 482]{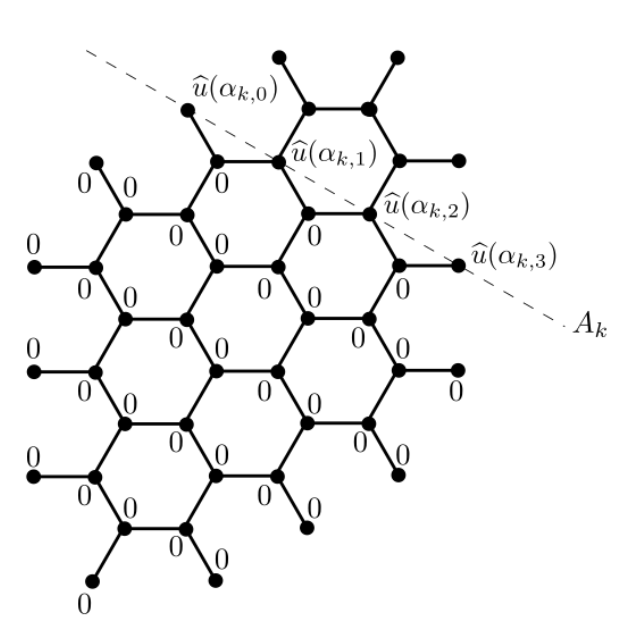}
\caption{Line $A_k$}
\label{LineAk}
\end{figure}

Now, for $0 \leq k \leq N$, let us consider a {\it diagonal} line $A_k$ (see  Figure \ref{LineAk})  :
\begin{equation}
A_k = \{x_1 + ix_2\, ; \, x_1 + \sqrt3x_2 = a_k\},
\label{S6Dinagonalline}
\end{equation}
where $a_k$ is chosen so that $A_k$ passes through
\begin{equation}
\alpha_k =  \alpha_0 + k (1+ \omega )  \in 
(\partial\Omega)_T.
\end{equation} 
The vertices on $A_k\cap\Omega$ are written as
\begin{equation}
\alpha_{k,\ell} = \alpha_k + \ell (1 + \omega^5 ), \quad \ell = 0, 1, 2, \cdots.
\end{equation}

\begin{lemma}
\label{Lemmaspecialboundarydata}
Let $A_k \cap\partial\Omega = \{\alpha_{k,0},\alpha_{k,m}\}$. Then, there exists a unique solution $\widehat u$ to the equation
\begin{equation}
\big(- \widehat\Delta_{\mathcal V,\lambda} + \widehat Q_{\mathcal V,\lambda}\big)\widehat u = 0 \quad {\it in} \quad \stackrel{\circ}\Omega,
\label{S7BVP}
\end{equation}
with partial Dirichlet data $\widehat f$ such that
\begin{equation}
\left\{
\begin{split}
& \widehat f(\alpha_{k,0}) = 1, \\
& \widehat f(z) = 0 \quad {\it for} \quad 
z \in \partial\Omega\setminus\big((\partial\Omega)_R
\cup\alpha_{k,0}\cup\alpha_{k,m}\big)
\end{split}
\right.
\end{equation}
and partial Neumann data $\widehat g = 0$ on $(\partial\Omega)_L$. It satisfies
\begin{equation}
\widehat u(x_1 + ix_2) = 0 \quad {\it if} \quad 
x_1 + \sqrt3 x_2 < a_k.
\label{S7Lemmu=0belowAk}
\end{equation}
\end{lemma}

An important feature is that $\widehat u$ vanishes below the line $A_k$. By using this property, we reconstructed the vertex potentials and defectes of the hexagonal lattice in \cite{AnIsoMo17(1)}. 
We make use of the same idea.

\begin{figure}[hbtp]
\centering
\includegraphics[width=8cm, bb= 0 0 579 425]{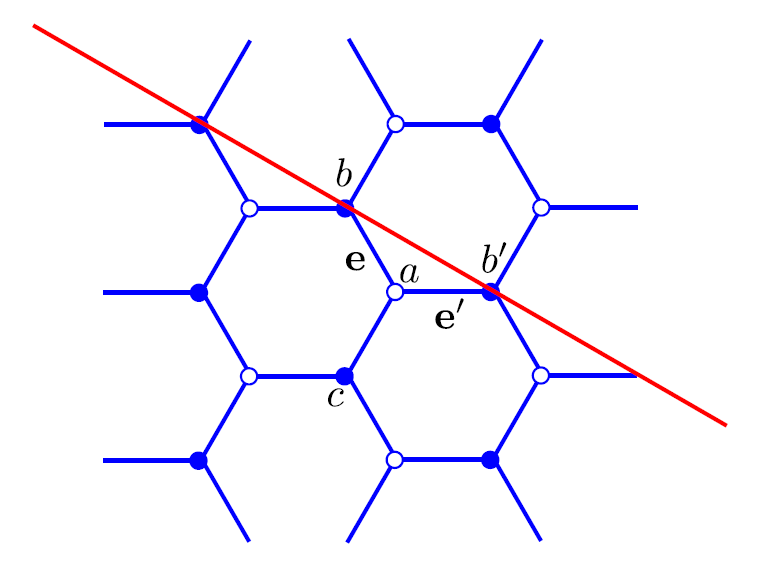}
\caption{$\widehat u(b)$ and $\widehat u(b')$}
\label{S7hex_edge}
\end{figure}

Let $\widehat u$ be a solution of the equation
\begin{equation}
 (- \widehat\Delta_{\mathcal V,\lambda} + \widehat Q_{\mathcal V,\lambda})\widehat u = 0, \quad {\rm in} \quad \stackrel{\circ}\Omega,
\label{S7Equationtobeevaluated}
\end{equation}
which vanishes in the region $x_1 + \sqrt{3}x_2 < a_k$. Let $a, b, b', c \in \mathcal V$ and ${\bf e}, {\bf e}' \in \mathcal E$ be as in  Figure
\ref{S7hex_edge}. 
Then, evaluating the equation (\ref{S7Equationtobeevaluated}) at $v = a$ and using (\ref{S3DefineVertexLaplacian}), (\ref{S3DefinewidehatQVlambda}),
we obtain
\begin{equation}
\frac{1}{\psi_{ba}(1,\lambda)}\widehat u(b) + \frac{1}{\psi_{b'a}(1,\lambda)}\widehat u(b') = 0.
\label{S7equationbypsiwv}
\end{equation}
Here, for any edge ${\bf e} \in \mathcal E$, we associate an edge $[{\bf e}]$ without orientation and a function $\phi_{[{\bf e}]}(z,\lambda)$ satisfying
\begin{equation}
\left\{
\begin{split}
& \left(- \frac{d^2}{dz^2} + q_{{\bf e}}(z) - \lambda\right)\phi_{[{\bf e}]}(z,\lambda) = 0, \quad {\rm for} \quad  0 < z < 1,\\
& \phi_{[{\bf e}]}(0,\lambda) = 0, \quad \phi_{[{\bf e}]}'(0,\lambda) = 1.
\end{split}
\right.
\nonumber
\end{equation}
By the assumption (Q-3), $\phi_{[{\bf e}]}(z,\lambda)$ is determined by ${\bf e}$ and  independent of the orientation of ${\bf e}$. 
Then, the equation (\ref{S7equationbypsiwv}) is rewritten as
\begin{equation}
\widehat u(b) = - \frac{\phi_{[{\bf e}]}(1,\lambda)}{\phi_{[{\bf e}']}(1,\lambda)}\widehat u(b').
\label{S7ubandub'}
\end{equation}
Let ${\bf e}_{k,1}, {\bf e}'_{k,1}, {\bf e}_{k,2}, {\bf e}'_{k,2}, \cdots$ be the series of edges just below $A_k$ starting from the vertex $\alpha_k$, and put
\begin{equation}
f_{k,m}(\lambda) = - \frac{\phi_{[{\bf e}_{k,m}]}(1,\lambda)}{\phi_{[{\bf e}'_{k,m}]}(1,\lambda)}.
\label{s7equationforfkm}
\end{equation}
Then, we obtain the following lemma.

\begin{lemma}
The solution $\widehat u$ in  Lemma \ref{Lemmaspecialboundarydata} satisfies
$$
\widehat u(\alpha_{k,\ell}) = f_{k,1}(\lambda)\cdots f_{k,\ell}(\lambda).
$$
\end{lemma}

\subsection{Reconstruction procedure}
We now prove Theorem \ref{Maintheorem1} by showing the reconstruction algorithm of the potential $q_{\bf e}(z)$.

\medskip
\noindent
{\it 1st step}. We first take a sufficiently large hexagonal parallelogram $\Omega$ as in Figure \ref{S6HexaParallel} which contains all the supports of the potential $q_{\bf e}(z)$.

\medskip
\noindent
{\it 2nd step}. For an arbitrary $k$, draw a line $A_k$ as in  Figure \ref{LineAk} and take the boundary data $\widehat f$ having the properties in Lemma \ref{Lemmaspecialboundarydata}. 

\medskip
\noindent
{\it 3rd step}. Compute the values of the associated solution $\widehat u$ to the boundary value problem in Lemma \ref{Lemmaspecialboundarydata} at the points $\alpha_{k,\ell}$, $\ell = 0, 1, 2, \cdots$.

\medskip
\noindent
{\it 4th step}. Look at Figure \ref{S6HexaParallel}. Two edges ${\bf e}$ and ${\bf e'}$ between $A_k$ and $A_k'$ are said to be $A_k'$-adjacent if they have a vertex in common on $A_k'$ (see Figure \ref{S7hex_edge}).
 Take two $A_k'$-adjacent edges ${\bf e}$ and ${\bf e}'$ between $A_k$ and $A_k'$, and use the formula (\ref{s7equationforfkm}) to compute the ratio of $\phi_{[{\bf e}]}(1,\lambda)$ and $\phi_{[{\bf e}']}(1,\lambda)$.

\medskip
\noindent
{\it 5th step}. Rotate the whole system by the angle $\pi$ and take a hexagonal parallelogram congruent to the previous one. Then, the roles of $A_k$ and $A_k'$ are exchanged. One can then compute the  ratio of 
$\phi_{[{\bf e}]}(1,\lambda)$ and $\phi_{[{\bf e}']}(1,\lambda)$ for $A_k'$-adjacent pairs in the sense after the rotation, which are $A_k$-adjacent before the rotation.

\medskip
After the 4th and 5th steps, for all pairs ${\bf e}$ and ${\bf e}'$ which are either $A_k$-adjacent or $A_k'$-adjacent, one has computed the ratio of 
$\phi_{[{\bf e}]}(1,\lambda)$ and $\phi_{[{\bf e}']}(1,\lambda)$.

\begin{figure}[htbp]
  \centering
   \includegraphics[width=90mm, bb=0 0 633 440]{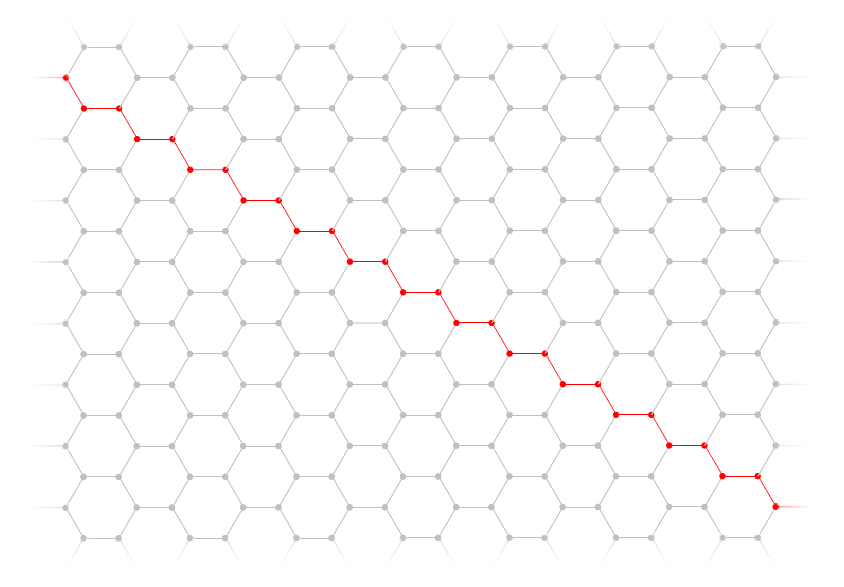}
  \caption{Zigzag line in the hexagonal lattice}
  \label{fig:line2}
\end{figure}

\medskip
\noindent
{\it 6th step}. Take a zigzag line on the hexagonal lattice (see Figure \ref{fig:line2}), and take any two edges ${\bf e}$ and ${\bf e}'$ on it. They are between $A_k$ and $A_k'$ for some $k$. Then, using the 4th and 5th steps, one can compute the ratio of $\phi_{[{\bf e}]}(1,\lambda)$ and $\phi_{[{\bf e}']}(1,\lambda)$ by computing the ratio for two successive edges between ${\bf e}$ and ${\bf e}'$.

\medskip
\noindent
{\it 7th step}.  For a sufficiently remote edge ${\bf e}'$, one knows $\phi_{[{\bf e}']}(1,\lambda)$ since $q_{{\bf e}'}(z) = 0$ on ${\bf e}'$. One can thus compute $\phi_{[{\bf e}]}(1,\lambda)$ for any edge ${\bf e}$.
Then, by the analytic continuation, one can compute the zeros of $\phi_{[{\bf e}]}(1,\lambda)$ for any edge ${\bf e}$. 

\medskip
\noindent
{\it 8th step}. Note that the zeros of  $\phi_{[{\bf e}]}(1,\lambda)$ are the Dirichlet eigenvalues for the operator $- (d/dz)^2 + q_{\bf e}(z)$ on $(0,1)$. Since the potential is symmetric, by Borg's theorem (see e.g. \cite{PoTru}, p. 117) these eigenvalues determine the potential $q_{\bf e}(z)$. 

\medskip
We have now completed the proof of Theorem \ref{Maintheorem1}.

\medskip
Note that for the 1st step, we need a-priori knowledge of the size of the support of the potential $q_{\mathcal E}(z)$. The knowledge of the D-N map is used in the 2nd step (in the proof of  Lemma \ref{S6partialDNdata}). In the 3rd step, one uses the equation (\ref{S7BVP}) and the fact that $\widehat u = 0$ below $A_k$.

\medskip
The proof of Theorem \ref{Maintheorem2} requires no essential change. Instead of $\frac{\sin\sqrt{\lambda}z}{\sqrt{\lambda}}$ and $\frac{\sin\sqrt{\lambda}(1-z)}{\sqrt{\lambda}}$, we have only to use the corresponding solutions to the Schr{\"o}dinger equation 
$\big(- (d/dz)^2 + q_0(z) - \lambda)\varphi = 0$.



\begin{thebibliography}{99}


\bibitem{AgHo76}
S. Agmon and L. H\"ormander, \textit{Asymptotic properties of solutions of differential
equations with simple characteristics}, J. d'Anal. Math., \textbf{30} (1976),
1-38.

\bibitem{Ando12}
K. Ando, \textit{Inverse scattering theory for discrete Schr{\"o}dinger operators on the hexagonal lattice}, Ann.  Henri Poincar{\'e} \textbf{14} (2013), 347-383.


\bibitem{AnIsKorMo20}
K. Ando, H. Isozaki, E. Korotyaev and H. Morioka, \textit{Inverse scattering on the quantum graph for graphene}, preprint (2020).

\bibitem{AnIsoMo17}
K. Ando, H. Isozaki and H. Morioka, \textit{Spectral properties of Schr{\"o}dinger operators on perturbed lattices}, Ann. Henri Poincar{\'e} \textbf{17} (2016), 2103-2171.

\bibitem{AnIsoMo17(1)}
K. Ando, H. Isozaki and H. Morioka, \textit{Inverse scattering for Schr{\"o}dinger operators on perturbed lattices}, Ann. Henri Poincar{\'e} \textbf{19} (2018), 3397-3455.

\bibitem{AnIsoMo17(2)}
K. Ando, H. Isozaki and H. Morioka, \textit{Correction to : Inverse scattering for Schr{\"o}dinger operators on perturbed lattices}, Ann. Henri Poincar{\'e} \textbf{20} (2019), 337-338.

\bibitem{AvdBelMat11}
S. Avdonin, B. P. Belinskiy, and J. V. Matthews, \textit{Dynamical inverse problem on a metric tree}, Inverse Porblems \textbf{27} (2011),  075011.

\bibitem{Bel04}
M. I. Belishev, \textit{Boundary spectral inverse  problem on a class of graphs (trees) by the BC method}, Inverse Problems \textbf{20} (2004), 647-672.


\bibitem{Below85}
J. von Below, \textit{A characteristsic equation associated to an eigenvalue problem on $c^2$-networks}, Linear Algebra Appl. \textbf{71} (1985), 309-325.

\bibitem{BK13}
G. Berkolaiko and P. Kuchment, \textit{Introduction to Quantum Graphs}, Mathematical Surveys and Mnonographs \textbf{186}, AMS (2013).

\bibitem{Borg}
G. Borg, \textit{Eine Umkehrung der Sturm-Liouvilleschen Eigenwertaufgabe. Bestimmung der Differentialgleichung durch die Eigenwerte}, Acta Math. \textbf{78} (1946), 1-96.

\bibitem{BroWei09}
B. M. Brown and R. Weikard, \textit{A Borg-Levinson theorem for trees}, Proc. Royal Soc. Lond. Ser. A Math. Phys. Eng. Sci. \textbf{461} 2062  (2005), 3231-3243.  

\bibitem{BPG08}
J. Br{\"u}ning, V. Geyley and K. Pankrashkin, \textit{Spectra of self-adjoint extensions and applications to solvable Schr{\"o}dinger operators}, Rev. Math. Phys. \textbf{20} (2008), 1-70.

\bibitem{C97}
C. Cattaneo, \textit{The spectrum of the continuous Laplacian on a graph}, Monatsh. Math. \textbf{124} (1997), 215-235.

\bibitem{ChExTu11}
T. Chen, P. Exner and O. Turek, \textit{Inverse scattering for quantum graph vertices}, Phys. Rev. A (2011), 86:062715.


\bibitem{Ch97}
F. Chung, \textit{Spectral Graph Theory}, AMS. Providence, Rhodse Island (1997).

\bibitem{Col98}
Y. Colin de Verdi{\`e}re, \textit{Spectre de graphes}, Cours sp{\'e}cialis{\'e}s \textbf{4}, S. M. F., Paris, (1998).

\bibitem{CurtMor00}
E. B. Curtis and J. A. Morrow, \textit{Inverse Problems for Electrical Networks}, On Applied Mathematics, World Scientific, (2000).

\bibitem{CDGT88}
D. Cvetkovic, M. Doob, I. Gutman and A. Torgasev, \textit{A recent result in the theory of graph spectra}, Annals of Discrete Mathematics \textbf{36}, North-Holland Publishing Co., Amsterdam (1988).

\bibitem{CDS95} D. Cvetkovic, M. Doob and H. Saks, \textit{Spectra of graphs, Theory and applications}, 3rd edition, Johann Ambrosius Barth, Heidelberg (1995).



\bibitem{EKMN17}
P. Exner, A. Kostenko, M. Malamud and H. Neidhardt, \textit{Spectral theory for infinite quantum graph}, Ann. Henri Poincar{\'e} \textbf{19} (2018), 3457-3510.

\bibitem{Es} 
M. S. Eskina, \textit{The direct and the inverse scattering problem for a partial difference equation}, Soviet Math. Doklady, \textbf{7} (1966), 193-197. 


\bibitem{GutSmil01}
B. Gutkin and U. Smilansky, \textit{Can one hear the shape of a graph?} J. Phys. A \textbf{34} (2001), 6061-6068.

\bibitem{IsKo12} H. Isozaki and E. Korotyaev, \textit{Inverse problems, trace formulae for discrete Schr\"{o}dinger operators},   Ann.  Henri Poincar{\'e}, \textbf{13} (2012), 751-788.

\bibitem{IsMo}
H. Isozaki and H. Morioka, \textit{Inverse scattering at a fixed energy for discrete Schr{\"o}dinger operators on the square lattice}, Ann. l'Inst. Fourier \textbf{65} (2015), 1153-1200.

\bibitem{KoLo07}
E. Korotyaev and I. Lobanov, \textit{Schr{\"o}dinger operators on zigzag nanotubes}, Ann. Henri Poincar{\'e} \textbf {8} (2007), 1151-1076.

\bibitem{KoSa14}
E. Korotyaev and N. Saburova, \textit{Schr{\"o}dinger operators on periodic discrete graphs}, J. Math. Anal. Appl. \textbf{420} (2014), 576-611. 

\bibitem{KoSa15a}
E. Korotyaev and N. Saburova, \textit{Spectral band localization for Schr{\"o}dinger operators on periodic  graphs}, Proc. Amer. Math. Soc. \textbf{143} (2015), 3951-3967. 

\bibitem{KoSa15}
E. Korotyaev and N. Saburova, \textit{Scattering on metric graphs}, arXiv:1507.06441v1 [math.SP] 23 Jul 2015.


\bibitem{KoSa15b}
E. Korotyaev and N. Saburova, \textit{Estimates of bands for Laplacians on periodic equilateral metric graphs},  Proc. Amer. Math. Soc. \textbf{114} (2016), 1605-1617.

\bibitem{KoSa15c}
E. Korotyaev and N. Saburova, \textit{Effective masses for Laplacians on periodic graphs}, J. Math. Anal. Appl. \textbf{436} (2016), 104-130.

\bibitem{KostSchr99}
V. Kostrykin and R. Schrader, \textit{Kirchhoff's rule for quantum wires}, J. Phys. A \textbf{32} (1999), 595-630.


\bibitem{Kuch13}
P. Kuchment, \textit{Quantum graph spectra of a graphyne structure}, NanoNMTA, \textbf{2} (2013), 107-123.

\bibitem{KuchPost}
P. Kuchment and O. Post, \textit{On the spectra of carbon nano-structures}, arXiv:math-ph/0612021v4.

\bibitem{KuVa}
P. Kuchment and B. Vainberg, \textit{On absence of embedded eigenvalues for Schr{\"o}dinger operators with perturbed periodic potentials}, Comm. PDE, \textbf{25} (2000), 1809-1826.

\bibitem{Kura08}
P. Kurasov, \textit{Schr{\"o}dinger operators on graphs and geometry. I. Essentially bounded potentials}, J. Funct. Anal. \textbf{254} (2008), 934-953.

\bibitem{Lev49}
N. Levinson, \textit{The inverse Sturm-Liouville problem}, Mat. Tidsskr. B. (1949), 25-30.

\bibitem{MochiTroosh12}
K. Mochizuki and I. Yu. Trooshin, 
\textit{On the scattering on a loop-shaped graph}, 
Progress of Math. \textbf{301} (2012), 227-245.

\bibitem{Nakamura14}
S. Nakamura, \textit{Modified wave operators for discrete Scr{\"o}dinger operators with long-range perturbations}, J. Math. Phys. \textbf{55} (2014), 112101.

\bibitem{Niikuni16}
H. Niikuni, \textit{Spectral band structure of periodic Schr{\"o}dinger operators with two potentials on the degenerate zigzag nanotube}, J. Appl. Math. Comput. (2016) 50:453-482.

\bibitem{Pank06}
K. Pankraskin, \textit{Spactra of Schr{\"o}dinger operators on equilateral quantum graphs}, Lett. Math. Phys. \textbf{77} (2006), 139-154.

\bibitem{ParRich18}
D. Parra and S. Richard, \textit{Spectral and scattering theory for Schr{\"o}dinger operators on perturbed topological crystals}, 
Rev. Math. Phys. \textbf{30} (2018), Article No. 1850009, pp 1-39.

\bibitem{Paul36}
L. Pauling, \textit{The diamagentic anisotropy of aromatic molecules}, J. Chem, Phys. \textbf{4} (1936), 673-674.


\bibitem{Pivo00}
V. Pivovarchik, \textit{Inverse problem for the Sturm-Liouville equation on a simple graph}, SIAM J. Math. Anal. \textbf{32} (2000), 801-819.

\bibitem{Post12}
O. Post, \textit{Spectal Analysis on Graph-like Spaces}, Lecture Notes in Mathematics \textbf{2039}, Springer, Heidelberg (2012).


\bibitem{PoTru}
J. P{\"o}schel and E. Trubowitz, \textit{Inverse Spectral Theory}, 
Academic Press, Boston, (1987).


\bibitem{Sha} W. Shaban and B. Vainberg, \textit{Radiation conditions for the difference Schr\"{o}dinger operators}, Applicable Analysis, \textbf{80} (2001), 525-556.

\bibitem{Tadano16}
Y. Tadano, \textit{Long-range scattering for discrete Schr{\"o}dinger operators}, Ann. Henri Poincar{\'e} \textbf{20} (2019), 1439-1469.

\bibitem{VisComMirSor11}
F. Visco-Comandini, M. Mirrahimi, and M. Sorine, \textit{ Some inverse scattering problems on star-shaped graphs}, J. Math. Anal. Appl. \textbf{387}  (2011), 343-358.

\bibitem{Yurk05}
V. Yurko, \textit{Inverse spectral problems for Sturm-Liouville operators on graphs}, Inverse Problems \textbf{21} (2005), 1075-1086.

\end{thebibliography}
\end{document}